\documentclass[a4paper,fleqn,usenatbib,useAMS]{mnras}

\usepackage{graphicx}
\usepackage{dcolumn}
\usepackage{color}
\usepackage{booktabs}
\usepackage{amsmath}
\usepackage{amssymb}
\usepackage{slashed}

\usepackage[T1]{fontenc}
\usepackage{ae,aecompl}

\usepackage{eso-pic}% http://ctan.org/pkg/eso-pic

\usepackage{txfonts}
\newcommand{\be}{\begin{equation}}
\newcommand{\ee}{\end{equation}}
\newcommand{\ba}{\begin{eqnarray}}
\newcommand{\ea}{\end{eqnarray}}

\newcommand{\nside}{\ifmmode N_{\mathrm{side}}\else $N_{\mathrm{side}}$\fi}
\newcommand{\npix}{\ifmmode n_{\mathrm{pix}}\else $n_{\mathrm{pix}}$\fi}
\newcommand{\lmin}{\ifmmode \ell_{\mathrm{min}}\else $\ell_{\mathrm{min}}$\fi}
\newcommand{\lmax}{\ifmmode \ell_{\mathrm{max}}\else $\ell_{\mathrm{max}}$\fi}

\newcommand{\WMAP}{{\slshape WMAP~}}
\newcommand{\LCDM}{$ \Lambda $CDM~}
\newcommand{\LCDMc}{$ \Lambda $CDM}

\newcommand{\Planck}{{\slshape Planck~}}
\newcommand{\Planckc}{{\slshape Planck}}

\newcommand{\HEALPIX}{\textsc {Healpix}}
\newcommand{\Healpix}{{\HEALPIX}}

\newcommand{\zphot}{z_{\mathrm{phot}}}

\title[CMB lensing tomography with DES-SV]{CMB lensing tomography with the DES Science Verification galaxies} 

\author[T. Giannantonio, P. Fosalba et al.] {T.~Giannantonio$^{1,2,3,\star}$, P.~Fosalba$^{4,\dagger}$,
R. Cawthon$^{6,12}$,Y. Omori$^9$, M. Crocce$^4$, F. Elsner$^{5}$, 
\newauthor
B. Leistedt$^{5}$,  S. Dodelson$^{6,12,16}$
% ...
\\
 \smallskip {\textit{Affiliations are listed at the end of the paper.}} $^{\star}${t.giannantonio@ast.cam.ac.uk}, $^{\dagger}${fosalba@ice.cat}
 }

\date{Last updated \today}

\pubyear{2016}

\begin{document}
\label{firstpage}
\pagerange{\pageref{firstpage}--\pageref{lastpage}}

\maketitle

\begin {abstract} 
  We measure the cross-correlation between the galaxy density in the Dark Energy Survey (DES) Science Verification data and the lensing of the cosmic microwave background (CMB) as reconstructed with the \Planck satellite and the South Pole Telescope (SPT).
  When using the DES main galaxy sample over the full redshift range $0.2 < \zphot < 1.2$, a cross-correlation signal is detected at
  $6 \sigma$ and $4\sigma$ with SPT and \Planck respectively.
 We then divide the DES galaxies into five photometric redshift bins, finding  significant ($>$$2 \sigma$) detections in all bins.
 Comparing to the fiducial \Planck cosmology, we find the redshift evolution of the signal matches expectations, although the amplitude is consistently lower than predicted across redshift bins.
  We test for possible systematics that could affect our result and find no evidence for significant contamination.
 Finally, we demonstrate how these measurements can be used to constrain the growth of structure across cosmic time.
 We find the data are fit by a model in which the amplitude of structure in the $z<1.2$ universe is $0.73 \pm 0.16$ times as large as predicted in the \LCDM \Planck cosmology, a $1.7\sigma$ deviation.
\end {abstract}


\AddToShipoutPictureBG*{%
  \AtPageUpperLeft{%
    \hspace{0.75\paperwidth}%
    \raisebox{-4\baselineskip}{%
      \makebox[0pt][l]{\textnormal{DES 2015-0048}}
}}}%

\AddToShipoutPictureBG*{%
  \AtPageUpperLeft{%
    \hspace{0.75\paperwidth}%
    \raisebox{-5\baselineskip}{%
      \makebox[0pt][l]{\textnormal{Fermilab PUB-15-308-AE}}
}}}%

%\clearpage

\section {Introduction} \label {sec:intro}
The cosmic microwave background (CMB) radiation, released at the time of hydrogen recombination, provides a view of the Universe when it was only 380,000 years old.
However, this image has been slightly altered since the last-scattering surface, as the CMB photons had to travel through an inhomogeneous distribution of matter before reaching us today.
 Beyond the simple background cooling due to the Hubble expansion, the intervening large-scale structure (LSS) of the Universe can alter the energies and paths of the CMB photons, producing a range of effects beyond the primary CMB power spectrum; these are collectively known as secondary CMB anisotropies. 

The CMB photons freely stream through neutral hydrogen after recombination, 
but they can undergo Compton scattering once again at late times in the re-ionised intergalactic medium or in the hot, ionised gas in the potential wells of massive clusters of galaxies. This latter phenomenon is known as the Sunyaev-Zel'dovich
(SZ) effect \citep[see e.g.][]{1980ARA&A..18..537S, 2002ARA&A..40..643C}. When travelling in and out of gravitational potential wells, they may gain a net energy when the potentials are evolving in time (integrated Sachs-Wolfe, ISW effect) \citep{1967ApJ...147...73S, 1968Natur.217..511R, Crittenden:1996a, 2003ApJ...597L..89F, 2004Natur.427...45B, 2004MNRAS.350L..37F, 2006MNRAS.372L..23C, 2006PhRvD..74f3520G, Giannantonio:2008a, 2012MNRAS.426.2581G, Giannantonio2014}. Finally, as they travel through the LSS, the CMB photons are gravitationally deflected by the mass distribution along their way, distorting the image we eventually observe. Here we focus on this last effect, CMB lensing.

As described in the review by \citet{2006PhR...429....1L}, the typical gravitational deflections of the CMB photons are of order a few arcminutes \citep{1989MNRAS.239..195C}. These deflections, integrated along the entire line of sight, alter the CMB anisotropies we observe in a number of ways. First, lensing smooths out the peaks and troughs in the temperature and polarisation angular power spectra on arcminute scales \citep{1996ApJ...463....1S}. Lensing leads to power leakage from large into smaller angular scales \citep{1990MNRAS.243..353L}, and from $E$- to $B$-mode polarisation \citep{1998PhRvD..58b3003Z}. Lensing also introduces non-vanishing higher-order statistics of the temperature and polarisation fields, which  can be used to reconstruct the lensing potential \citep{2003PhRvD..67h3002O,Hirata2003}, provided a sufficiently high-resolution and low-noise map is available.
Such reconstructed maps of the lensing potentials contain the integrated information of the entire matter distribution in the Universe, out to the surface of last scattering. In order to interpret this information to optimally constrain cosmology, and in particular the evolution of structure formation, it is desirable to study the lensing contribution as a function of redshift: this can be achieved by cross-correlating the full reconstructed CMB lensing maps and tracers of matter at known redshift, such as galaxy surveys \citep{2006PhR...429....1L}. By cross-correlating the CMB lensing potential with the LSS, we can measure the growth of structure as a function of time in redshift bins; this measurement can be used for example to help identify the mechanism driving the current epoch of cosmic acceleration. In addition, CMB lensing-galaxy correlations can be used to improve the control of systematics in weak lensing analyses \citep{2013arXiv1311.2338D}.

The power of the cross-correlation technique is made evident by the early works in this field: while the CMB lensing potential itself was only weakly detectable (at $<$$2\sigma$) from the \WMAP temperature maps, due to their comparatively low resolution and high noise \citep{Smidt2011, Feng2012}, the first significant detection of CMB lensing was achieved by \citet{2007PhRvD..76d3510S} at the $3.4 \sigma$ level by cross-correlating \WMAP data with radio-galaxies from the NRAO VLA Sky Survey \citep[NVSS,][]{Condon1998}. This was later extended by \citet{Hirata2008} using multiple galaxy catalogues, in a first attempt at studying the redshift evolution of the signal, finding a lower combined evidence of $2.5 \sigma$.
The field is now flourishing: CMB lensing has been detected not only indirectly from the smearing of the CMB temperature power spectrum \citep{Das2011,Keisler2011,Story2013,PlanckXVI2014}, but also directly at high significance from the non-Gaussianity of the CMB temperature field using high-resolution data from ACT \citep{2011PhRvL.107b1301D,Das1014}, SPT \citep{2012ApJ...756..142V, 2014arXiv1412.4760S} and \Planck \citep{2013arXiv1303.5077P, 2015arXiv150201591P}. The latest analyses of these experiments achieved detections of CMB lensing at the 
$4.6\sigma$, $14\sigma$, and $40\sigma$   
levels respectively; the different significance levels depend on the different beam resolutions, detector noise levels, and sky coverage fractions. With respect to the last, the \Planck satellite has a clear advantage, thanks to its large sky coverage, even in the 
galaxy-masked maps, while the small-scale resolution and noise are superior for the ground-based surveys. 

CMB lensing has also been detected through its impact on the B-mode signal in CMB polarisation data with BICEP2 \citep{2014PhRvL.112x1101A}, with a joint BICEP2-\Planck polarisation analysis \citep{2015PhRvL.114j1301B}, the Keck Array \citep{2015ApJ...811..126A}, POLARBEAR \citep{2014ApJ...794..171T}, and SPT \citep{2015ApJ...807..151K}, as well as from the four-point function of POLARBEAR polarisation data \citep{2013arXiv1312.6646P}.

The ACT lensing data, reconstructed over six regions within the SDSS Stripe 82 covering a total of 320 deg$^2$, has been used  by \citet{Sherwin2012} for cross-correlation with optically-selected, photometric quasars from SDSS \citep{Bovy2011}, finding a detection of significance $3.8\sigma$. The ACTPol data, including information from CMB polarisation, were cross-correlated  by \citet{vanEngelen2014} with cosmic infrared background (CIB) maps reconstructed by \Planckc, finding a detection at the $9\sigma$ level.

The SPT lensing maps were cross-correlated by \citet{Bleem2012} over four distinct fields of $\sim 50 $ deg$^2$ each with optically-selected galaxies from the Blanco Cosmology Survey \citep{Desai2012,Bleem2015}, IR sources from SPT Spitzer Deep Field \citep{Ashby2009}, and from the \textit{WISE} all-sky IR survey \citep{Wright2010, Geach2013}. Significant correlations ($>$$4\sigma$) were found in all cases, although the interpretation was complicated by the large uncertainties on the redshift of these sources.
Additionally, \citet{Holder2013} detected the correlation between the SPT lensing maps and the diffuse CIB maps measured by \textit{Herschel}/SPIRE \citep{Griffin2010}, finding positive detections at significances between $6.7\sigma$ and $8.8\sigma$ in three sub-mm frequency bands. Cross-correlation between SPTPol and the CIB was detected at $7.7\sigma$ by \citet{Hanson2013}.
The CMB lensing-CIB correlation was also detected with POLARBEAR data \citep{2014PhRvL.112m1302A}.
These works further demonstrate that the CIB is well-suited for CMB lensing cross-correlations, due to its broad and deep redshift distribution, leading to a significant overlap with the CMB lensing kernel; on the other hand, the interpretation of the results is more challenging than for resolved sources, due to the relative uncertainty on the CIB redshift distribution. 

The \Planck team took immediate advantage of their data \citep{2013arXiv1303.5077P}, by cross correlating their CMB lensing map with four tracers of the LSS: NVSS, SDSS LRGs \citep{Ross2011}, SDSS clusters \citep{Koester2007}, and the \textit{WISE} sub-mm satellite survey. These cross-correlations were measured at high significance: $7\sigma$ for \textit{WISE} and clusters, $10 \sigma$ for the SDSS LRGs, and $20\sigma$ for NVSS, thanks to the dramatic extension of sky coverage with respect to previous CMB lensing data. These results were also confirmed by \citet{2013arXiv1312.5154G} and extended to the final photometric SDSS main galaxies \citep{Aihara2011}, SDSS photometric quasars, X-ray background \citep{1987PhR...146..215B}, and the 2MASS IR survey \citep{Skrutskie2006}. The cross-correlation between \Planck lensing and quasars from \textit{WISE} and SDSS was measured by \citet{Geach2013}, \citet{DiPompeo2015}.
A further study of the cross-correlation between \Planck lensing and \textit{Herschel} was performed recently by \citet{Bianchini2014}, while \citet{Omori2015} measured at $>$$5\sigma$ the cross-correlation with the Canada-France-Hawaii Telescope Lensing Survey (CFHTLenS) galaxy number density.

However, none of the existing galaxy surveys has the depth and density of sources over a contiguous area required for a comprehensive tomographic analysis of the CMB lensing signal; this is finally possible with the Dark Energy Survey (DES) and SPT \citep[see e.g.][]{Vallinotto:2013eva}, and it is the main focus of this work.
The DES finished its second of five years of operations in March 2015, and will eventually image 5000 square degrees in the Southern Hemisphere from the Blanco Telescope in Chile, in the bands $g$, $r$, $i$, $z$, and $Y$ using the Dark Energy Camera \citep{Flaugher2015}.
Its depth makes it well-suited for measuring CMB lensing tomography, because it allows the survey to detect a larger fraction of the CMB lensing signal, whose contribution peaks at redshifts $z > 1$.
In this paper, we cross-correlate the initial DES Science Verification (SV) data with the CMB lensing maps reconstructed by the \Planck and SPT surveys, and report a detection of the correlation in broad agreement with the expectations under the assumption of a concordance \LCDM model, with a significance of $6 \sigma$ and $4 \sigma$ for SPT and \Planck respectively. The DES SV data consist of near full-depth imaging of $\sim 300$ deg$^2$, of which we use the  $\sim 200 $ deg$^2$ of the SPT-E field, which is reduced to $131$ deg$^2$ after masking. The SPT lensing data we use were derived by \citet{2012ApJ...756..142V} from the 2500 deg$^2$ SPT-SZ survey \citep{Story2013}, which fully overlaps by design with the DES footprint, while the \Planck public data \citep{2013arXiv1303.5077P,2015arXiv150201591P} cover the entire extra-galactic sky.
Motivated by the high significance of the SPT detection, we measure this cross-correlation in redshift bins, reconstructing the time evolution of CMB lensing.

The plan of this paper is as follows: after briefly reviewing the theoretical expectations in Section~\ref{sec:theory}, we present the data in Section~\ref{sec:data} and the mocks we use to estimate the covariances in Section~\ref{sec:mocks}; we then report our results in Section~\ref{sec:results}, tests for possible systematics in Section~\ref{sec:systematics}, and present some basic cosmological implications in Section~\ref{sec:implications}, before concluding in Section~\ref{sec:conclusion}.

\section {Theory} \label{sec:theory}
\subsection {Power spectra}
Gravitational lensing deflects the primordial CMB temperature anisotropies, so that the temperature we observe in a direction $\hat{\mathbf{n}}$ corresponds to the primordial unlensed anisotropy in the direction $\hat{\mathbf{n}} + \nabla \varphi (\hat{\mathbf{n}})$. Here $\varphi(\hat{\mathbf{n}})$ is the CMB lensing potential, defined in a flat universe as \citep{2006PhR...429....1L}  
\be
\varphi (\hat{\mathbf{n}}) = - \int_0^{\chi_*} d \chi \, \frac{\chi_* - \chi}{\chi_* \chi} \, \left[\Phi + \Psi \right](\chi \hat{\mathbf{n}}, \eta_0 - \chi) \, ,
\ee
where $\chi$ is the comoving distance, asterisks denote quantities evaluated at the last-scattering surface, $\eta_0$ is the conformal time today, and $\Phi, \Psi$ are the matter and light gravitational potentials, which are effectively equal in the standard \LCDM model in linear theory.
The convergence field $\kappa(\hat{\mathbf{n}})$ can be used in place of the lensing potential $\varphi(\hat{\mathbf{n}})$; the two are related in multipole space as
\be \label{eq:convergence}
\kappa_{\ell m} =  \frac {\ell (\ell + 1)}{2} \varphi_{\ell m}  \, .
\ee
By applying the Poisson equation, the CMB convergence in a direction  $\hat{\mathbf{n}}$ can be rewritten as a function of the matter overdensity $\delta$ \citep[see e.g.][]{Bleem2012}:
\be
\kappa (\hat{\mathbf{n}}) =  \frac{3 \Omega_m H_0^2}{2} \int_0^{\chi_*} d \chi \,\frac{\chi^2}{a(\chi)} \, \frac{\chi_* - \chi}{\chi_* \chi} \, \delta (\chi \hat{\mathbf{n}}, \eta_0 - \chi) \, ,
\ee
where $H_0$ is the Hubble parameter today, $\Omega_m$ the matter energy density and $a(\chi)$ is the scale factor.

In the local bias model \citep{Fry1993}, the smoothed galaxy overdensity $\delta_g$ is related to the smoothed matter overdensity $\delta$ by a Taylor expansion, so that if the bias is assumed to be deterministic, 
$ \delta_g (\mathbf{x}, z) = \sum_{i = 0}^{\infty}  b_i(z) \, \delta^i (\mathbf{x}, z) / i! $.
In the present analysis of DES data we will consider only scales where the linear bias suffices, as demonstrated by \citet{CrocceACF}. In this case, $\delta_g (\mathbf{x}, z) = b(z) \, \delta (\mathbf{x}, z)$.
A galaxy catalogue with redshift distribution $dn/dz(z)$ thus provides an estimate of the projected overdensity in a direction  $\hat{\mathbf{n}}$  as
\be
\delta_g (\hat{\mathbf{n}}) = \int_0^{\infty} b(z) \, \frac{dn}{dz}(z) \, \delta( \chi \hat{\mathbf{n}}, z) \, dz \, ,
\ee
where $b(z)$ is the galaxy bias (assumed here linear, deterministic and scale-independent) and $\delta$ the total matter overdensity field.

The two-point statistics of the galaxy-galaxy and galaxy-CMB lensing correlations can be written in harmonic space as
\begin{align} \label{eq:cl_defs}
C_\ell^{gg} =& \frac{2}{\pi} \int_0^{\infty} dk \, k^2 \, P(k) \, W_\ell^g (k) \, W_\ell^g (k) \, \\
C_\ell^{\kappa g} =& \frac{2}{\pi} \int_0^{\infty} dk \, k^2 \, P(k) \, W_\ell^{\kappa} (k) \, W_\ell^g (k) \, , 
\end{align}
where $P(k)$ is the matter power spectrum at $z=0$, and the kernels for galaxies and CMB lensing convergence are in the standard model ($\Phi = \Psi$) for a flat universe  \citep{2006PhR...429....1L, Bleem2012, Sherwin2012}:
\begin{align}
W_\ell^{g}(k) =& \int_0^{\infty} dz \, b (z) \frac{dn}{dz}(z) \, D(z) \, j_\ell[k \chi(z)] \\
W_\ell^{\kappa}(k) =&  \frac{3 \Omega_m H_0^2}{2} \int_0^{\infty} dz \, \frac{\chi_* - \chi}{\chi_* \chi}(z) \, D(z) \, j_\ell[k \chi(z)] \, ,
\end{align}
where $D(z)$ is the linear growth function defined so that $\delta (z) = D(z) \, \delta(z = 0)$, $j_\ell$ are the spherical Bessel functions, and we have assumed $c = 1$;
the lensing potential power spectra can be readily obtained using Eq.~(\ref{eq:convergence}).
The equivalent expressions in real space can be derived with a Legendre transformation.
We will indicate in the following the two-point statistics of generic fields $a, b$ as $C_\ell^{ab}$, $w^{ab}(\vartheta)$, related by
\be \label{eq:wtheta}
w^{ab}(\vartheta) = \sum_{\ell=0}^{\infty} \left( \frac{2 \ell + 1}{4 \pi} \right) \, P_\ell (\cos \vartheta) \, C_\ell^{ab} \, ,
\ee
where $P_\ell$ are the Legendre polynomials, and in practice the sum is limited to $\ell_{\max}$, chosen to be sufficiently high to ensure convergence.

From the definitions of Eqs.~(\ref{eq:cl_defs}), it is clear that to first approximation, valid in the limit of a narrow redshift range for local and deterministic linear bias,
\be
C_\ell^{gg} (z) \propto b^2 (z) \, D^2 (z) \,, \:\:\:\:\:\:\:\:\:\:\:\:\:\:  C_\ell^{\kappa g} (z) \propto b (z) \, D^2 (z) \, ,
\ee
so that a joint measurement of these two quantities can break the degeneracy between bias and linear growth \citep[see e.g.][]{Gaztanaga2012}. We  develop this idea in Section~\ref{sec:implications} below.

\subsection {Stochasticity}
Alternatively, it is possible to assume cosmology to be fixed, and to  use the data to constrain
galaxy bias instead.
Non-linear bias is expected at small scales, as well as a stochastic component due to the discrete sampling and the physical processes affecting halo collapse and galaxy formation.
  As bias non-linearities on the scales considered have been excluded for our sample by \cite{CrocceACF}, we will consider stochasticity $\epsilon$, which changes the biasing law to \citep{Tegmark1998,Pen1998,Dekel1999}
\be
\delta_g (\mathbf{x}, z) = b(z) \, \delta (\mathbf{x}, z) + \epsilon (\mathbf{x}, z) \, ,
\ee
which leads to the power spectra
\begin{align}
C_\ell^{gg} (z) &\simeq b^2(z) \, C_\ell^{mm} (z) +  C_\ell^{\epsilon \epsilon} (z) \, \\
C_\ell^{\kappa g} (z) &\simeq b(z) \, C_\ell^{\kappa m} (z)  \, , 
\end{align}
where $ C_\ell^{mm} $ and $C_\ell^{\epsilon \epsilon}$ are the matter and stochasticity power spectra respectively.
It is clear that, in the absence of non-linearity, a measurement of the correlation coefficient 
$ r \equiv  C_\ell^{gm} /  \left(\sqrt{C_\ell^{mm} \, C_\ell^{gg}} \right) $
constrains the stochastic component, as
\be \label{eq:stoc}
r  = \left[ 1 + \frac{C_l^{\epsilon \epsilon}}{b^2 \, C_l^{mm}} \right]^{-\frac{1}{2}} \simeq 1 - \frac{C_l^{\epsilon \epsilon}}{2 \, b^2 \, C_l^{mm}} \, .
\ee
Notice that if stochasticity is present, the bias inferred from the measured galaxy auto-correlation $ b_{\text{auto}} =  \sqrt{C_\ell^{gg} / C_\ell^{mm}}$ will absorb the stochastic component, and it will thus be different from what is obtained from the galaxy-CMB lensing cross-correlation $b_{\text{cross}} = C_\ell^{\kappa g} / C_\ell^{mm} $; the mismatch is simply given by $ r = b_{\text{cross}} / b_{\text{auto}} $. In the following, we will assume no stochasticity throughout, thus assuming $b_{\text{cross}} = b_{\text{auto}} = b $, except from Section~\ref{sec:stochasticity}, where we discuss the possible interpretation of our results as a measurement of stochasticity.

Stochasticity has been studied with $N$-body simulations and constrained with observations. Recent simulation studies report a negligible stochastic component, except on the smallest scales \citep{Baldauf2010, Cai2011, Manera2011}. Observational constraints have been obtained by combining galaxy clustering with weak gravitational lensing data, using the methods by \citet{Schneider1998}, \citet{vanWaerbeke1998}. The most recent results were obtained by \citet{Jullo2012} using the COSMOS survey, finding no evidence for stochasticity; this is however in tension with the significant stochasticity found by \citet{Hoekstra2002} using the Red-Sequence Cluster Survey and the VIROS-DESCART survey, by \citet{Sheldon2004} with the Sloan Digital Sky Survey, and by \citet{Simon2007} with the GaBoDS survey. The current and upcoming DES clustering and weak lensing data, including CMB lensing, are well-suited to obtain better constraints on this issue.

We calculate all theoretical power spectra and correlation functions using a full Boltzmann code implemented in \textsc{camb} \citep{Lewis:2000a, Challinor2011}, including the (small) effect of redshift-space distortions. We include the effects of non-linear matter clustering using the \textsc{Halofit} formalism \citep{2003MNRAS.341.1311S, 2012ApJ...761..152T}. We have tested from the slopes of the number counts that the effect of cosmic magnification \citep[see e.g.][]{vanWaerbeke2010} is negligible for all cases considered in this paper, so that we neglect this contribution.
Unless otherwise specified, we assume a fiducial \Planck 2013 (+ \WMAP polarisation + ACT/SPT + BAO) best-fit flat \LCDMc+$\nu$ (1 massive neutrino) cosmology of parameters: $\omega_b  = 0.0222$, $\omega_c = 0.119$, $\omega_{\nu} = 0.00064$, $h = 0.678$, $\tau = 0.0952$, $A_s = 2.21 \times10^{-9}$, $n_s = 0.961 $ at a pivot scale $\bar k = 0.05$ Mpc$^{-1}$, corresponding to $\sigma_8 = 0.829$, where $h \equiv H_0 / 100$ km s$^{-1}$ Mpc$^{-1}$ and $\omega_i \equiv \Omega_i h^2 $ for each species $i$ \citep{PlanckXVI2014}. (We have checked that assuming a \Planck 2015 cosmology has negligible impact on the results.)

\subsection{Expected signal-to-noise}
We first estimate the signal-to-noise expected for the detection of the CMB lensing -- galaxies cross-correlation with current and upcoming data.
We include the uncertainties from cosmic variance and the noise, $N_\ell$, which is due to  shot noise for the galaxy counts and to the primary CMB, instrumental, and atmospheric noise for the CMB lensing maps.

\begin{figure}
\begin{center}
  \includegraphics[width=\linewidth, angle=0]{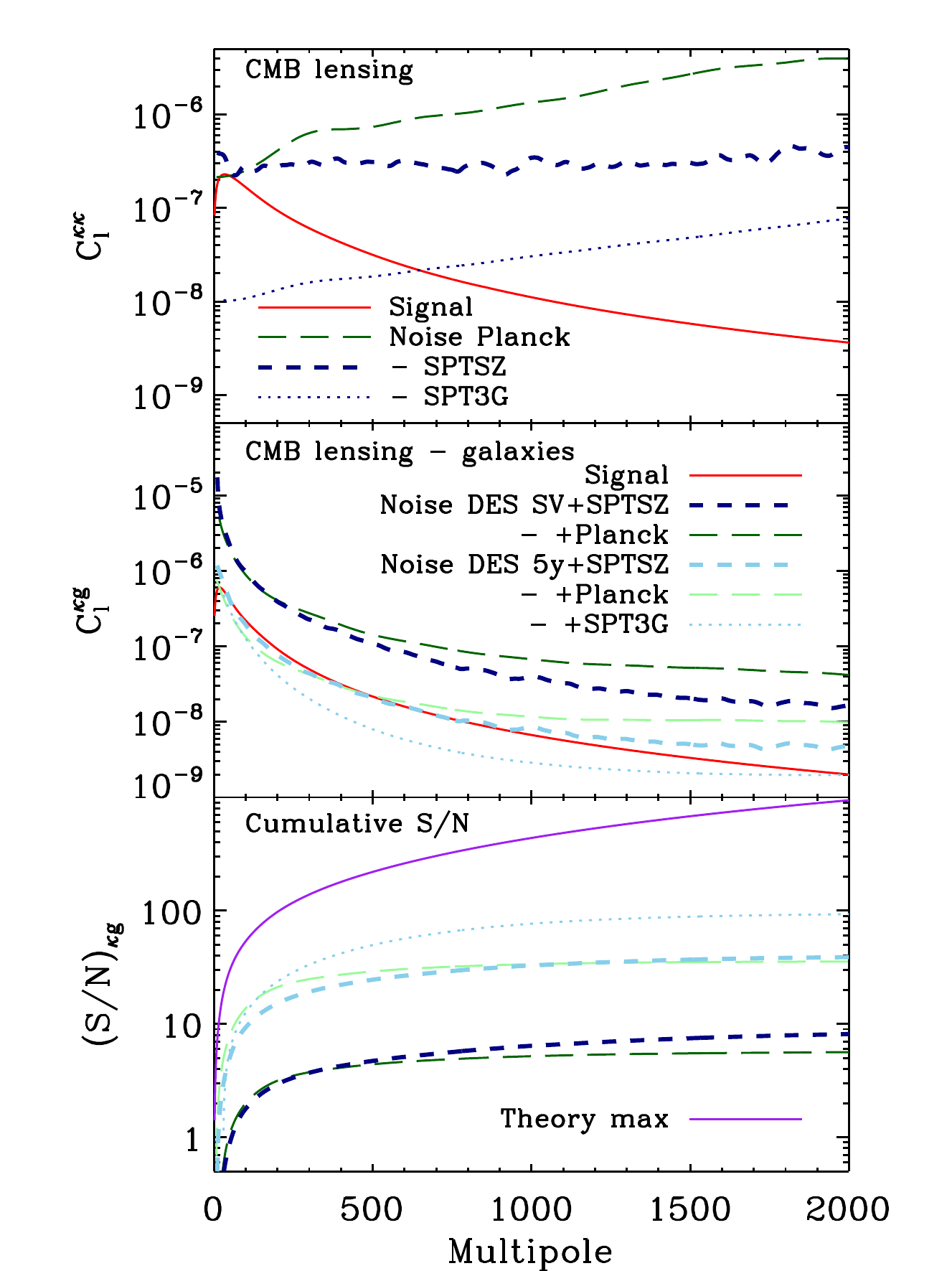}
  \caption{Signal-to-noise forecasts for the DES-CMB lensing correlations, for a range of different CMB and DES data sets. \emph{Top panel:} The theoretical CMB lensing auto-spectrum compared with the noise of \Planck 2015 and SPT-SZ as well as the projected noise of the upcoming SPT-3G survey.  \emph{Central panel:} The theoretical CMB lensing-galaxy cross-spectrum compared with the analytical errors estimated for the \Planck and SPT cases, considering DES SV (\emph{top,  darker colours}) and 5-year data (\emph{bottom, lighter colours}). The errors are large as they are shown per individual multipole $\ell$, and are correspondingly reduced once binned. \emph{Bottom panel:} The cumulative signal-to-noise ratio of the CMB lensing-galaxy cross-correlations for the same cases, compared with the theoretical maximum fixed by cosmic variance. Note that a 5-10 $\sigma$ detection is expected for SV data with the information coming from $\ell < 2000$. For the full DES 5-year data, the measurement with \Planck is expected to yield a similar significance to SPT-SZ, given the larger overlapping area. SPT-3G will achieve the most accurate measurement.}
\label{fig:SN}
\end{center}
\end{figure}

The top panel of Fig.~\ref{fig:SN} shows the noise $N_\ell^{\kappa \kappa}$ compared with the CMB lensing auto-spectrum $C_\ell^{\kappa \kappa}$, for currently available CMB lensing data. We compare the \Planck 2015 CMB lensing noise \citep{2015arXiv150201591P} with the mean noise level of the SPT-SZ lensing maps we use. The effective SPT-SZ noise is lower than \Planckc's on most scales ($\ell > 100$), while on larger scales the 2015 \Planck lensing data have higher sensitivity.
We also show how future CMB data from the SPT-3G survey \citep{2014SPIE.9153E..1PB} are expected to lower the lensing noise by an order of magnitude.
For SPT we have assumed a minimum mode $\ell_{\min} = 30$, given the smaller sky coverage of this survey, while for \Planck we use $\ell_{\min} = 8$, as specified by the public data provided.
The SPT-3G forecast is based on a minimum-variance lensing reconstruction up to $\ell=3000$, without explicitly considering the effect of foreground contaminations.

In the second panel of Fig.~\ref{fig:SN} we show the theoretical prediction for the cross-spectrum $C_\ell^{\kappa g}$ and the corresponding theoretical noise per multipole \citep[see e.g.][]{2011MNRAS.415.2193R}
\begin{equation} \label{eq:errcl}
 \sigma \left(C_\ell^{\kappa g} \right) =  \frac{1}{\sqrt{f_{\mathrm{sky}} (2 \ell + 1)}}  \,
 \left[ \left(C_\ell^{\kappa g} \right)^2 + \left(C_\ell^{\kappa \kappa} + N_\ell^{\kappa \kappa}  \right) \left(C_\ell^{gg} + N_\ell^{gg} \right) \right]^{1/2} \, , 
\end{equation}
where $f_{\mathrm{sky}}$ is the overlapping sky fraction of the surveys, the CMB lensing noise $N_\ell^{\kappa \kappa} $ is discussed above, and the galaxy noise is $N_\ell^{gg} = 1/ n$, where $ n$ is the galaxy density per steradian.
For the signal to noise projection on the DES-SV area, we use the specifications of the real galaxy catalogue described below in Section~\ref{sec:data}: we assume the real redshift distribution of the full sample,  a galaxy number density of $5.39$ arcmin$^{-2}$, and a sky coverage of 131 deg$^2$, fully overlapping both SPT and \Planckc. For the forecasts of the DES 5-year survey, we instead assume that galaxies follow the simple redshift distribution by \citet{1995MNRAS.273..277S} with the original proposed specifications of DES \citep{2005astro.ph.10346T}, i.e. a median redshift $\bar z = 0.7 $ and a galaxy number density of $10$ arcmin$^{-2}$. We further assume a sky coverage of 5000 deg$^2$ (fully overlapping \Planckc, but of which only 50\% overlaps SPT). We finally assume constant bias, equal to 1 at all scales. In reality, galaxy bias will be estimated from the galaxy auto-correlations; we show below in Section~\ref{sec:results} that the bias for the DES main galaxy sample is only marginally larger than 1.
We can see that the noise per multipole for DES-SV is large compared with the theory, but this is significantly reduced once binning is used. The noise per multipole is reduced to the same level of the signal with DES 5-year data; in this case the noise level of \Planck is lower than SPT-SZ at low $\ell$, given the larger area overlap with the full DES footprint.

Finally, the expected signal-to-noise is
\begin{equation} \label{eq:sncl}
\left( \frac{S}{N} \right)^2_{\kappa g} =  f_{\mathrm{sky}} \sum_{\ell=\ell_{\min}}^{\ell_{\max}} (2 \ell + 1) \,   
 \frac{\left(C_\ell^{\kappa g} \right)^2}{\left(C_\ell^{\kappa g} \right)^2 + \left(C_\ell^{\kappa \kappa} + N_\ell^{\kappa \kappa}  \right) \left(C_\ell^{gg} + N_\ell^{gg} \right) } \, . 
\end{equation}
We show in the third panel of Fig.~\ref{fig:SN} the cumulative signal-to-noise using different assumptions for the CMB and galaxy data. Here we can see that using DES SV data only, a S/N $ \simeq $ 8 (5) is expected using current SPT (\Planckc) data, thus motivating the analysis in this study.
Beyond the current analysis, we can see that the theoretical maximum S/N determined by cosmic variance is significantly larger than what is possible at present; we further discuss in Section~\ref{sec:conclusion} the prospects for future improvements of this measurement.

\section {Data} \label{sec:data}

\subsection{Galaxy catalogue}

The DES Science Verification (SV) data include imaging of $\sim 300$ square degrees over multiple disconnected fields; the largest contiguous areas are the SPT-E and SPT-W fields, covering $\sim 200$ and $\sim 50$ deg$^2$ respectively, which overlap the SPT-SZ survey. We consider here the larger SPT-E field only.

 The SV area was imaged over 78 nights from November 2012 until February 2013, and includes $\sim 4 \cdot 10^7$ unique co-add objects. The raw data were processed as described by \citet{RykoffSV}, \citet{CrocceACF}.
From the DES-SV final (`Gold') main galaxy catalogue \citep{RykoffSV}, we use the `Benchmark' galaxy selection introduced by \citet{CrocceACF}, which we also briefly describe here.

The `Gold' catalogue covers 254.4 deg$^2$ with dec $> -61$ deg after masking, thus removing the Large Magellanic Cloud and R Doradus regions, unsuitable for extra-galactic science. Only regions with at least one CCD coverage in each band (except $Y$) were included.
Star-galaxy separation is achieved with a cut in the \texttt{wavg\_spread\_model} quantity \citep{CrocceACF}.
The `Gold' catalogue includes a total of 25,227,559 galaxies over the whole SV area.
From them, we select the `Benchmark' galaxy sample over the SPT-E field by imposing
the following cuts:
\begin{itemize}
\item $18.0 < i < 22.5$  (completeness, $10\sigma$ detection); 
\item $ -1 < g - r < 3 $ and  $ -1 < r - i < 2 $ and $ -1 < i - z < 2 $ (remove strong colours from diffraction artefacts);
\item \texttt{wavg\_spread\_model(i)} $ > 0.003$  (star-galaxy separation);
\item $60 < \mathrm{r.a.} < 95$ and $-61 < \mathrm{dec} < -40$  (SPT-E field).
\end{itemize}
Notice that we use two different choices of magnitude definition for the completeness cut (\texttt{slr\_mag\_auto}) and for the colour cuts (\texttt{mag\_detmodel}); see details in \citet{RykoffSV}.
We have checked that using a different magnitude definition for the completeness cuts does not change the results significantly; likewise, the galaxy-CMB lensing cross-correlation results remain consistent if using a different classifier for star-galaxy separation \citep[\texttt{modest\_class}, ][]{RykoffSV}.
Finally, note that our declination cut at $ \mathrm{dec} > -61 $ is marginally less conservative than the cut applied by \citet{CrocceACF} at  $ \mathrm{dec} > -60 $.

Photometric redshifts of DES galaxies were estimated using a variety of techniques \citep{PhotozDES_2014}. 
We consider here the machine learning `Trees for photometric redshifts' (TPZ) \citep{2013MNRAS.432.1483C} and the template-based `Bayesian photometric redshifts' \citep[BPZ,][]{2000ApJ...536..571B} methods.
TPZ was shown to perform well compared with a validation sample of known redshifts \citep{PhotozDES_2014}, and we therefore use this method for our main results.
We show however in Section~\ref{sec:photoz} below that using BPZ does not change our results significantly.
Briefly, TPZ is a machine-learning algorithm using prediction trees and a random forest method that was shown to minimise the number of catastrophic outliers with respect to other techniques. The TPZ implementation we use does not include information from $Y$-band observations.
In addition to the above-mentioned cuts, we discard the tails of the photometric redshift distribution, by selecting galaxies with maximum likelihood photo-$z$ $ 0.2 < \zphot < 1.2$ only, which reduces the sample by $\sim 5\%$. This leaves us with 3,207,934 objects.
Our selection agrees with \citet{CrocceACF} except from the small difference in the declination cut, so that the results of the two papers can be directly compared.

\begin{figure}
\begin{center}
\includegraphics[width=\linewidth, angle=0]{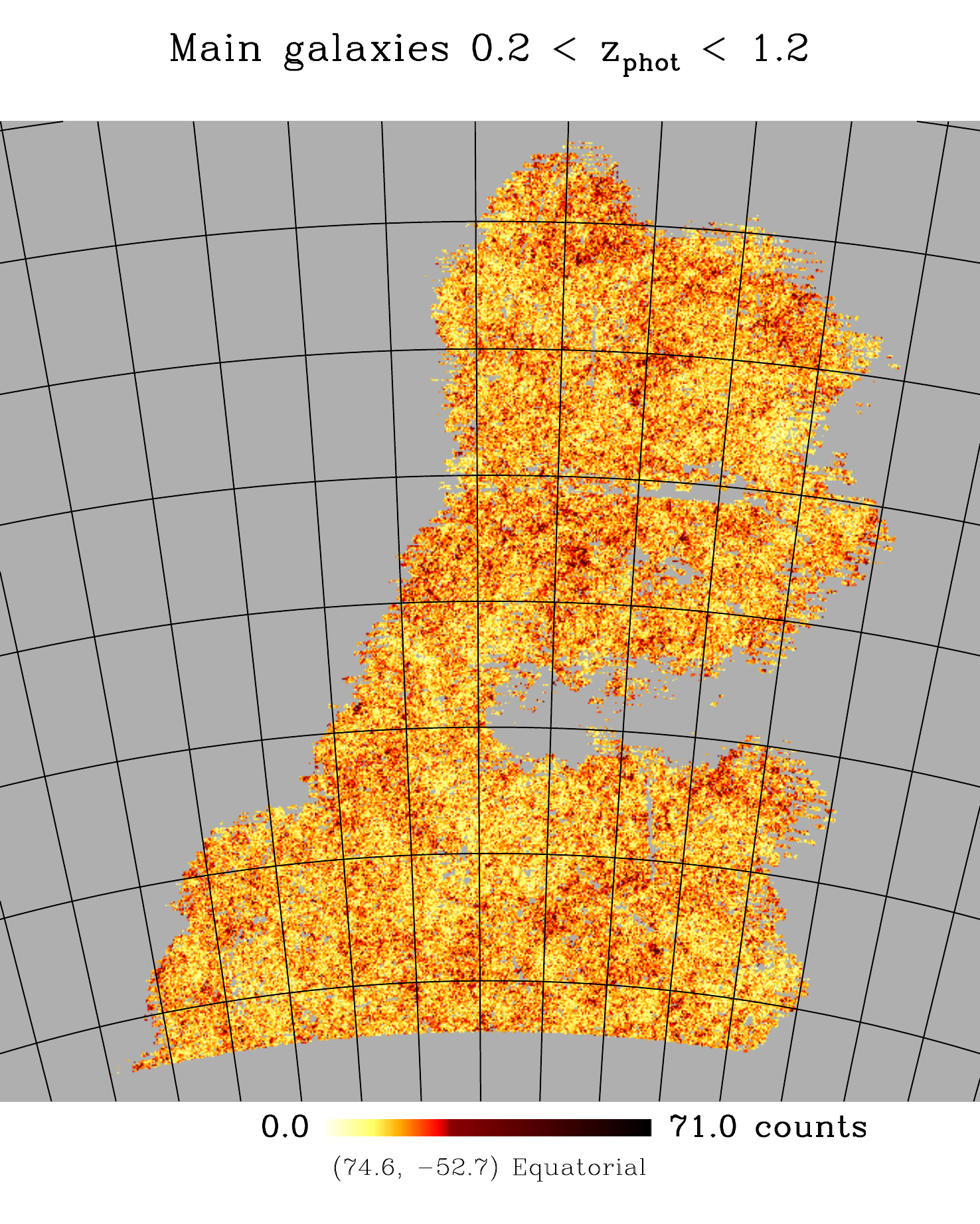}
\caption{Map of the main galaxies used for our analysis in the SPT-E field, pixellated on the \Healpix~$N_{\mathrm{side}} = 2048$  scheme (pixel side: $1.7'$) in Equatorial coordinates, after masking. The colour scale indicates the number of galaxy counts in each pixel. The grid lines are 2.5 deg apart. Grey areas indicate masked data or areas outside the SV footprint. The coordinates $(74.6, -52.7)$ indicate the position of the map centre.}
\label{fig:maps}
\end{center}
\end{figure}

We then pixelise the data on the sky using the \Healpix~scheme \citep{2005ApJ...622..759G} at resolution $N_{\mathrm{side}} = 2048$ (the corresponding pixel side is $d_{\mathrm{pix}} \sim 1.7'$), which is sufficient to capture all the information in both the SPT and \Planck lensing data.
The mask is constructed by excluding regions of photometry shallower than the completeness cut at $ i < 22.5$;
in addition, pixels are discarded unless $ > 80 \%$ of their area has detections. 
After masking, the SPT-E field is left with 2,544,276 objects.
The sky fraction covered is $ f_{\mathrm{sky}} = 3.176 \cdot 10^{-3} $, corresponding to 131.02 deg$^2$, with number density $ n = 6.37 \cdot 10^7 $ sr$^{-1}$ , or $5.39$ arcmin$^{-2}$.
Future DES catalogues will be denser as the magnitude limit is pushed faintward.

We refer to Fig.~2 by \citet{CrocceACF} for the stacked probability distribution of the photometric redshifts of the `Benchmark' main galaxies, for both TPZ and BPZ methods.
In addition to the full sample, we also use five redshift bins of width $\Delta \zphot = 0.2$ that we use in the tomographic analysis below. Also in this case, the cuts are applied on the maximum-likelihood photo-$z$; in all cases, the stacked photo-$z$ PDF has tails outside the cut boundaries.
The number of galaxies in each bin is: 509,456; 818,376; 673,881; 424,437; 118,126 from low to high $z$ respectively.
While the number of galaxies in the last bin is significantly lower than in the others, we choose the current binning in order to explore the clustering and the CMB lensing correlation up to the highest redshifts that are accessible to DES.

We show the masked map of the DES galaxy sample we use in our analysis in Fig.~\ref{fig:maps}.

\begin{figure}
\begin{center}
\includegraphics[width=0.49\linewidth, angle=0]{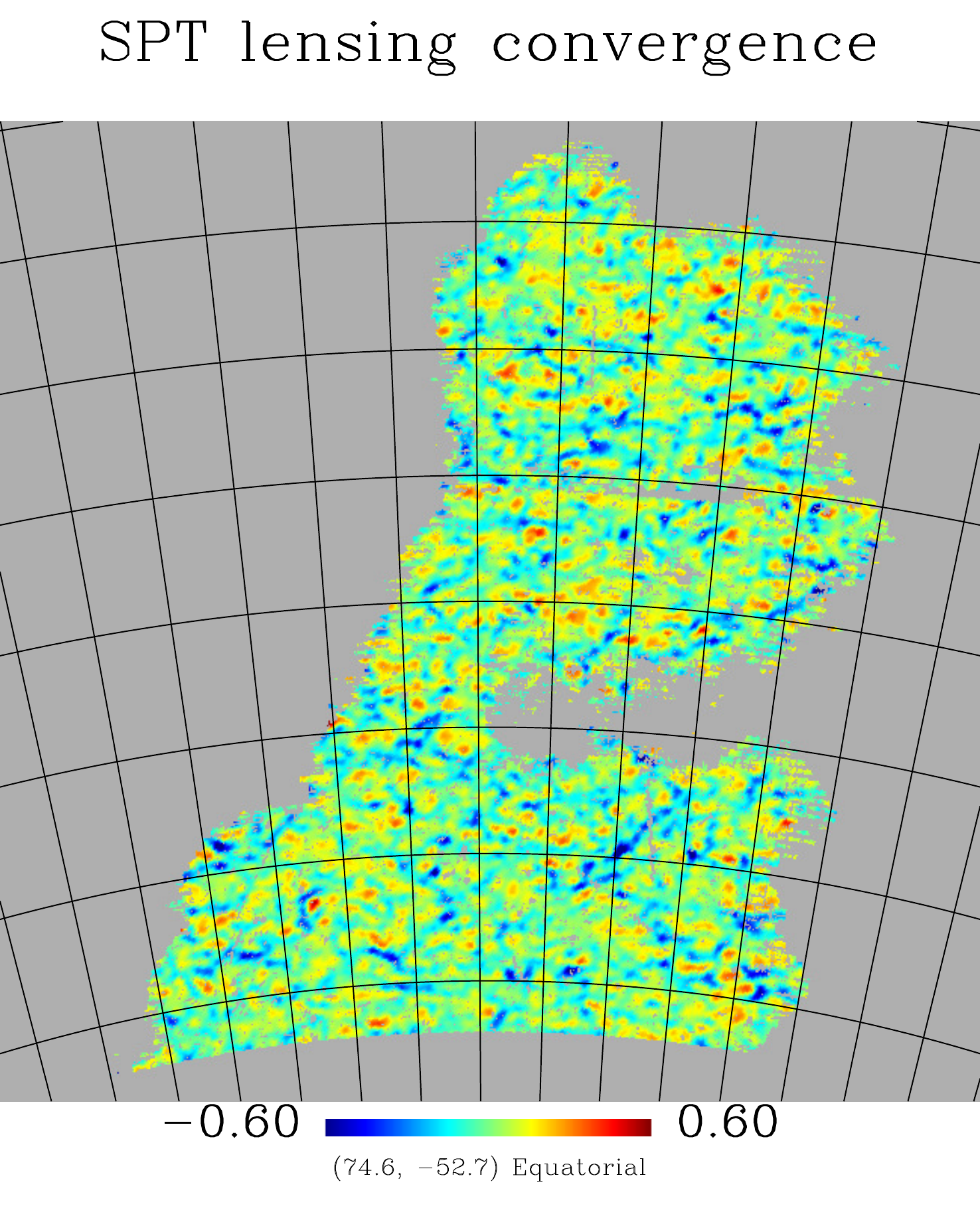}
\includegraphics[width=0.49\linewidth, angle=0]{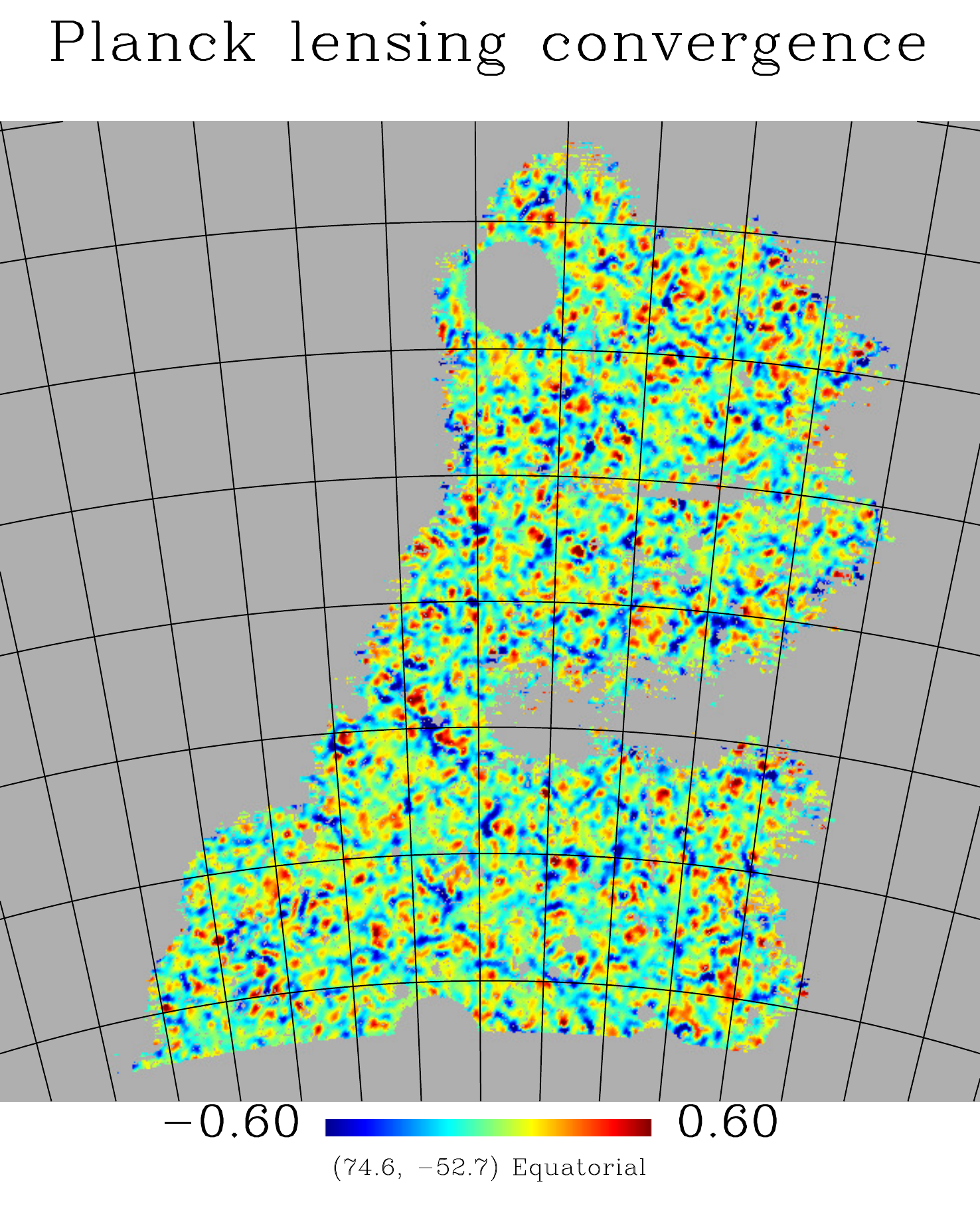}
\caption{Maps of the CMB lensing convergence in the SPT-E field for SPT (\emph{left}) and \Planck (\emph{right}), pixellated on the \Healpix~$N_{\mathrm{side}} = 2048$  scheme (pixel side: $1.7'$) in Equatorial coordinates, smoothed on angular scales of 10 arcmin to improve visualisation. The grid lines are 2.5 deg apart. Grey areas indicate masked data. The DES mask has been applied for clarity, but we do not impose it onto the CMB data in our cross-correlation estimation. The \Planck lensing map also includes the \Planck lensing mask. \Planck shows higher amplitude variations, but this is due to higher noise caused by the lower spatial resolution of its map.}
\label{fig:CMBmaps}
\end{center}
\end{figure}

\subsection{CMB lensing maps}

We consider the lensing convergence maps reconstructed from observations of the CMB temperature anisotropies by the South Pole telescope (SPT) and by the \Planck satellite shown in Fig.~\ref{fig:CMBmaps}. 
 For each experiment, we also use simulated CMB observations to characterise the noise properties in the cross-correlation analysis with the DES data. 
We present in Fig.~\ref{fig:clstpzKK} below the angular power spectra of the CMB lensing maps together with their noise properties inferred from the mocks.

\subsubsection{The South Pole Telescope lensing maps}

The SPT-SZ survey was assembled from hundreds of individual observations of each of 19 contiguous fields that together covered the full survey area. 
For the SPT-E field a $25^\circ \times 25^\circ$ 150 GHz map was made by forming an inverse-variance weighted coadd of all the overlapping observations. A lensing map was constructed from this CMB map following the procedures described in detail by 
\citet{2012ApJ...756..142V}, which we briefly outline below.

Individual sources detected with signal-to-noise greater than 15 (in any of the 3 SPT frequencies) were masked, with the masked regions in the CMB map filled in using Wiener interpolation. These maps were filtered in Fourier space (using the flat-sky approximation) with an anisotropic filter that removed Fourier modes along the scan direction with $\ell_x<500$, and an isotropic filter that removed modes with $\ell>4000$.
A flat-sky lensing map was generated from the filtered maps using a quadratic estimator technique \citep{2003PhRvD..67h3002O}.
This map was then projected into spherical coordinates for the cross-correlation.

The details of point source masking, anisotropic noise, non-stationary noise, and spatially varying beams are sufficiently complex that calibrations and noise estimates were obtained from simulated data. Starting with 100 mock lensed skies and 100 mock unlensed skies, synthetic time-streams were generated, masked and filtered identically to the data. By cross-correlating the 100 lensed output reconstructions with the known input lensing potential, a lensing transfer function was estimated. This lensing transfer function was applied to both the data and the output from the unlensed simulations, which provided 100 noise realisations to characterise the noise properties of the cross-correlations.

We use multipoles between $30 < \ell < 2000$ in the SPT lensing map, as including higher multipoles negligibly changes the overall signal-to-noise in the SPT lensing data.

\subsubsection{The \Planck lensing maps}

\begin{figure}
\begin{center}
\includegraphics[width=0.6\linewidth, angle=90]{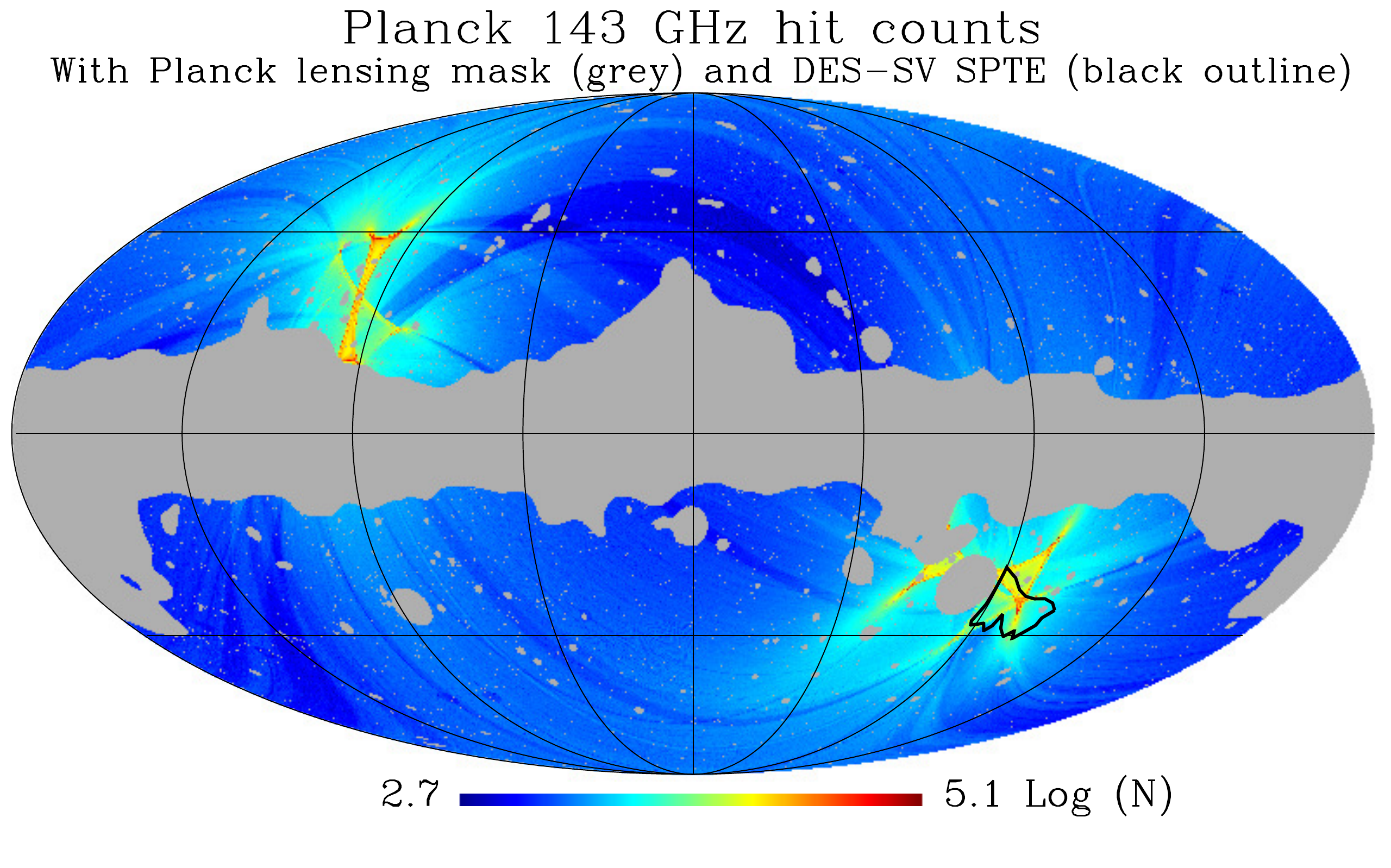}
\caption{Map of the hit counts in the 143 GHz \Planck channel. The stripy structure reflects the scanning strategy of the \Planck satellite, with nodes near the Ecliptic poles. We also show the \Planck lensing mask (\emph{grey area}) and the outline of the DES-SV SPT-E area (\emph{black}), on which we perform the present analysis.
  The SPT-E region's noise properties are clearly atypical, as the field lies near the South Ecliptic Pole, where the \Planck hit count is approximately five times higher than the full-sky average.}
\label{fig:PlanckHits}
\end{center}
\end{figure}

We use the 2015 CMB lensing map provided by the \Planck collaboration \citep{2015arXiv150201591P}. As in the SPT case, this was derived with the quadratic estimator by \citet{2003PhRvD..67h3002O}, which was extended to use the combined information from both CMB temperature and polarisation, thus reducing the noise with respect to the 2013 lensing map \citep{2013arXiv1303.5077P}. The \Planck lensing map is provided as a table of spherical harmonic coefficients up to $\ell_{\max} = 2048$. The reconstructed map covers the full sky, but the comparatively low sensitivity of \Planck means that, averaged over the full sky, the maps are noise-dominated on most scales, as it can be seen in the first panel of Fig.~\ref{fig:SN}. 
However, we can see in Fig.~\ref{fig:PlanckHits} that \Planck observed the DES-SV SPT-E area on which we perform the current analysis with significantly better than average accuracy, as the SPT-E area lies near the South Ecliptic Pole that was repeatedly scanned at every rotation of the \Planck satellite. We therefore expect that the typical lensing noise over this region will be significantly reduced with respect to its full-sky level; we confirm this in Section~\ref{sec:pseudo_cl} below.
For the current analysis we apply the \Planck mask provided with the lensing map shown in grey in Fig.~\ref{fig:PlanckHits}, which masks the Galactic plane and resolved point sources. 
As this map does not include multipoles smaller than $\ell <8$, these modes are also removed from the modelling. 

The \Planck collaboration also provided 100 realisations of the CMB lensing sky, either including only the cosmological signal based on the fiducial \LCDM model with relative fluctuations, or together with the experimental \Planck noise. We can therefore reconstruct 100 noise realisations by taking the difference between the two.

\section {Mocks} \label{sec:mocks}

Here we describe the two approaches to build galaxy mocks that we use to estimate covariance matrices in our analysis, as described in Section \ref{sec:covar}.
We use both methods to demonstrate robustness, but we use the $N$-body mocks for our nominal results.

\subsection {Monte Carlo Gaussian mocks}\label{sec:mc}

The first method to build simulated DES-SV galaxy mocks is based on a Monte Carlo (MC) procedure.
In this approach we generate Gaussian random realisations of the maps we use: galaxies, SPT lensing, and \Planck lensing, all with their (average) noise properties. We produce 1000 random realisations using the \texttt{synfast} code from the \Healpix~package, using random seeds and based on the non-linear \Planck best-fit fiducial theory described above, assuming a linear, constant bias $b = 1.15$, which we find to be consistent with our auto-correlation clustering measurements.
In addition to the fiducial cosmological power spectrum, the mock CMB lensing maps also include fluctuations from the effective noise of SPT and \Planckc, so that we can generate a larger number of MC mocks than the number of realistic noise realisations available. We discuss below in detail in Section~\ref{sec:pseudo_cl} how the effective CMB lensing noise levels over the DES-SV area compare with the simplified average noise presented above in Section~\ref{sec:theory}.

The random maps are generated in such a way to include their correlations as described e.g. by \citet{1998NewA....3..275B}, \citet{cabre07}, \citet{Giannantonio:2008a}.
For the galaxy mocks, after generating random overdensity maps with the correct statistical properties, we transform them into number count maps assuming the actual galaxy number density of the real data. At this point, we add the appropriate galaxy shot noise on each pixel by random sampling from a Poisson distribution of expected value equal to the pixel occupation number.
We finally smooth all mock maps with the same Gaussian beam used for the data: $\vartheta_{\mathrm{FWHM}} = 5.4'$ for the maps intended for the DES auto-correlation and DES-SPT cross-correlation, and $\vartheta_{\mathrm{FWHM}} = 10.8'$ for the DES-\Planck cross-correlation respectively.

We expect the MC method to yield covariances that are similar to a purely analytic Gaussian estimate (see Eq.~\ref{eq:errcl}), except for the effect of the angular survey mask, which adds non-trivial correlations between angular multipoles. Such analytic and MC covariances are expected to be more accurate in their diagonal elements and on large (linear) scales where the fields are close to Gaussian distributed \citep{cabre07}.

\subsection {$\mathbf{N}$-body mocks}\label{sec:nbody}

The second method to produce simulated DES-SV galaxy mocks uses $N$-body outputs from the MICE Grand Challenge $N$-body light-cone simulation (MICE-GC hereafter). 
The MICE simulations are based on the following fiducial cosmology:  $\omega_b  = 0.02156$, $\omega_c = 0.10094$, $\omega_{\nu} = 0$, $h = 0.7$, $A_s = 2.44 \times10^{-9}$, $n_s = 0.95 $, which has a lower matter content than currently preferred \Planck results.
For further details about this simulation see \citet{MICEI}, \citet{MICEII},  \citet{MICEIII}. 

We generate CMB lensing mocks by using an all-sky lensing potential map of the MICE-GC. This map includes lenses at $0 < z < 100$ and sources at the last scattering surface ($z\simeq1100$). We have checked that lenses at $z>100$ give a negligible contribution. The lensing map has been pixelised in the \Healpix~scheme at $\nside=8192$ ($0.43'$ pixels) and downgraded to the required resolution of our analysis ($\nside = 2048$) for covariance estimation.

As for the mock galaxy number density map, we match as closely as possible the `Benchmark' main galaxy sample. For this purpose we have used the dark-matter counts in the light-cone (i.e. unbiased galaxies) as a good approximation to the overall DES-SV galaxy population, given the low bias recovered from the main galaxies. We have then weighted the dark-matter counts with the redshift distribution and bias of our DES-SV galaxy sample in the range $0.2 < \zphot < 1.2$; we assume $b = 1.15$.
The resulting mock galaxy number density is projected onto a \Healpix~map of $\nside=8192$, and downgraded in the same way as the lensing map described above.
We add Poisson noise matching that of the SV galaxy sample to the $N$-body mocks as described for the MC mocks above.

From the full sky we produce 100 non-overlapping rotations of the SPT-E mask. This procedure yields 100 effectively independent realisations of the galaxy and CMB lensing fields, as described and validated in Appendix~\ref{sec:appendix_rotations}.
Onto each CMB lensing mock we then add one mock CMB lensing noise realisation, as provided by the SPT and \Planck collaborations. 
Finally, we apply a Gaussian smoothing of $\vartheta_{\mathrm{FWHM}} = 5.4'$ or $10.8'$ to all mock maps, as we do to the data.

\section {Results} \label{sec:results}

We present here the results of the clustering analysis of the DES-SV galaxies and their correlations with the CMB lensing data. For robustness, we set up two independent analysis methods, measuring all quantities in real and harmonic spaces.

As both SPT and \Planck lensing data only contain meaningful information at multipoles $\ell < \ell_{\max}$, we enforce a cutoff in our analysis by applying a Gaussian smoothing to all data maps and mocks. 
For the DES auto-correlation and DES-SPT cross-correlation, we choose a beam size $\vartheta_{\mathrm{FWHM}} = 5.4'$, which corresponds to a multipole $\ell_{\mathrm{FWHM}} \sim \pi / \vartheta_{\mathrm{FWHM}} = 2000$.  
For the DES-\Planck cross-correlation, given the lower resolution and sensitivity, we use instead $\vartheta_{\mathrm{FWHM}} = 10.8'$, corresponding to $\ell_{\mathrm{FWHM}} \sim \pi / \vartheta_{\mathrm{FWHM}} = 1000$.  We explore in Section~\ref{sec:sysCMB} below the robustness of the results for different choices of $\ell_{\max}$.
The theoretical power spectrum predictions are thus suppressed by a Gaussian beam $B^2_\ell = e^{-\ell (\ell+1) \sigma^2}$, where $\sigma = \vartheta_{\mathrm{FWHM}} / \sqrt{8 \ln 2}$; we have checked that indeed this beam suppresses the signal by $>85\%$ at $\ell = \ell_{\mathrm{FWHM}}$.
As we do not enforce on the maps a sharp cut-off at $\ell < \ell_{\max}$, a small fraction of heavily suppressed power from higher multipoles is retained in the maps and consistently in the theoretical predictions.

A sharp cutoff at $\ell = \ell_{\mathrm{FWHM}}$ would also be a reasonable choice for the harmonic space analysis, and we have indeed confirmed that our results remain consistent with this choice; the real space analysis on the other hand is ill-behaved for sharp cutoffs in $\ell$, so that the Gaussian smoothing is a better strategy for maintaining consistency between the two methods.

Finally, we note that we apply the Gaussian smoothing on masked data consistently by smoothing both the masked map and the mask itself, and then dividing the smoothed masked map by the smoothed mask. This method removes the effect of the mask from the smoothing procedure, and we tested it is equivalent to first applying the smoothing on full-sky mock data and later masking them. 

\subsection {Real space}
\label{sec:realspace}

We begin with the real space analysis, where we measure the projected two-point correlation functions $w(\vartheta)$ of the pixellated maps. 

\subsubsection {Correlation function estimators}

Given the observed number of objects $n_i$ in each pixel $i = 1, ..., N_{\mathrm{pix}}$, and given a binary coverage mask $f_i = \{0, 1\}$, we first estimate the average number density per pixel $\bar n$. We can then use for the correlation between two galaxy density maps $a, b$ the estimator
\be
\hat w^{ab}(\vartheta) = \frac{1}{N^{ab}_{\vartheta}} \sum_{i,j=1}^{N_\mathrm{pix}} \frac {f^a_i (n^a_i - \bar n^a) \, f^b_j (n^b_j - \bar n^b)}{\bar n^a \, \bar n^b} \, \Theta_{i,j} \, ,
\ee
where $ \Theta_{i,j} $ is 1 if the pixel pair $i,j$ is at angular separation $\vartheta$ within the bin size $\Delta \vartheta$, and 0 otherwise, and the number of 
pixel pairs at angular separation $\vartheta$ is $N^{ab}_{\vartheta}  = \sum_{i,j=1}^{N_\mathrm{pix}} f^a_i f^b_j \,  \, \Theta_{i,j}$.
The CMB lensing maps $\kappa_i$ have zero mean, so that the correlation between a galaxy density map and a convergence map can be estimated as
\be
\hat w^{\kappa g}(\vartheta) = \frac{1}{N^{\kappa g}_{\vartheta}} \sum_{i,j=1}^{N_\mathrm{pix}} \frac {f^g_i (n^g_i - \bar n^g)}{\bar n^g} \, f^{\kappa}_j \kappa_j \, \, \Theta_{i,j} \, ,
\ee
where the coverages of galaxies and CMB lensing $ f^g_i , f^{\kappa}_j$ are both binary masks defining the sky area used.

We use this estimator to measure the correlations in $p = 12$ angular bins, equally spaced in logarithm between 0.04 deg ($=$ 2.4 arcmin)  and 5 deg. We have tested with mock data and analytical covariances that this binning optimally recovers the maximum possible information available for maps smoothed at $\vartheta_{\mathrm{FWHM}} = 5.4'$, as the addition of extra bins does not increase the signal-to-noise any further.
Notice that the smallest angle we consider is $\vartheta_{\min} = 2.4 \, \text{arcmin} < \vartheta_{\mathrm{FWHM}} $, as the imposed cut-off on the maps is Gaussian and not top-hat.
Due to the Gaussian smoothing, the shot noise contribution to the galaxy auto-correlation functions affects angular scales at separations $\vartheta > 0$ deg: we describe in detail in Appendix~\ref{sec:shotnoise} how we model the shot noise component, which we subtract from all measured auto-correlation functions.

\subsubsection {Covariance matrix}\label{sec:covar}

We estimate the covariances with several different methods: MC realisations, $N$-body mocks, an analytic method, and jack-knife techniques.  We describe these methods and demonstrate their consistency in Appendix~\ref{sec:appendcovariance}.
Differently from the MC and analytic covariances, the $N$-body method fully reproduces the anisotropic nature of the CMB lensing noise, and it also includes the non-Gaussian contributions to the covariance matrix produced by non-linear clustering, while being more stable than the JK estimator.
 We thus deem the $N$-body method to be our most realistic noise estimator, and we use this for our main results.

We estimate the covariance matrix from the mocks as follows.
We first measure the correlation functions of the mock maps, using the same estimator and keeping the same angular binning as done for the data. 
We use $N = 1000$ MC and  100 $N$-body realisations, and the covariance matrix is then estimated from the scatter of the mock correlations in each angular bin $i$: $\hat w_{\alpha,i}^{ab} \equiv \hat w^{ab}_\alpha(\vartheta_i)$, where $\alpha$ labels a given realisation:
\be \label{eq:covmat}
 \mathcal{\hat C}_{ij}^{ab} =\frac{1}{N} \sum_{\alpha=1}^{N} \left( \hat w_{\alpha,i}^{ab} - \bar w_i^{ab} \right) \left( \hat w_{\alpha,j}^{ab} - \bar w_j^{ab} \right) \, ,
\ee
where $\bar w_i^{ab}$ is the mean correlation function over all realisations in the bin $i$.
Notice that, for all covariance estimators based on multiple realisations $N$, the unbiased estimator for the inverse covariance matrix
is not simply $\left(\mathcal{\hat C}_{ij}^{ab} \right)^{-1}$, but
\citep{2007A&A...464..399H}
\be \label{eq:hartlap}
\widehat {\left(\mathcal{C}_{ij}^{ab} \right)^{-1}}     =  \beta  \, \left(\mathcal{\hat C}_{ij}^{ab} \right)^{-1}   \, ,
\ee
where $ \beta = (N - p - 2) / (N - 1) $ and $p$ is the number of angular bins; $\beta$ tends to one in the limit of large $N$.
We can also define the correlation matrices as
\be \label{eq:correlationmat}
\mathcal{R}_{ij}^{ab} \equiv \frac {\mathcal{C}_{ij}^{ab}} {\sqrt{\mathcal{C}_{ii}^{ab} \mathcal{C}_{jj}^{ab}}} \, ,
\ee
which we show below in this section and in Appendix~\ref{sec:appendcovariance}.  
Note that even a covariance matrix that is diagonal in harmonic space corresponds to a real-space correlation matrix with significant off-diagonal components.

If we assume the likelihood distribution to be Gaussian, 
the above estimate of the inverse covariance matrix (Eq. \ref{eq:hartlap})
 can then be used to calculate the likelihood distribution of some parameters $\mathbf{x}$ given the data $ \hat w_i^{ab}$ as
\begin{multline}
\label{eq:lik}
 \mathcal{L(\mathbf{x})} = (2 \pi)^{-p/2} \left[ \mathrm{det} \,  \mathcal{\hat C}_{ij}^{ab} \right]^{-1/2}  \\
 \times \exp \left\{ - \frac{1}{2} \, \sum_{i,j=1}^p  \, \widehat {\left(\mathcal{C}_{ij}^{ab} \right)^{-1}} \,   \left[ \hat w_{i}^{ab} - w_{i}^{ab}(\mathbf{x})  \right] \, \left[ \hat w_{j}^{ab} - w_{j}^{ab}(\mathbf{x})  \right]     \right\} \, , 
\end{multline}
where $w_{i}^{ab}(\mathbf{x})$ are the binned theoretical correlation functions predicted from the parameters $\mathbf{x}$.
As a consequence of the central limit theorem, the Gaussian likelihood is a good approximation at all but the largest angular scales, whose contribution to our measurement is negligible.
The effect of the uncertainty on the data covariance itself
onto the final parameters variance can be estimated \citep{Taylor2013, Dodelson2013, Percival2014}. We have tested that this contribution is small throughout this work;
the central values of fit parameters are unchanged while error bars are affected at the $<10\%$ level.

In the following, we use a theory template based on the fiducial (fid) \Planck cosmology, and we fit its amplitude. We therefore have for the auto- and cross-correlations:
\begin{equation}
 w_{i}^{gg} =  b^2 \left(w_{i}^{gg} \right)_{\mathrm{fid}}   \, , \:\:\:\:\:\:\:\: w_{i}^{\kappa g} =  A \, \left( w_{i}^{\kappa g} \right)_{\mathrm{fid}}  \, .
\end{equation}
The amplitude of the auto-correlations is given by the galaxy bias $b^2$.
The amplitude of the cross-correlations $A$ depends on both the galaxy bias and the actual amplitude of the CMB lensing signal
$A_\mathrm{Lens}$, so that $A = b \, A_\mathrm{Lens}$. If the underlying true cosmology matches our fiducial \LCDM model, so that $\langle A_\mathrm{Lens} \rangle  = 1$, the expectation value for the amplitude should 
be equal to the galaxy bias from the auto-correlation  $\langle A \rangle = b $,
if the same scales are considered; if instead the scales considered do not match precisely, we expect
 this  to hold only approximately. 
$A$ and $b$ are the parameters that we fit from our measurements on data and mocks below by calculating the likelihood (Eq.~\ref{eq:lik}) over a grid of parameter values.

\subsubsection{Real-space results: full sample}
\label{sec:acf}

\begin{figure}
\begin{center}
\includegraphics[width=\linewidth, angle=0]{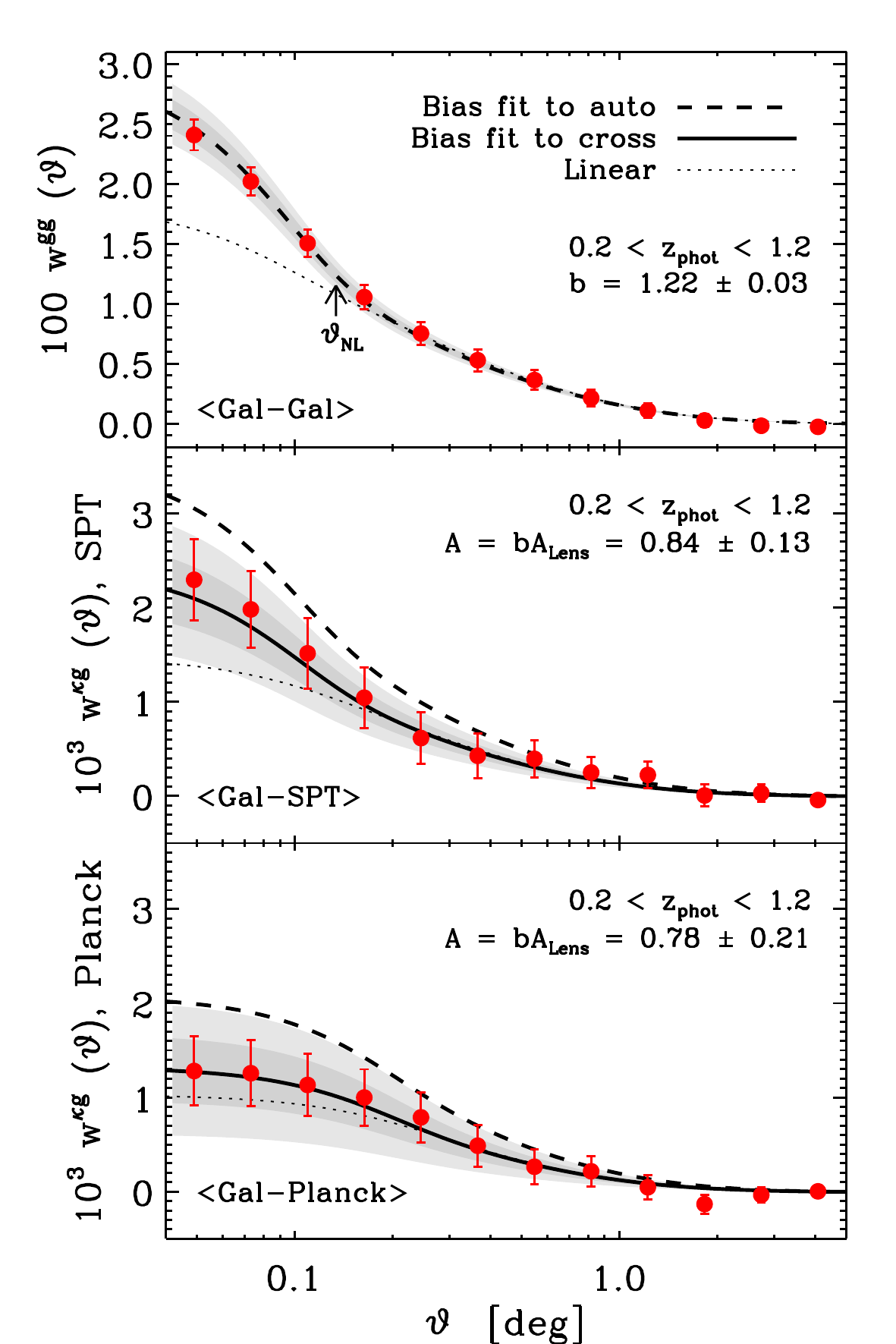}
\caption{Measured two-point correlation functions of DES-SV main galaxies and their correlations with CMB lensing maps. The red dots show the measured results using our full galaxy catalogue. 
The top panel shows the galaxy auto-correlation, the central panel is the correlation with SPT lensing convergence, while the bottom panel shows the same with \Planckc. 
The thick lines show the theoretical expectations from our \Planck fiducial cosmology, rescaled by the best-fit bias $b$ to the auto-correlation (\emph{dashed}) and best-fit amplitude $A = b A_{\mathrm{Lens}}$ to the cross-correlation functions (\emph{solid}). The thin dotted lines refer to linear theory; the scale below which linear and non-linear theories differ by $> 20 \%$, $\vartheta_{\mathrm{NL}}$, is marked in the first panel. The dark and light gray bands represent the $1$ and $2\sigma$ uncertainties on the best fit respectively.
The error bars are from the $N$-body covariance, and they are highly correlated.
The correlation shapes for DES-SPT and DES-\Planck correlations differ because the \Planck map is smoothed on larger scales.
}
\label{fig:results_w}
\end{center}
\end{figure}

\begin{figure}
\begin{flushleft}
\includegraphics[width=0.32\linewidth, angle=0]{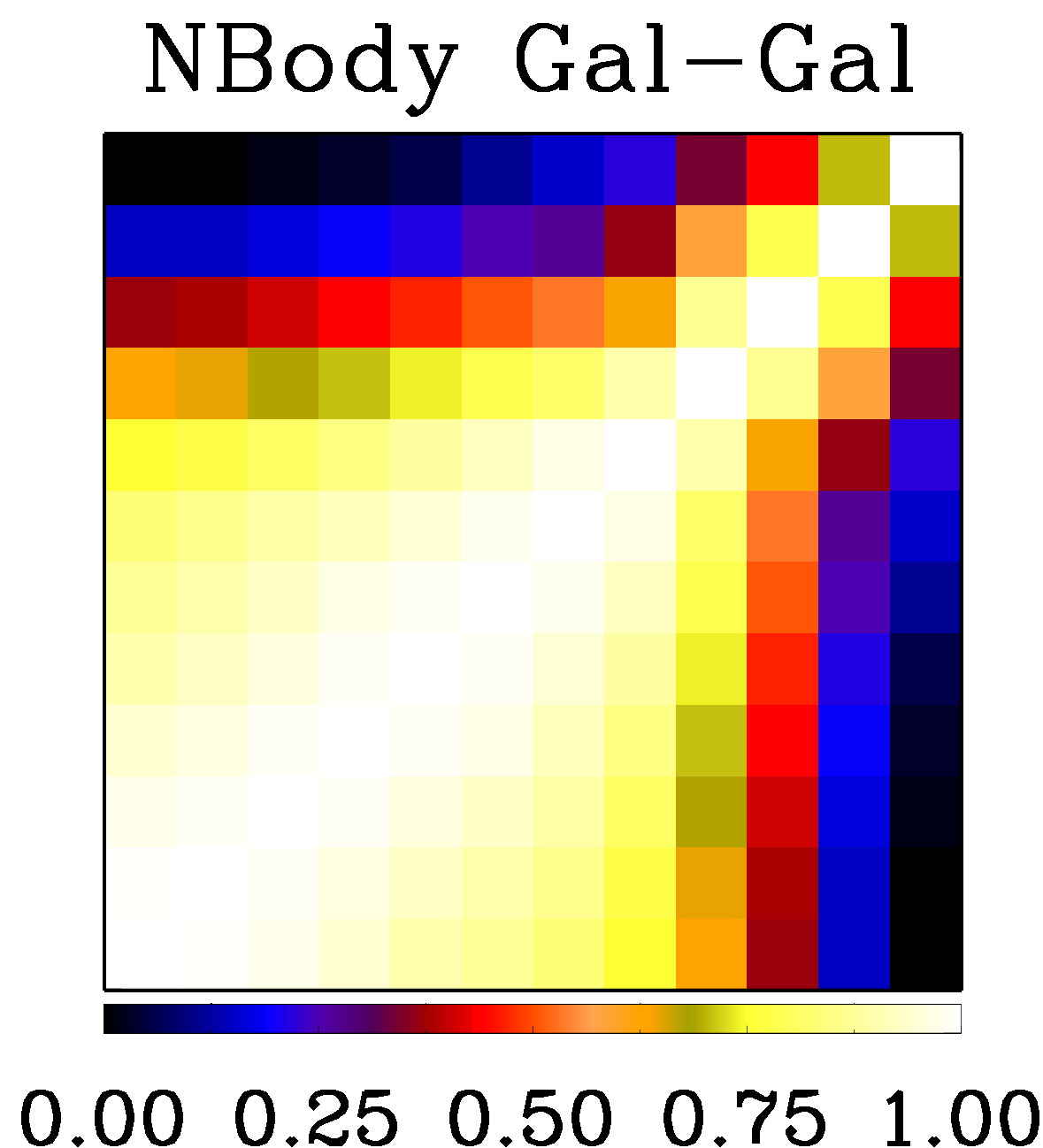}
\includegraphics[width=0.32\linewidth, angle=0]{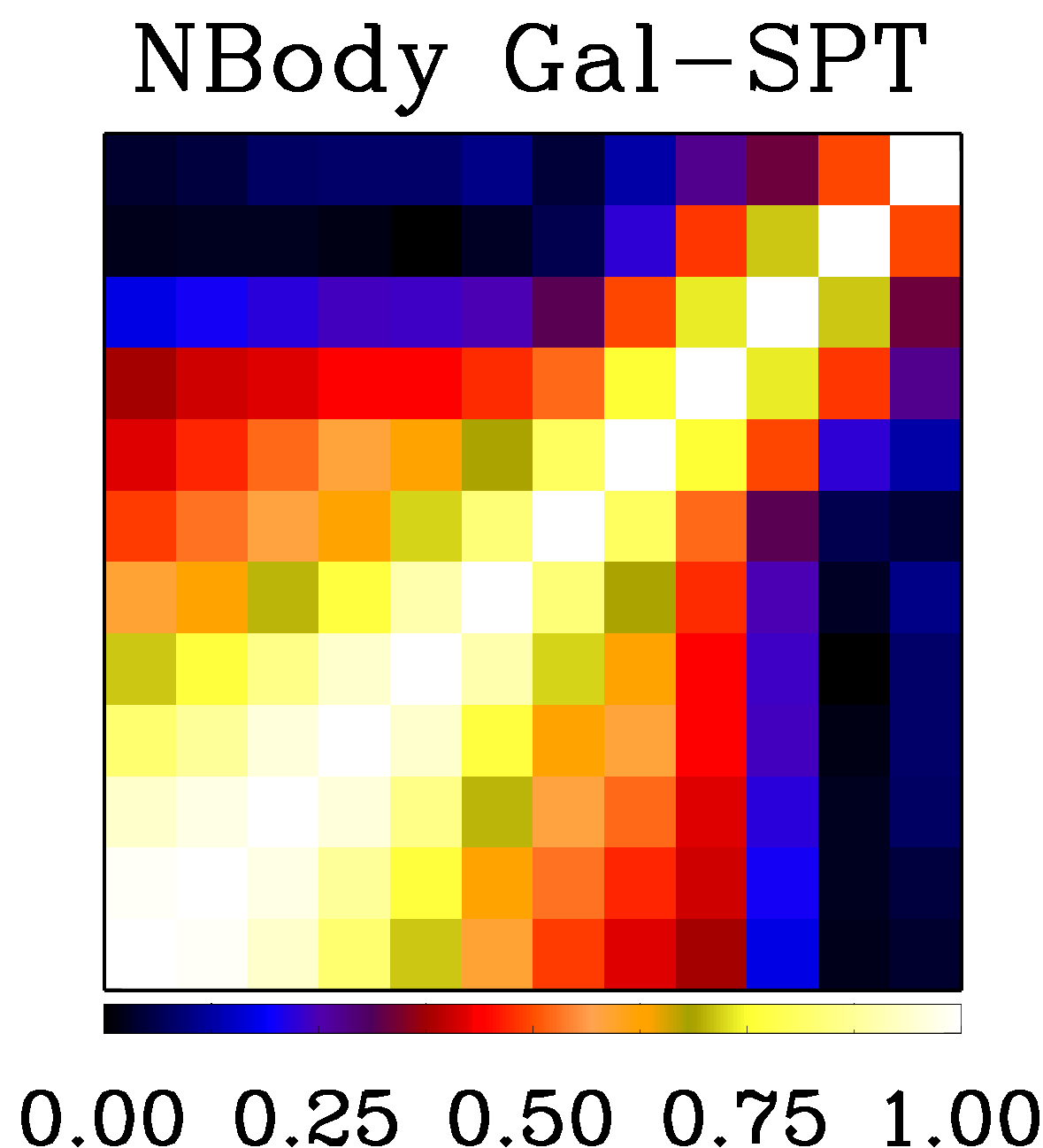}
\includegraphics[width=0.32\linewidth, angle=0]{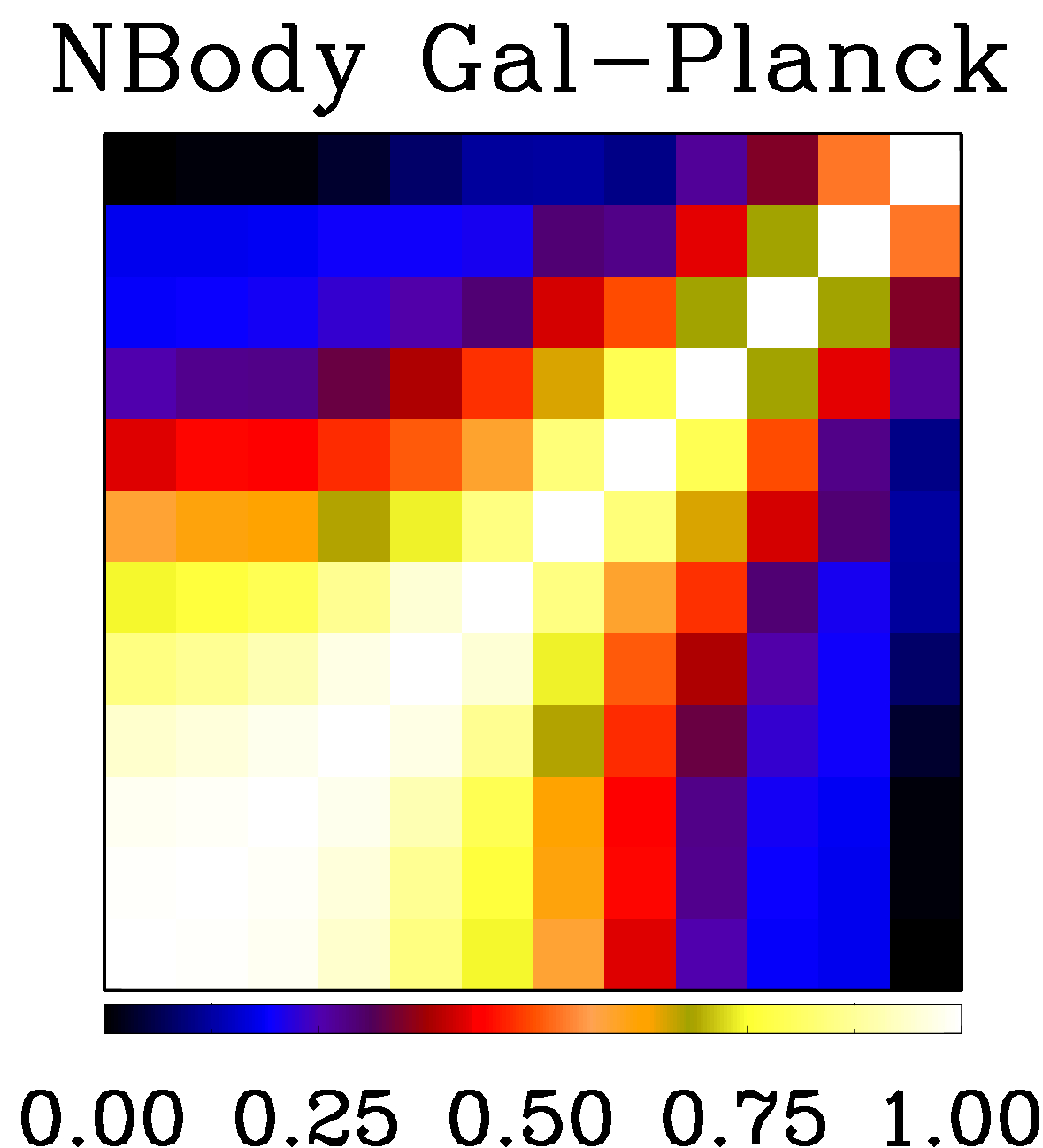}
\caption{Correlation matrices for the three cases we consider, estimated with the $N$-body method.
  The matrices refer to galaxy-galaxy, galaxy--SPT, and galaxy--\Planck lensing respectively. The angular range is from 2.4 arcmin to 5 deg as in Fig.~\ref{fig:results_w}.
  We see that the galaxy-CMB lensing correlation matrix is more diagonal than the galaxy-galaxy case, as the auto-correlation theory is more non-linear, and thus more non-Gaussian and less diagonal. Furthermore, all matrices become less diagonal in the first few angular bins due to the introduction of the Gaussian smoothing to the maps, which effectively blurs information on scales  $\vartheta < \vartheta_{\mathrm{FWHM}} = 5.4'$ (DES-SPT) and  $ 10.8'$ (DES-\Planckc).}
\label{fig:corrmat}
\end{flushleft}
\end{figure}

\begin{table*}
\begin{center}
\begin{tabular}{c c c c c c c c}
\toprule
               \multicolumn{2}{c} {Full sample,   $0.2 < \zphot < 1.2$}  &     \multicolumn{3}{c} {Real space }              &   \multicolumn{3}{c} {Harmonic space}    \\
\midrule 
 Correlation & Covariance  &  $b \pm \sigma_b$ &  S/N  & $\chi^2 / $ d.o.f.  & $b \pm \sigma_b$ &  S/N  & $\chi^2 / $ d.o.f.  \\ 
 \midrule
 Gal-Gal         &$N$-body& $1.22 \pm 0.03 $  & 41  &  3.8 / 8  &  $ 1.22 \pm 0.04  $  & 34 & 2.7 / 3 \\
\midrule 
 Correlation & Covariance  &  $A \pm \sigma_A$ &  S/N  & $\chi^2 / $ d.o.f.   &  $A \pm \sigma_A$ &  S/N  & $\chi^2 / $ d.o.f.   \\ 
\midrule
Gal-SPT         & $N$-body& $0.84 \pm 0.13  $  & 6.3  &  8.4 / 11  &  $ 0.84 \pm 0.15  $   & 5.6 & 8.7 / 19  \\
%\midrule 
Gal-\Planck     &         & $0.78 \pm 0.21  $  & 3.7  &  11 / 10 &  $0.81 \pm 0.20  $   & 3.8 & 7.7 / 9  \\
\bottomrule
\end{tabular}
\caption{Summary of the results for the main galaxy sample for real (\emph{left}) and harmonic (\emph{right}) spaces:  best-fit linear bias $b$ and correlation amplitudes $A = b A_{\mathrm{Lens}}$ for the three correlation functions and the $N$-body covariance estimator. The results are consistent between each other and with respect to the theoretical expectations for our fiducial model, but the cross-correlation amplitude is lower than the auto-correlation by $2-3 \sigma$. The recovered $\chi^2$ per degree of freedom indicates the models and covariance estimators are in all cases appropriate for the data.}
\label{tab:A}
\end{center}
\end{table*}

We show in Fig.~\ref{fig:results_w} the measured two-point correlation functions  in real space of the DES-SV main galaxies in the SPT-E field. The three panels show, from top to bottom, the galaxy auto-correlation function, and the cross-correlation functions with SPT and \Planck CMB lensing.

We compare the measurements with the predictions from our fiducial cosmology, where we use the non-linear matter power spectrum from the \textsc{Halofit} formalism \citep{2003MNRAS.341.1311S, 2012ApJ...761..152T}.
We fit the amplitudes of auto- and cross-correlations given this model, binned consistently with the data, with simple one-parameter likelihood fits.

\begin{figure*}
\begin{center}
\includegraphics[width=0.45\linewidth, angle=0]{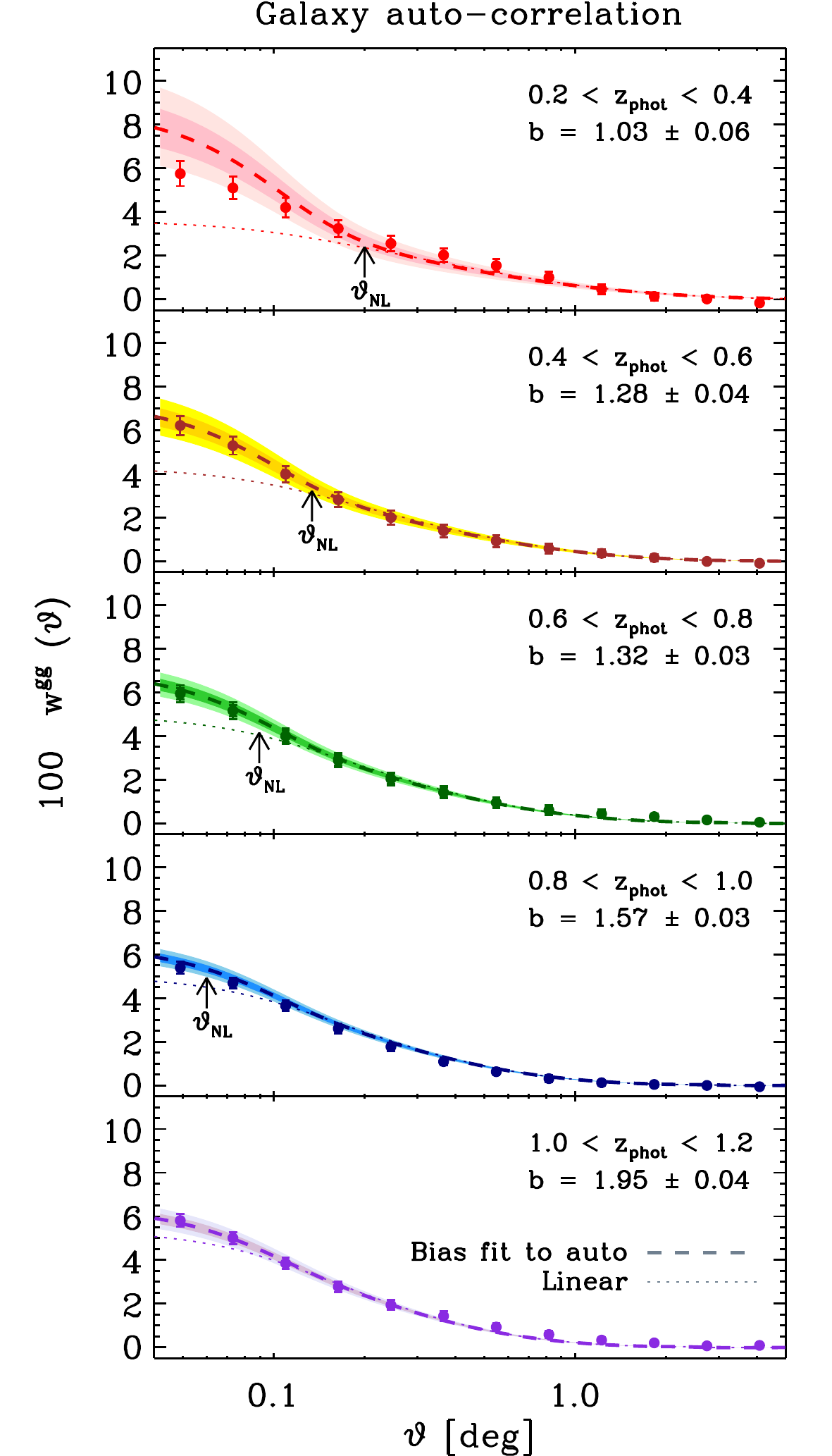}
\includegraphics[width=0.45\linewidth, angle=0]{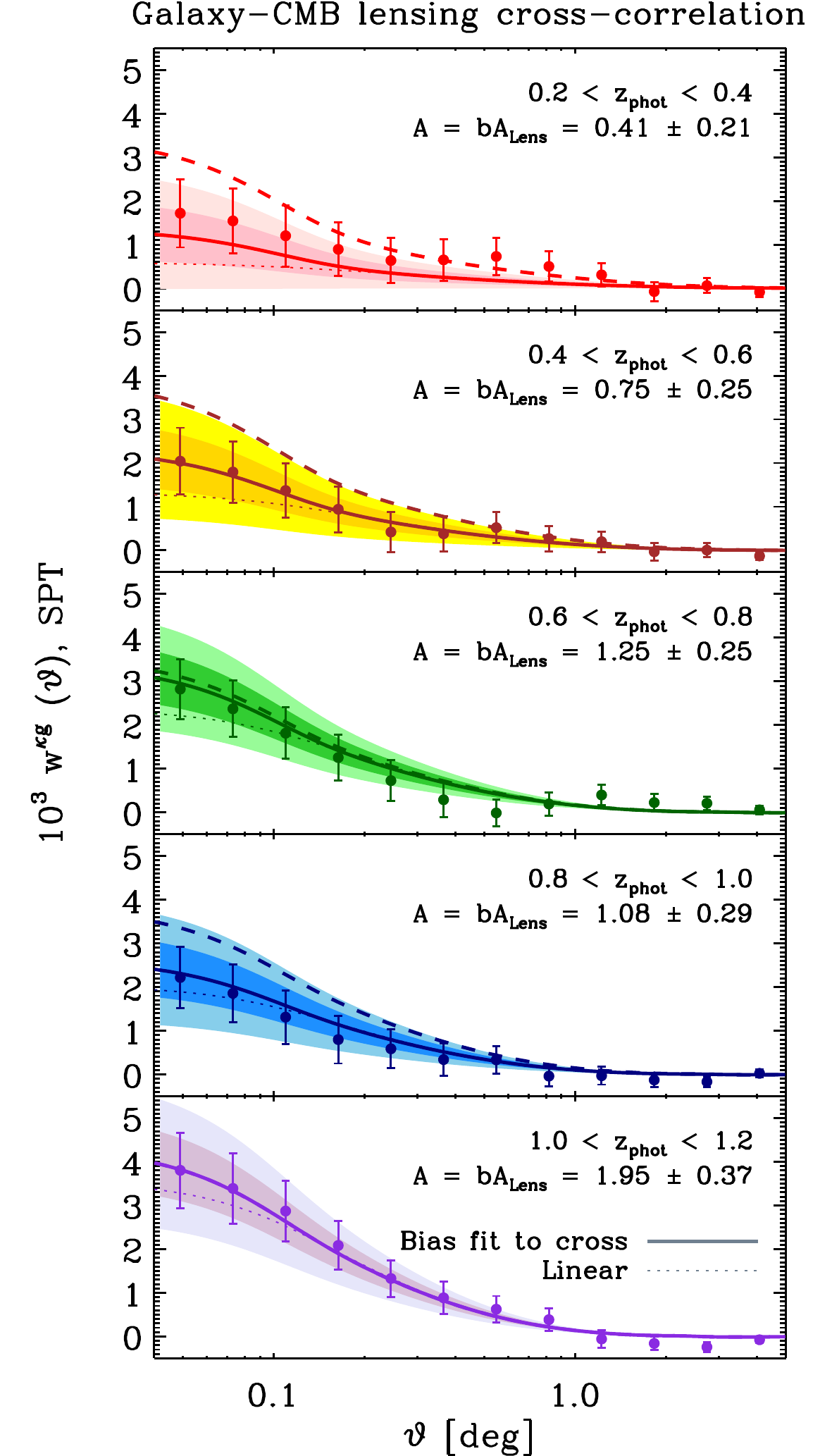}
\caption{Measured auto- (\emph{left}) and cross-correlation functions (\emph{right}) of DES-SV main galaxies as a function of photometric redshift. The panels refer to thin photo-$z$ bins, from low to high redshift. The error bars are derived from the $N$-body covariance matrix. The lines show the fiducial \Planck cosmology rescaled by the best-fit linear bias or amplitude obtained from the auto- (\emph{dashed}) and from the cross-correlations (\emph{solid}); for each case, the linear theory is shown with thin dotted lines. The best-fit bias values and their $1\sigma$ errors are also shown in each panel; the coloured bands represent $1$ and $2\sigma$ uncertainties on the best fits. When fitting the auto-correlation bias, the points at $\vartheta < \vartheta_{\mathrm{NL}}$ have been excluded from the fit, consistently with \citet{CrocceACF}, as they lie in the non-linear regime where the non-linear corrections are $>20\%$. All points are included in the cross-correlation fits. The auto-correlation results are presented and discussed in more detail by \citet{CrocceACF}, including a further discussion on the anomalous behaviour of the lowest-redshift bin at small angular scales.}
\label{fig:realtomography}
\end{center}
\end{figure*}

In the case of the auto-correlation we determine the galaxy bias $b$, assumed constant and linear. Given the comparatively large effect of non-linearities compared with the statistical error bars, and in order to obtain a physically meaningful value for the linear galaxy bias, we restrict the fit to the bins at angular scales $\vartheta > \vartheta_{\mathrm{NL}}$, where $\vartheta_{\mathrm{NL}}$ is defined as the scale where the non-linear auto-correlation function diverges from the linear theory by $> 20\%$. 
In the case of the cross-correlations, our main purpose is instead to extract as much signal as possible, and the theoretical uncertainties due to non-linearities are much smaller than the statistical errors. For these reasons, we fit in this case the overall amplitude $A$ to the galaxy-CMB lensing cross-correlation functions at all scales. For the DES-\Planck correlation, we exclude the first angular bin, as it is $\sim 100\%$ correlated with the second bin due to the larger smoothing applied.

We can see in Fig.~\ref{fig:results_w} that the galaxy auto-correlation is in agreement with our fiducial \LCDM model with a linear bias $b = 1.22 \pm 0.03$ ($N$-body covariance).
The physically crude approximation of an effective average bias across the full redshift range is actually able to correctly model the observed auto-correlation of the full galaxy sample; we study in our tomographic analysis below the actual redshift evolution of galaxy bias.
The CMB lensing cross-correlations prefer a 
lower amplitude: $A = 0.84 \pm 0.13$ and $A = 0.78 \pm 0.21$ using the SPT and \Planck maps, respectively. 
These results are quoted for our most reliable covariance matrix ($N$-body), which we show in Fig.~\ref{fig:corrmat} for the three correlation functions considered; we present in Appendix \ref{sec:appendcovariance} a detailed comparison of the four covariance matrix estimators, where we demonstrate consistency and robustness of both diagonal and off-diagonal elements.

We estimate the significance of the detections by evaluating the best fits of the linear bias $b \pm \sigma_b$ and amplitude $A \pm \sigma_A$ for the auto- and cross-correlations obtained with a simple one-parameter $\chi^2$ fit from the measured correlation functions.
We show a summary of the results in the left section of Table~\ref{tab:A},
from which we can already anticipate that the real and harmonic-space results presented in Section~\ref{sec:pseudo_cl} below yield consistent results in all cases.
For both SPT and \Planckc, the cross-correlation amplitude is lower than the auto-correlation by $2-3 \sigma$. We later discuss possible explanations for this result: in Section~\ref{sec:systematics} we discuss systematic uncertainties, and in Section~\ref{sec:implications} we discuss possible cosmological interpretations.
If we define the final significance of the detection to be $A/\sigma_A$, we find it to be $\sim 6 \sigma$ for the DES-SPT and $\sim 4 \sigma$ for the DES-\Planck cases respectively. These numbers should be compared with the (ideal) theoretical signal-to-noise levels to be expected from Eq.~(\ref{eq:errcl}), which are $\sim 8$ and $\sim 5$ respectively. Hence our results are consistent with the expectations; the lower significance recovered is mainly due to the actual best fit being lower than expected in the fiducial model, and to the more realistic $N$-body covariance matrix we use.
Finally, we see that our best fits are in most cases good fits, as the $\chi^2$ per degree of freedom is generally close to (or below) unity, which confirms that our estimate of the covariance is realistic given the scatter observed in the data.

\subsubsection{Redshift tomography in real space}
Given the significance of the recovered detection in the DES-SPT case, we then study the evolution of the correlations as a function of redshift.
We measure the DES-SPT cross-correlations in each of the photo-$z$ bins, and we present the results
in Fig.~\ref{fig:realtomography}. 
The covariances are estimated with the most reliable $N$-body method only, constructed for each redshift bin from its photo-$z$ redshift distribution, and assuming in each case a constant bias equal to the best fit to that bin's auto-correlation (we cross-checked that analytic covariances yield consistent results on the scales we consider).

We fit from each bin auto-correlation the best-fit bias $b$, considering only quasi-linear scales $\vartheta > \vartheta_{\mathrm{NL}}$,  where non-linearities are less than 20\% of the total auto-correlation function; we see that $\vartheta_{\mathrm{NL}}$ decreases in redshift as expected, allowing us to consider all data points in the highest-redshift bin. We fit the cross-correlation amplitude $A = b A_{\mathrm{Lens}}$ from the DES-SPT lensing cross-correlations, using in this case all the available scales, as discussed above for the full sample.

We can see that the auto-correlation observations are in agreement with our fiducial model and a set of constant linear bias parameters that increase with redshift. The bias values we obtain are fully consistent with the main results by \citet{CrocceACF}, thus validating both analyses.
In the cross-correlation case, we also find an agreement with the same model, although the uncertainties and the scatter are larger than what we find for the full sample, especially at low redshift.
Both auto- and cross-correlations agree less well with the expectations in the first bin at $0.2 < \zphot < 0.4$; see \citet{CrocceACF} for a more detailed discussion of the possible residual systematics in this bin.

We summarise in Table~\ref{tab:summary} the best-fit biases and amplitudes of the cross-correlations with their errors, assumed Gaussian. We see that we do recover a significant correlation (at $> 2 \sigma$) in all bins and $> 3 \sigma$ in all but the lowest redshift bin; however, the best-fit cross-correlation amplitude recovered fluctuates significantly with respect to the expectation, and with respect to the best-fit bias. We see that the trend of obtaining $A(z) < b(z)$ is recovered in most redshift bins, confirming  what we find for the full sample.
We also show that the reduced $\chi^2$ associated with the best-fit bias and amplitudes are close to 1 in most cases, indicating that our estimate of the covariances is realistic, and that our best-fit model is consistent with the observations. The only notable exceptions are the galaxy auto-correlations in the first and last redshift bins.
We discuss below in Section~\ref{sec:implications} the cosmological implications of these results.

\begin{table*}
\begin{center}
\begin{tabular}{c c c c c c c c c}
\toprule
           &     \multicolumn{2}{c} {Redshift tomography}  &     \multicolumn{3}{c} {Real space }              &   \multicolumn{3}{c} {Harmonic space}    \\
\midrule
  Correlation  &Covariance &  Photo-$z$ bin        &    $ b \pm \sigma_b $   &  S/N & $\chi^2 / $ d.o.f. &    $ b \pm \sigma_b$   &  S/N & $\chi^2 / $ d.o.f. \\ 
\midrule    
  Gal-Gal      &$N$-body   &   $0.2 < \zphot < 0.4$    &    $  1.03 \pm 0.06  $   & 17  &  20 / 7 &  $ 1.14 \pm 0.05  $  & 22 & 1.4 / 1 \\
               &           & $0.4 < \zphot < 0.6 $   &    $  1.28 \pm 0.04  $   & 31  &  2.2 / 8 &  $ 1.29 \pm 0.05  $  & 28 & 0.6 / 3  \\
               &           & $0.6 < \zphot < 0.8 $   &    $  1.32 \pm 0.03  $   & 46 &  6.9 / 9 & $ 1.29 \pm 0.03  $  & 40 & 2.7 / 5  \\
               &           & $0.8 < \zphot < 1.0 $   &    $  1.57 \pm 0.03  $   & 59  &  4.3 / 10 &  $ 1.58 \pm 0.03  $ & 54 & 2.5 / 7  \\
               &           & $1.0 < \zphot < 1.2 $   &    $  1.95 \pm 0.04  $   & 50  &  29 / 11 & $ 1.98 \pm 0.05  $  & 44 & 26 / 9  \\
\midrule
  Correlation  &Covariance &  Photo-$z$ bin        &    $ A \pm \sigma_A $   &  S/N & $\chi^2 / $ d.o.f. &    $ A \pm \sigma_A$   &  S/N & $\chi^2 / $ d.o.f. \\ 
\midrule    
     Gal-SPT      &$N$-body    &  $0.2 < \zphot < 0.4$    &    $  0.41 \pm 0.21  $   & 2.0  &  10 / 11 &  $ 0.57 \pm 0.25  $  & 2.3 & 16 / 19  \\
               &           &  $0.4 < \zphot < 0.6 $   &    $  0.75 \pm 0.25  $   & 3.1  &  11 / 11 &  $ 0.91 \pm 0.22  $  & 4.2 & 24 / 19 \\
               &           &  $0.6 < \zphot < 0.8 $   &    $  1.25 \pm 0.25  $   & 5.1  &  9.5 / 11 &  $ 0.68 \pm 0.28  $ & 2.4 & 29 / 19  \\
               &           &  $0.8 < \zphot < 1.0 $   &    $  1.08 \pm 0.29  $   & 3.8  &  7.3 / 11 &  $ 1.02 \pm 0.31  $ & 3.3 & 22 / 19 \\
               &           &  $1.0 < \zphot < 1.2 $   &    $  1.95 \pm 0.37  $   & 5.3  &  9.3 / 11 &  $ 1.83 \pm 0.42  $ & 4.4 & 23 / 19 \\
\bottomrule   
\end{tabular}
\caption{Summary of the main results of the redshift tomography in real and harmonic spaces. The top half of the table shows the best-fit biases $b$ to the DES auto-correlations, while the lower half illustrates the best fits to the DES-SPT lensing cross-correlation amplitudes $A = b A_{\mathrm{Lens}}$. All results are shown for the $N$-body covariance matrix. The real- and harmonic-space results are in good agreement with few exceptions, such as most notably the third bin cross-correlation, which we discuss in Section~\ref{sec:tomoCl} below. The reduced $\chi^2$ values are consistent with 1 in most cases, except the auto-correlations in the first and last redshift bins.}
\label{tab:summary}
\end{center}
\end{table*}

\subsection {Harmonic space analysis}

While measurements of the angular correlation function are formally
fully equivalent to the information contained in the power spectrum, there are fundamental differences
that warrant a detailed comparison. 
The harmonic space has some well-known advantages over real
space correlation estimators.
The covariance matrix, for a given survey mask, is more diagonal than in
real space, and measurements of the power spectrum in  
multipole bins are significantly less correlated, so that it is more straightforward 
to isolate clustering contributions at different physical scales, and to apply band-pass filters
if required.
Nonetheless, harmonic space estimators need to develop efficient
ways to deconvolve the mask, which is more difficult than in configuration space, thus making 
the analysis more expensive.
Different power spectrum estimators exist:
computationally expensive optimal estimators that extract all
information contained in the data \citep{1997PhRvD..55.5895T,
  1998PhRvD..57.2117B}, and pseudo-$C_\ell$ estimators that
are sub-optimal, but have a much lower computational complexity
\citep[e.g.,][]{2002ApJ...567....2H, 2004MNRAS.350..914C}.

\subsubsection {Power spectrum estimators}

In the following, we repeat our cross-correlation analysis in harmonic
space using two different estimators of the angular power spectra $C_{\ell}$: the pseudo-$C_{\ell}$ estimator \textsc{PolSpice} \citep{spice, 2004MNRAS.350..914C, Fosalba2004} for our main results of Sections~\ref{sec:clcovar}, \ref{sec:pseudo_cl}, \ref{sec:tomoCl}, and as a cross-check, a quadratic maximum likelihood estimator described in Section~\ref{sec:QML}. Masks and data remain the same as for the real-space analysis presented above.

We measure here the power spectra $C_\ell$ with the nearly-optimal and unbiased
pseudo-$C_{\ell}$ estimator implemented in the \textsc{PolSpice} code. 
This public code measures the two-point auto (or cross-) correlation functions $w(\vartheta)$ and the angular auto- (or cross-) power spectra $C_{\ell}$ from one (or two) sky map(s). It is based on the fast spherical harmonic transforms allowed by isolatitude pixelisations such as \Healpix; for $N_{\mathrm{pix}}$ pixels over the whole sky, and a $C_{\ell}$ computed up to $\ell = \ell_{\max}$, the \textsc{PolSpice} complexity scales like $N_{\mathrm{pix}}^{1/2} \, \ell_{\max}^2$ instead of $N_{\mathrm{pix}} \, \ell_{\max}^2$. The algorithm corrects for the effects of the masks and can deal with inhomogeneous weights given to the map pixels. In detail, \textsc{PolSpice} computes the (pseudo) $C_{\ell}$ of the map and weights/masks, calculates their (fast) Legendre transforms, i.e, the corresponding correlation functions, computes their ratio, applies apodisation if needed, and transforms back to harmonic space, where pixel deconvolution is simply applied to get the final $C_{\ell}$.

\subsubsection {Covariance matrix}\label{sec:clcovar}

Similarly to the real space case presented in Section~\ref{sec:covar}, we compute covariance matrices in harmonic space. This involves computing the covariance between different $C_\ell$  multipoles, by formally replacing the angular correlation function by the power spectrum in Eq.~(\ref{eq:covmat}) above.

We first estimate the covariance with the MC method. From our analysis we find that the covariance matrix of the galaxy-CMB lensing cross-correlation is approximately diagonal up to $\ell_{\max}=2000$, for a multipole bin width $\Delta \ell = 98$. We only sample scales down to $\ell_{\min}=30$ (i.e., $\vartheta < 6$ deg), as lower multipoles are poorly constrained by DES-SV data over the SPT-E area. This yields 20 multipole bins in the $\ell$ range used. 
\begin{figure}
 \begin{center}
\includegraphics[width=\linewidth, angle=0]{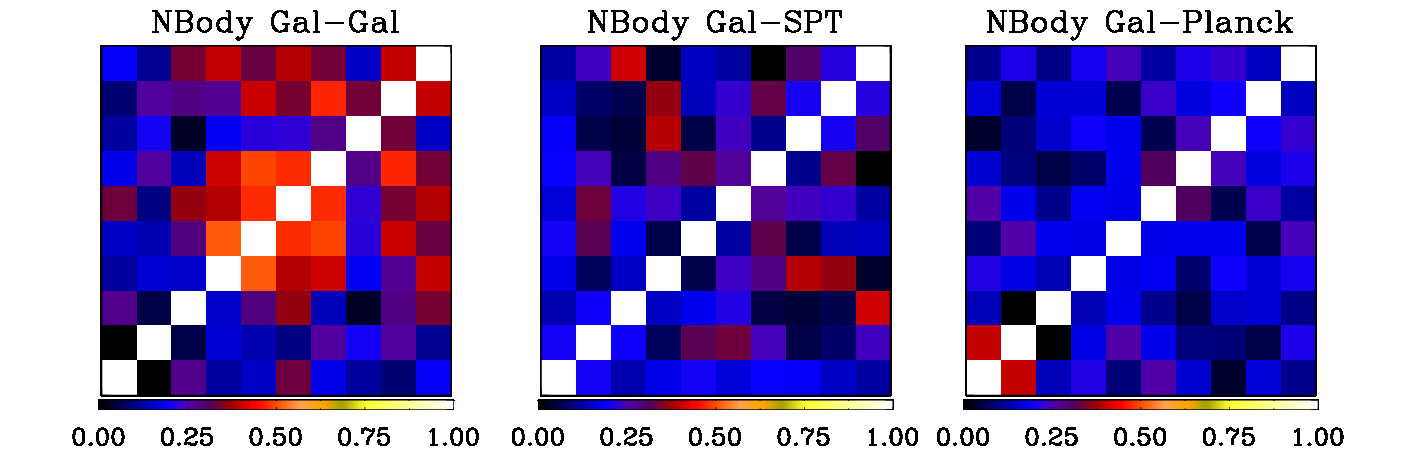}
%bincovtl_2.eps}
   \caption{Correlation matrices from  $N$-body  realisations in harmonic space: we show correlations among $C_\ell$ band-powers for the galaxy auto-, galaxy-SPT lensing and galaxy-\Planck cross-correlations, from left to right, respectively. 
We use ten linear multipole band-powers from $\ell_{min}=30$ to $\ell_{max}=2000$, with $\Delta_{\ell} = 197$, matching the bins of Fig.~\ref{fig:clstpz}.
} 
 \label{fig:Clcov}
 \end{center}
 \end{figure}

\begin{figure}
\begin{center}
\includegraphics[width=\linewidth, angle=0]{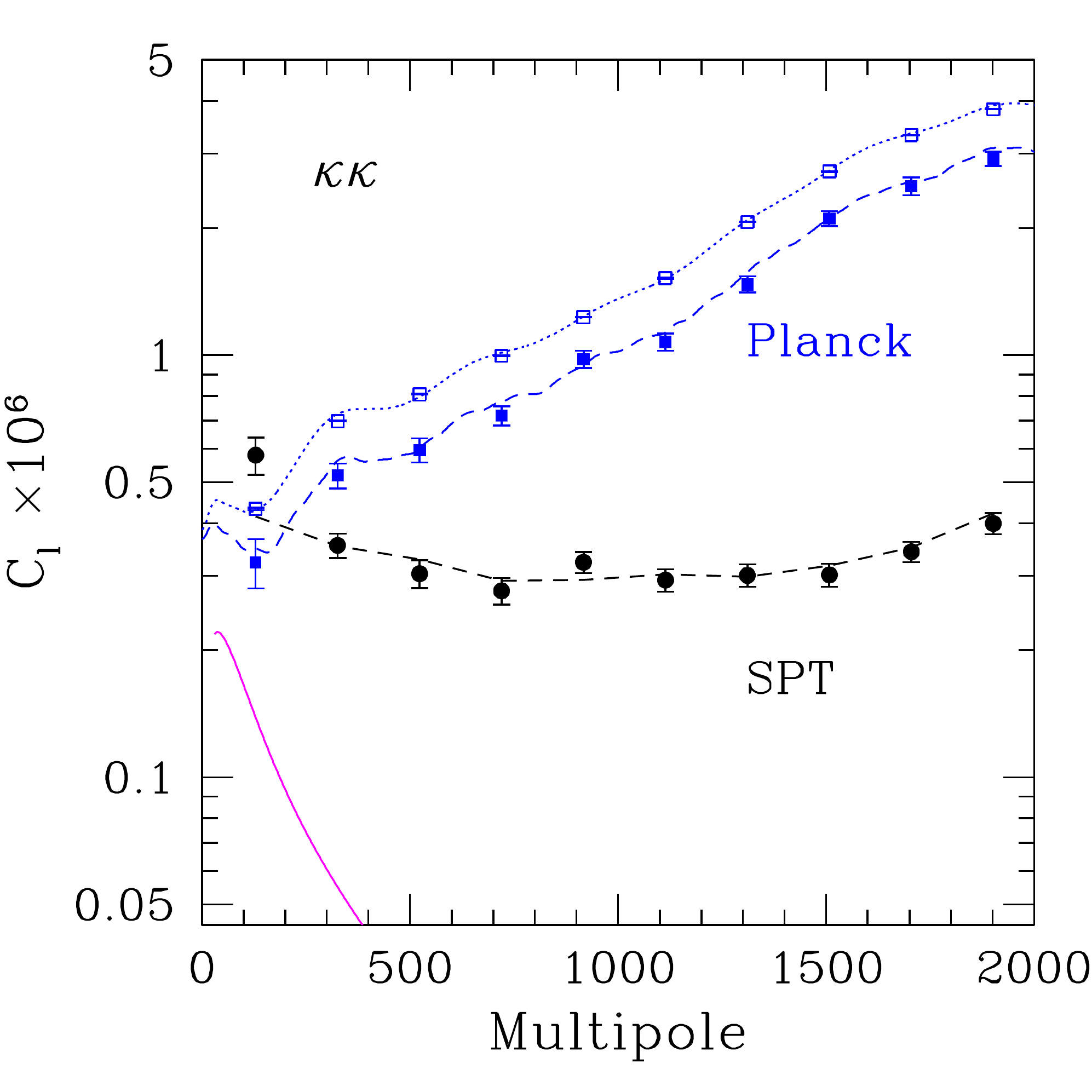}
\caption{Auto-spectra measured from the CMB lensing convergence maps (\emph{points with error bars}) from \Planck (\emph{blue squares}) and SPT (\emph{black circles}), compared with the fiducial cosmological signal (\emph{magenta solid line}). The dashed lines describe the average of 100 mock realisations that fully characterise the \Planck and SPT maps respectively. For the \Planck case we show two sets of data: measured over the full \Planck lensing mask (\emph{dotted line and empty points}), and over the intersection of the lensing mask with the DES-SV SPT-E mask (\emph{dashed line, full points}). We can see that the convenient position of the DES-SV SPT-E area next to the South Ecliptic Pole results in a 25\% noise reduction. No smoothing is applied to the maps for this figure.}
\label{fig:clstpzKK}
\end{center}
\end{figure}
\begin{figure}
\begin{center}
\includegraphics[width=\linewidth, angle=0]{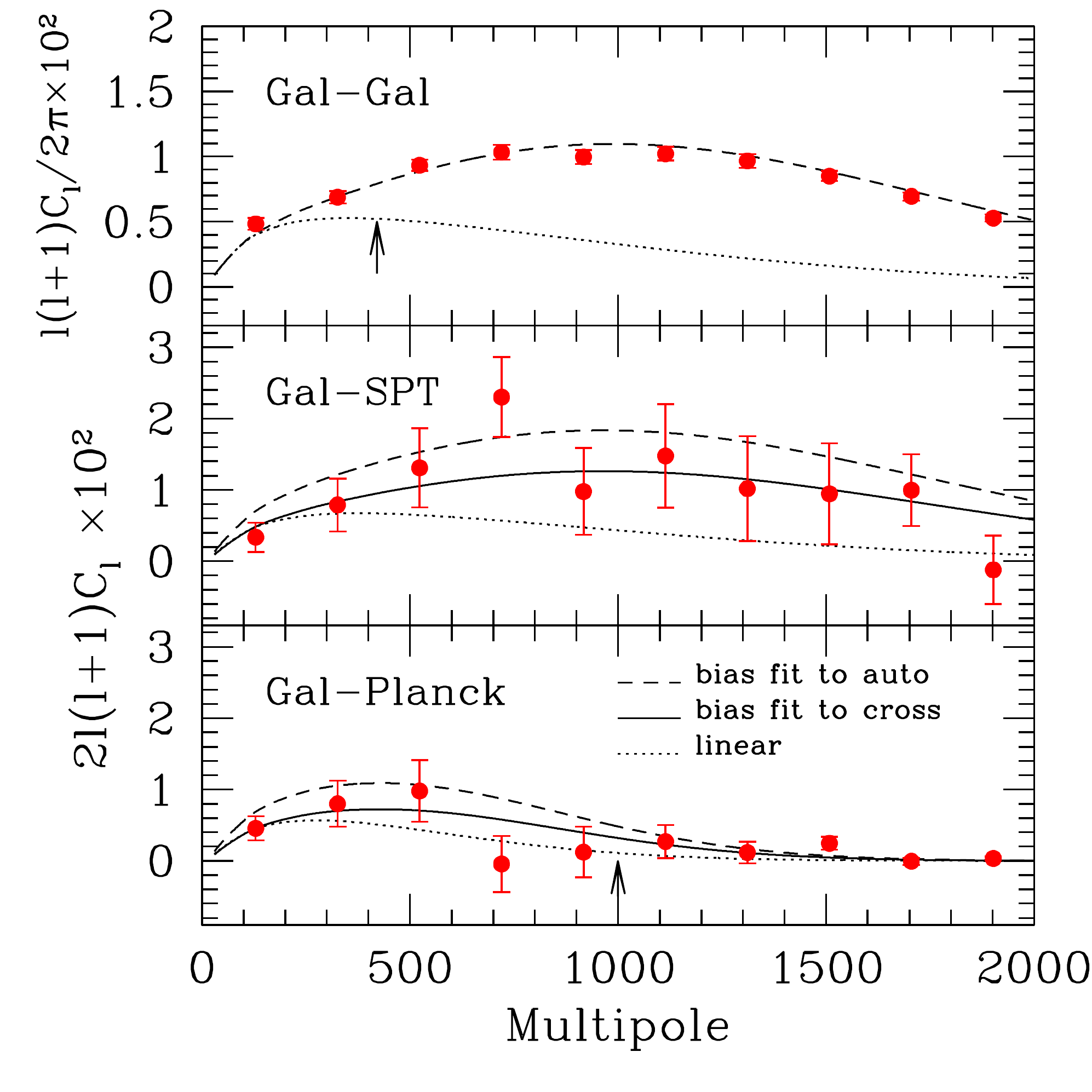}
\caption{Auto- and cross-correlations between our DES main galaxy sample and the CMB lensing convergence, in harmonic space. The first panel shows the galaxy auto-spectrum, while the second and third panels refer to the galaxy-SPT and galaxy-\Planck CMB lensing cross-spectra respectively. The lines show our fiducial cosmology rescaled by the best-fitting constant bias and amplitude to the auto-spectrum (\emph{dashed}) and to the cross-spectra (\emph{solid lines}). Dotted lines refer to linear theory. The arrow in the first panel indicates the multipole $\ell_{\mathrm{NL}}$ after which the full non-linear auto-correlation theory exceeds linear theory by $> 50\%$, which is our cutoff in the galaxy bias fit, while the arrow in the bottom panel indicates our cutoff scale for the DES-\Planck correlation at $\ell < 1000$. The amplitude of the cross-correlation is fit using $30 < \ell < 2000$ for DES-SPT and $30 < \ell < 1000$ for DES-\Planckc.  The error bars are the diagonal elements of the $N$-body covariance. The different shape of the DES-\Planck correlation is due to the stronger smoothing we apply.}
\label{fig:clstpz}
\end{center}
\end{figure}

We then estimate the covariance with the $N$-body method described in Section \ref{sec:nbody}, which provides us with 100 independent, realistic realisations of the galaxy and CMB lensing maps in the SPT-E field.
We derive the correlation matrices using the normalisation of Eq.~(\ref{eq:correlationmat}).
We show in Fig.~\ref{fig:Clcov} the 
resulting binned correlation matrices for galaxy-galaxy, galaxy-SPT and galaxy-\Planck lensing convergence. In particular, the covariances show band-powers of  $\ell (\ell+1)C_{\ell}^{gg} $ and $2\ell (\ell+1) C_\ell^{g\kappa}$.
We find that the galaxy-CMB lensing and CMB lensing-CMB lensing covariances stay block diagonal even when non-linear growth is taken into account using $N$-body simulations. The galaxy-galaxy covariance, however, displays large off-diagonal elements for $\ell \gtrsim 200$ (depending on the $z$ bin) due to non-linear mode coupling that induces a non-Gaussian contribution to the covariance, sourced by the gravitational matter trispectrum.
We show in Appendix~\ref{sec:appendcovariance} a comparison of the different covariance matrix estimators in harmonic space, where we demonstrate consistency of the results.

\subsubsection {Harmonic-space results: full sample}
\label{sec:pseudo_cl}

We first show in Fig.~\ref{fig:clstpzKK} the CMB lensing auto-spectrum for SPT and \Planckc. 
 As we showed from the covariance analysis in Section~\ref{sec:clcovar}, we can bin the data in multipole bins of width 
$\Delta \ell \simeq 100$, in order to get uncorrelated bandpower measurements. 
For plotting purposes, we use broader (uncorrelated) bins with $\Delta \ell \simeq 200$, in order to get smaller errors per bandpower. As the expected true spectrum is smooth, this step is not expected to destroy any information.
In Fig.~\ref{fig:clstpzKK} we can see that for both surveys, the convergence maps are noise-dominated at all scales, as the auto-spectrum is always larger than the fiducial cosmological signal shown in magenta; SPT has higher sensitivity at small scales $\ell>300$, while \Planck has an advantage on the largest angular scales.
In the case of SPT, we see that the convergence power spectrum of the data (black points) is well characterised by the mean SPT noise over the 100 anisotropic noise realisations (dashed black line). The small ($\sim 10\%$) errors of the lensing auto-power are due to the low level of scatter among lensing noise realisations.
In the case of \Planckc, we find that the convergence spectrum over the DES-SV area (solid blue square points) is $\sim 25\%$ lower than the spectrum over the full \Planck lensing mask (empty blue squares); this is confirmed by the mean of the 100 mock \Planck lensing realisations (dashed and dotted lines for the DES-SV area and full CMB lensing mask respectively). We can understand this given the especially convenient location of the DES-SV footprint, shown in Fig.~\ref{fig:PlanckHits}, which justifies the atypical noise properties over this area.
For our theoretical and MC covariances, we use the convergence noise levels observed from the mocks over the DES-SV area, as these are the most realistic noise estimations.

\begin{figure*}
\begin{center}
  \includegraphics[width=0.45\linewidth, angle=0]{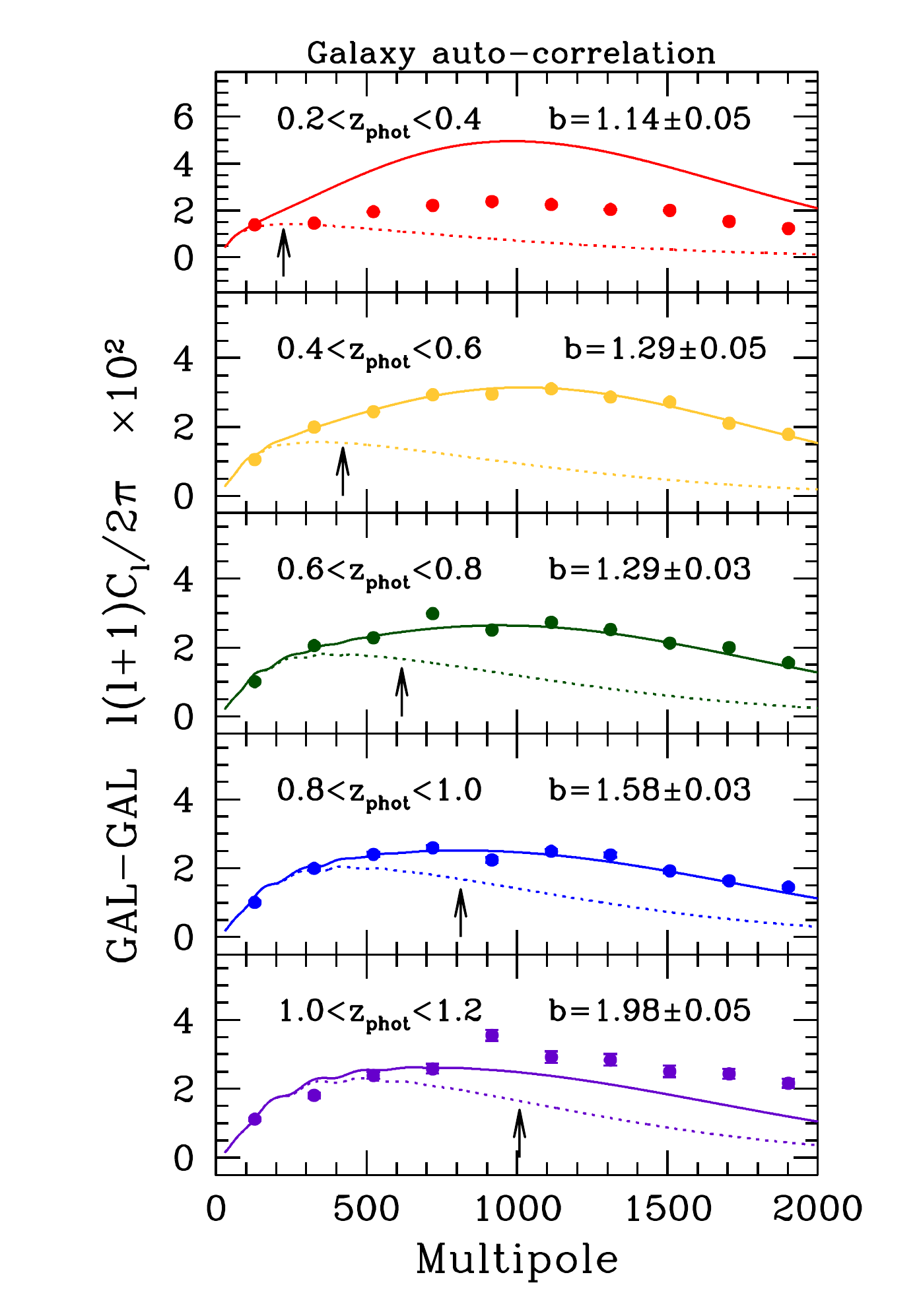}
  \includegraphics[width=0.45\linewidth, angle=0]{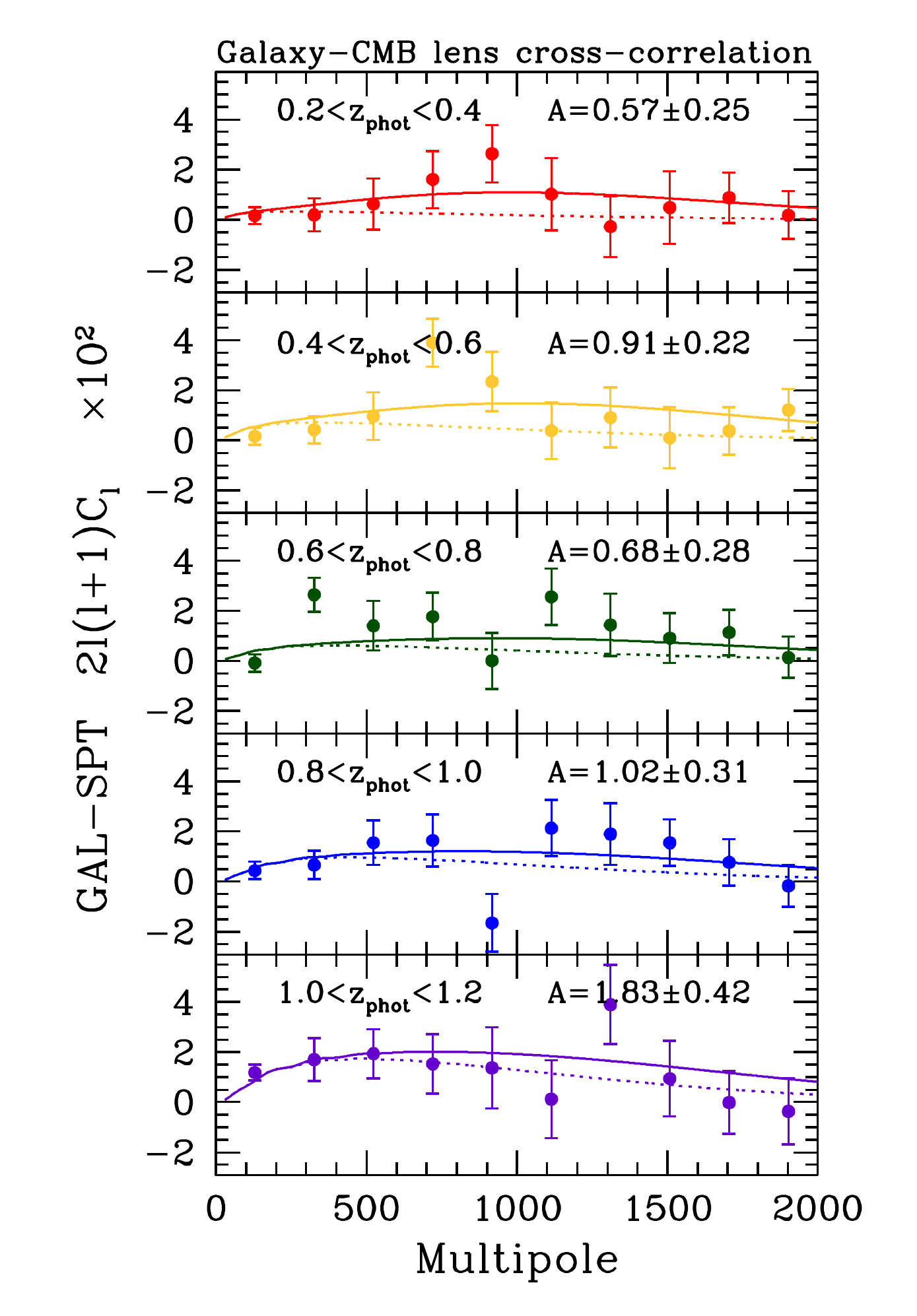}
\caption{Harmonic space redshift tomography of the auto- (\emph{left column}) and cross- (\emph{right column}) correlations. The panels from top to bottom describe the results of photo-$z$ bins of increasing redshift. The solid lines show our fiducial cosmology rescaled by the best-fit linear bias $b$ up to weakly non-linear scales $\ell_{\mathrm{NL}}$ marked with an arrow (for the auto-spectra) or the cross-correlation amplitudes $A = b A_{\mathrm{Lens}}$ over the whole range of scales (cross-spectra); the best-fit biases and amplitudes are reported in the captions with their $1\sigma$ errors. The dotted lines are linear theory predictions, and the error bars are from the full $N$-body covariance estimator. In agreement with the real-space analysis and with \citet{CrocceACF}, the auto-correlation in the lowest (and less significantly in the highest) redshift bins does not match the theoretical expectation on non-linear scales, which are discarded from our bias fits anyway.}
\label{fig:cltomo}
\end{center}
\end{figure*}

We then show in Fig.~\ref{fig:clstpz} the auto- and cross-correlations between the DES-SV main galaxy sample and 
the SPT and \Planck CMB lensing convergence,
with the diagonal errors from the $N$-body covariance.
Using the measured spectra and the $N$-body covariance matrices, we can estimate the best-fit amplitudes and corresponding detection significances for the cross-correlations.
As in the real-space analysis of Section~\ref{sec:realspace}, we apply a cut to the non-linear scales when fitting the galaxy bias $b$ from the auto-spectrum; in this case, the scale of non-linearity $\ell_{\mathrm{NL}}$ is defined so that the non-linear theory exceeds the linear model by $> 50\%$ at $\ell > \ell_{\mathrm{NL}}$; this scale is marked with an arrow in Fig.~\ref{fig:clstpz}.
This threshold is less stringent than the 20\% we use in the real-space analysis above; this is because, even with linear $\ell$ binning and logarithmic $\vartheta$ binning, the harmonic space analysis is more sensitive to non-linear scales than the real-space measurement, where information from all scales is mixed. Applying a 20\% threshold in harmonic space would leave only one data point in the galaxy auto-spectrum, while using a 50\% criterion in real space would lead to the inclusion of all data points.
 For the same motivations as above, we do not apply such a scale cutoff (beyond our Gaussian smoothing of the maps) when fitting the cross-correlation amplitudes $A$, so that we do not expect a perfect match between the two amplitudes.
The upper panel of Fig.~\ref{fig:clstpz} shows that the galaxy auto-power is best fit by our fiducial cosmology with linear galaxy bias $ b = 1.22 \pm 0.04 $, up to $\ell_{\mathrm{NL}}$ (dashed line) and assuming the $N$-body covariance. From the central panel, we see that the cross-correlation with SPT is best fit by a lower amplitude value, $ A = 0.84 \pm 0.15 $ (solid line), which is $\sim 2 \sigma$ smaller. Likewise, the bottom panel shows that the cross-correlation with \Planck is also lower than expected from the galaxy auto-spectrum: $ A = 0.81 \pm 0.20 $.

We summarise our harmonic-space results in detail in the right section of Table~\ref{tab:A}, where we show the results with the $N$-body covariance matrix.
The best-fit linear galaxy bias from the auto-spectrum is typically $\sim 2\sigma$ higher than the best-fit amplitude of the galaxy-CMB lensing cross-correlations, in agreement with what we find in real space. The cross-correlation significance of a detection is $\sim 6 \sigma$ for SPT and $\sim 4 \sigma$ for \Planckc; these numbers are in agreement with the real-space analysis results. We note that we do not expect a perfect agreement between the two analyses as they involve different estimators that weight physical scales in a different way; however, thanks to the Gaussian smoothing we apply to data and mocks, which effectively makes both estimators band-limited, we do manage to recover a good agreement. We test in Section~\ref{sec:consistency} below the consistency between the real- and harmonic-space estimators, and their degree of correlation.
We can see from the $\chi^2$ per degree of freedom that our best fits are good fits.
Finally, we point to Section~\ref{sec:sysCMB} for an analysis of the stability of the results with respect to different choices in the multipole range considered.

\subsubsection{Redshift tomography in harmonic space}
\label{sec:tomoCl}
In analogy to the real space results discussed above, we then measure the redshift tomography of the auto- and cross-spectra. The left column of Fig.~\ref{fig:cltomo} shows the auto-power of DES galaxies for the five photo-$z$ bins we consider.  The solid lines show the best-fit linear galaxy bias to the measured spectra, and the error bars are from the $N$-body estimator. In each case, we only include in the bias fit data points that are to the left of the non-linear scale $\ell_{\mathrm{NL}}$, marked with an arrow. We find that the recovered galaxy bias grows smoothly with redshift, as expected, in agreement with \citet{CrocceACF}. 
 We note that neighbouring photo-$z$ bins are significantly correlated, due to photo-$z$ errors.
In agreement with \citet{CrocceACF} and consistently with the real-space analysis, we find that for the lowest photo-$z$ bin (and more mildly for the highest one), the auto-correlation on non-linear scales disagrees with the theoretical expectations. While the harmonic-space analysis highlights this mismatch more significantly than what we see in Fig.~\ref{fig:realtomography} in real space, we remind the reader that these scales are not used for finding the best-fit bias. \citet{CrocceACF} attribute these discrepancies to possible non-linear bias in the lowest-redshift bin, and to systematic contaminations related to inaccurate photo-$z$ determination of blue galaxies in the absence of $u$-band photometry.

In the right column of Fig.~\ref{fig:cltomo} we show the corresponding cross-correlations with SPT lensing. Although the signal is clearly more noisy than the auto-spectra, we do find a 2-4$\sigma$ detection in every bin, and we also see that the cross-correlation amplitude grows with redshift, as expected, although typically we find $A(z) < b(z)$, confirming the general trend observed in the analysis of the full galaxy sample.
Note also that the scatter seen in the $C_\ell$s is larger than that of the corresponding two-point correlation functions shown in Fig.~\ref{fig:realtomography} above, since band-power measurements are much less correlated than the real-space angular bins used.

All results are summarised in detail in Table~\ref{tab:summary}, where they can be directly compared with their real-space counterparts. From this table, we can see that there is a good agreement between real- and harmonic-space analyses. One point that stands out as marginally inconsistent (at the $>2\sigma$ level) is the cross-correlation amplitude in the third photo-$z$ bin ($0.6 < \zphot < 0.8$), which is significantly lower in harmonic space than in real space. By inspecting the data in Fig.~\ref{fig:cltomo} for the third photo-$z$ bin (green), it is clear that this anomaly is driven by the low correlation observed in the first multipole bin at $\ell < 300$, also seen in real space in the form of an oscillating cross-correlation function at scales $\vartheta > 0.5 $ deg; we found that by discarding this point, we obtain an excellent agreement also in this bin, but we decide not to apply such a cut in our final result, to avoid any \emph{ad hoc} manipulation of the data.

\begin{figure}%
  \centerline{\resizebox{\hsize}{!}{\includegraphics*{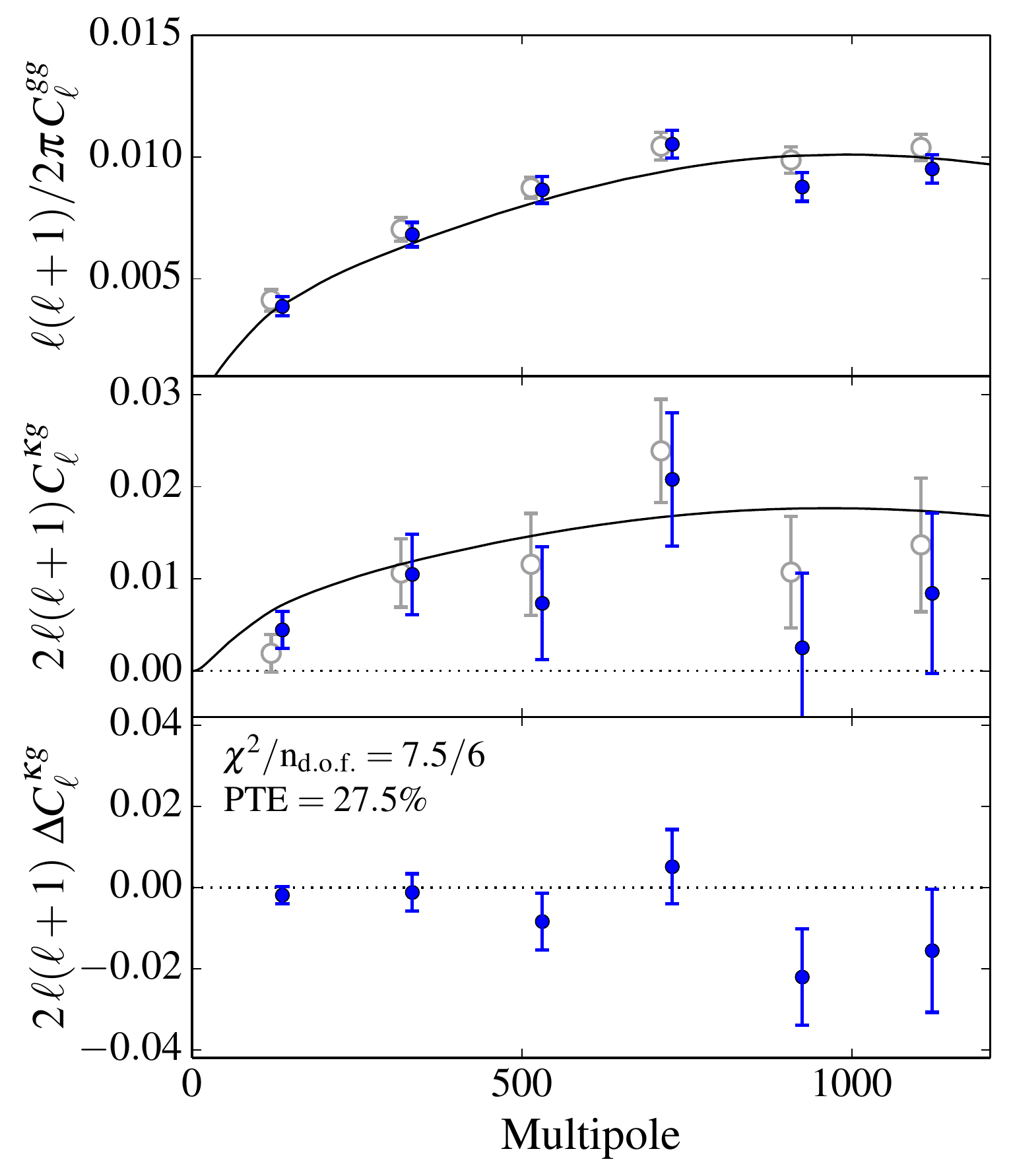}}}
%  \centerline{\resizebox{\hsize}{!}{\includegraphics*{figures/oqe_sim_test_spt_e.pdf}}}
  \caption{The power spectra derived with the optimal quadratic estimator are in
quantitative agreement with theoretical predictions and with the pseudo-$C_{\ell}$
estimator. We compare the
galaxy-galaxy (\emph{upper panel}) and the galaxy-CMB lensing potential power spectra (\emph{middle panel}) of SPT-E
(\emph{blue circles}) to the theoretical model (\emph{black solid
line}). For comparison, we include results of the pseudo-$C_{\ell}$
estimator (\emph{open grey symbols, plotted with a small offset in
$\ell$ for better visualisation}). Residuals of the cross-correlation
power spectrum, shown on linear scale (\emph{bottom panel}), are
consistent with zero within the error bars.}
  \label{fig:oqe_cl_results}
\end{figure}%

\subsubsection {Optimal quadratic estimator}
\label{sec:QML}

As a cross check of the harmonic space results with an independent
pipeline, we now present the analysis using the optimal estimator
introduced by \citet{1997PhRvD..55.5895T} in an implementation
described in \citet{2013MNRAS.435.1857L}. In short, it computes the
power spectrum estimates from a quadratic combination of the data
vector, normalised by the Fisher matrix. It can be shown that the
variance of the estimator saturates the Cram\'er-Rao bound, i.e. the
estimator is optimal. As well as producing the power spectrum estimates
themselves, the algorithm also allows the mathematically exact
calculation of the $C_\ell$ covariance matrix under the assumption of
a Gaussian signal. While some of the variance of the galaxy auto power
spectrum will not be captured (galaxy fields can more realistically be
described by a log-normal distribution), error bars of the galaxy-CMB
lensing convergence power spectrum are represented to sufficient
accuracy (see Section~\ref{sec:covar}).

Unfortunately, the computations involved in constructing the optimal
estimator scale with the third power of the number of pixels of the
input map, restricting its application to data vectors of moderate
size \citep{1999PhRvD..59b7302B}. To accommodate this requirement, we
therefore downgrade all maps to resolution $\nside = 512$. In
downgrading the mask, we set to zero all low-resolution pixels that
contain a masked pixel at the original resolution, thereby reducing
the sky fraction available.
To avoid aliasing, we further bandpass filter the data by
applying a top-hat kernel in harmonic space that restricts
fluctuations to the multipole range used in the analysis, $30 \le
\ell \le 1210$. Here, the lower limit reflects the restricted sky
coverage of the observed region, and the upper limit is a
conservative estimate up to which signals can be represented well
at the given resolution parameter \nside. 
 Since lowering the resolution comes at the expense
of a loss of information, we primarily use the method to demonstrate
robustness and independently verify the results presented in
Section~\ref{sec:pseudo_cl}. We summarise the details of the
implementation and test our analysis pipeline on simulations in
Appendix~\ref{sec:oqe_appendix}.

\begin{figure*}
\begin{center}
\includegraphics[width=0.4\linewidth, angle=0]{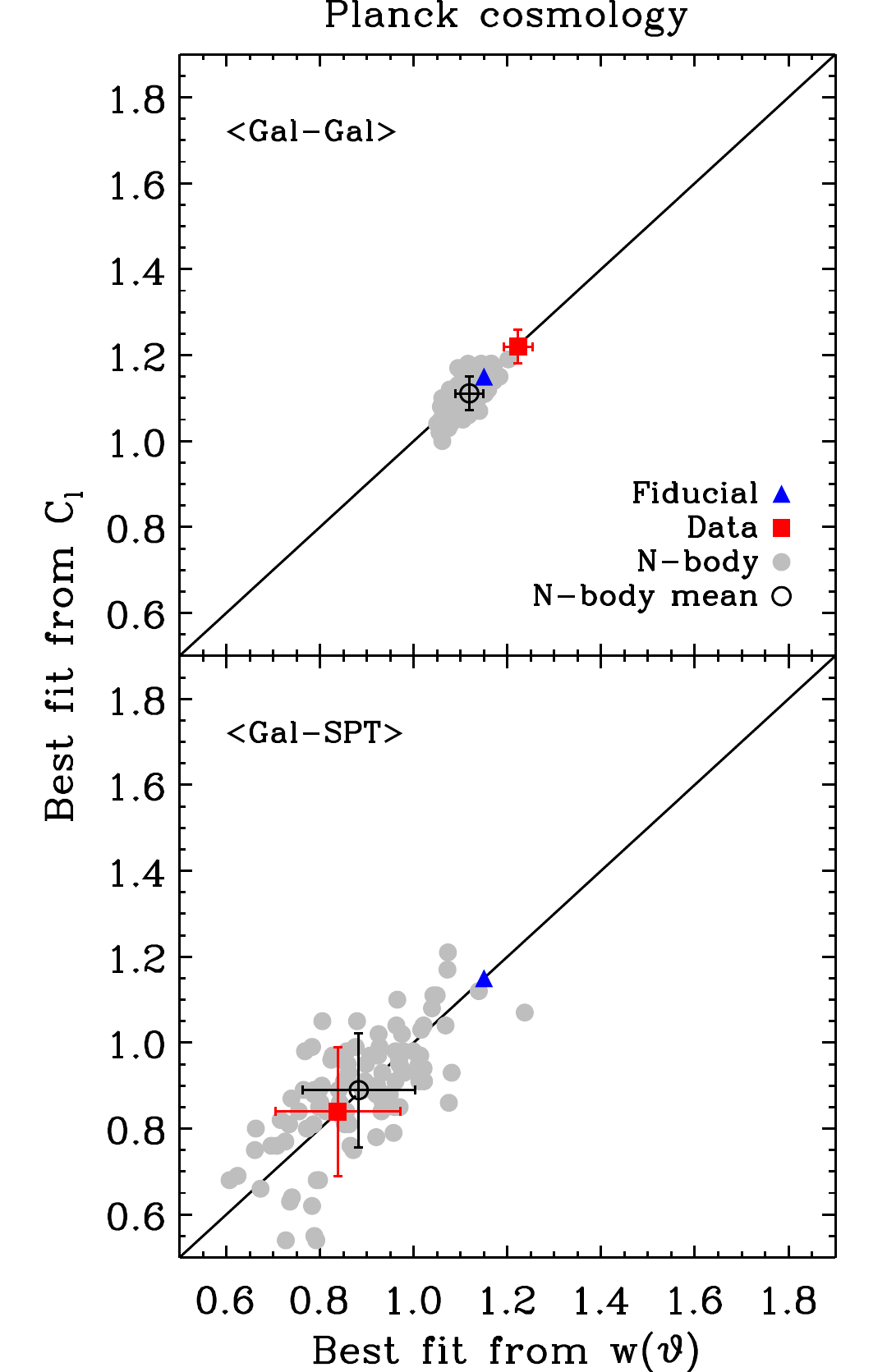}
\includegraphics[width=0.4\linewidth, angle=0]{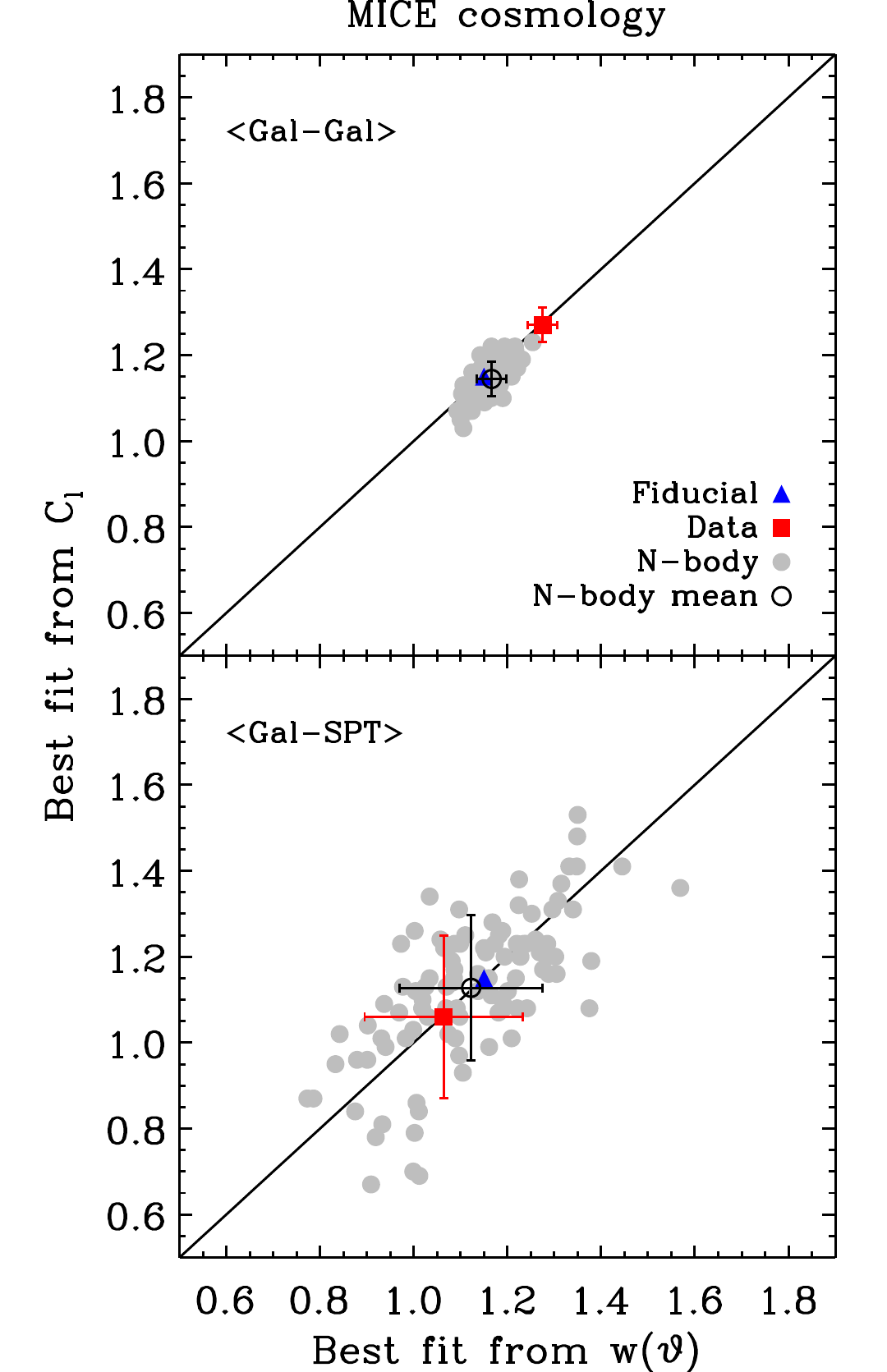}
\caption{Consistency of the results in real and harmonic spaces, assuming the fiducial \Planck cosmology (\emph{left}) and the MICE cosmology with which the mocks were generated (\emph{right}). In each case, the two scatter plots show a comparison of the best-fit bias $b$ from the galaxy auto-correlation (\emph{top}) and best-fit galaxy-CMB lensing cross-correlation amplitude $A = b A_{\mathrm{Lens}}$ (\emph{bottom}), obtained with the two methods. The red square points with error bars represent the results from real data, while the grey circles refer to the 100 $N$-body mocks. The black empty circles with error bars show the mean of the mocks and their standard deviation.  The blue triangle is the input value for the simulations ($b = A = 1.15$). The results from real data are largely consistent with the distribution of the mocks, although we see that the bias value assumed for the mocks is lower than the value recovered from DES galaxies auto-correlation, and higher than what measured from the galaxy-CMB lensing cross-correlation. The harmonic and real space estimators are correlated, but a significant scatter exists. The mock cross-correlation results are displaced from the fiducial input amplitude when they are interpreted with the \Planck cosmology, but they agree with the fiducial when interpreted assuming their own MICE cosmology. The data closely follow the behaviour of the mocks,
which in turn suggests the 
 data prefer a lower $\omega_m \sigma_8$ than expected in the \Planck cosmology. This is further discussed in 
 terms of the linear growth estimator, $D_G$, in Section~\ref{sec:implications}.}
\label{fig:scatter}
\end{center}
\end{figure*}

In Fig.~\ref{fig:oqe_cl_results}, we show results for the galaxy-galaxy and galaxy-CMB
 lensing convergence power
spectrum, convolved with the window function of a Gaussian kernel with
a FWHM of 5.4 arcmin. We treat the bias parameter as a free parameter
and obtain its numerical value from a fit to the galaxy-galaxy
 power spectrum assuming the best fit cosmological model of
\Planckc. Using analytic covariance matrices, we then compute $\chi^2$
values and reach a quantitative agreement with the theoretical model
in the multipole range probed. Compared to the results derived with
the pseudo-$C_\ell$ estimator in Section~\ref{sec:pseudo_cl}, we find a
good agreement, in particular in the intermediate multipole regime. On
large scales, the optimal estimator has an advantage since the effect
of the mask is taken into account in a mathematically exact way. At
high multipoles, however, the lower computational complexity of the
pseudo-$C_\ell$ estimator allows it to work at higher map and mask
resolution, yielding a larger effective sky area that can be
retained for analysis.

Finally, keeping the cosmological parameters fixed, we compute the
likelihood function of the galaxy bias parameter, finding a Gaussian
distribution with mean and standard deviation
 $b = 1.189 \pm 0.015 $. For the DES-SPT cross-correlation amplitude, we find
 $A \equiv b A_{\mathrm{Lens}}= 0.83 \pm 0.19$, i.e. a detection at the 4.5-$\sigma$ level. This result is in agreement with the pseudo-$C_\ell$ analysis of Section~\ref{sec:pseudo_cl}, although with a larger error bar due to the smaller range of scales probed. This further validates the robustness of our analysis.

\subsection{Consistency of the results} \label{sec:consistency}
In order to demonstrate the consistency of the results obtained with different estimators in real and harmonic spaces, we repeat our analysis, measuring auto- and cross-correlations of the 100 $N$-body mock realisations available for the DES galaxy and the SPT lensing convergence. 
We show in Fig.~\ref{fig:scatter} the best-fit bias obtained from the galaxy auto-correlations and the best-fit amplitude from the DES-SPT lensing cross-correlations, comparing the real data and $N$-body results in real and harmonic spaces.
As the $N$-body mocks we use were generated assuming the MICE cosmology, we repeat this test assuming both \Planck and MICE parameter values.

From these scatter plots, we first see that the harmonic and real-space estimators are correlated as expected, but the scatter between them is significant. Considering the 100 mock results, we obtain a Pearson correlation coefficient $r_P = 0.7$ for both auto- and  cross-correlations.
Further, we see that the results from the real data cross-correlations are largely consistent with being a random draw from the distribution of the $N$-body results; however in the auto-correlation case, the bias recovered from the real data ($b \simeq 1.2$) is marginally higher than what we assumed for producing the mocks ($b = 1.15$). While we could generate new mocks with a bias value matching the data more closely, we expect this to have only a minor effect on the covariance of the measurements: this is confirmed by the observation that the current mocks and a jack-knife method independent of bias yield consistent results, as shown in Appendix~\ref{sec:appendcovariance}. Furthermore, as the bias recovered from the cross-correlation of real data is actually $b < 1.15$, the current value can be seen as a compromise choice.
We then compare the results obtained assuming the two cosmologies. Here we see that the mocks and their mean fully agree with their input bias value when interpreted with their native MICE cosmology, as expected.
 Instead, the  values of $b$ and $A$ inferred from the mocks naturally deviate from the fiducial if the different \Planck cosmology is assumed.

If we focus again on the results from the real data, we notice that their behaviour is not dissimilar from the MICE mocks: the tension between $b$ and $A$ that is observed when the \Planck cosmology is assumed is significantly alleviated by the MICE model; we discuss this point further in Section~\ref{sec:implications} below.

\begin{figure}
\begin{center}
\includegraphics[width=0.45\linewidth, angle=0]{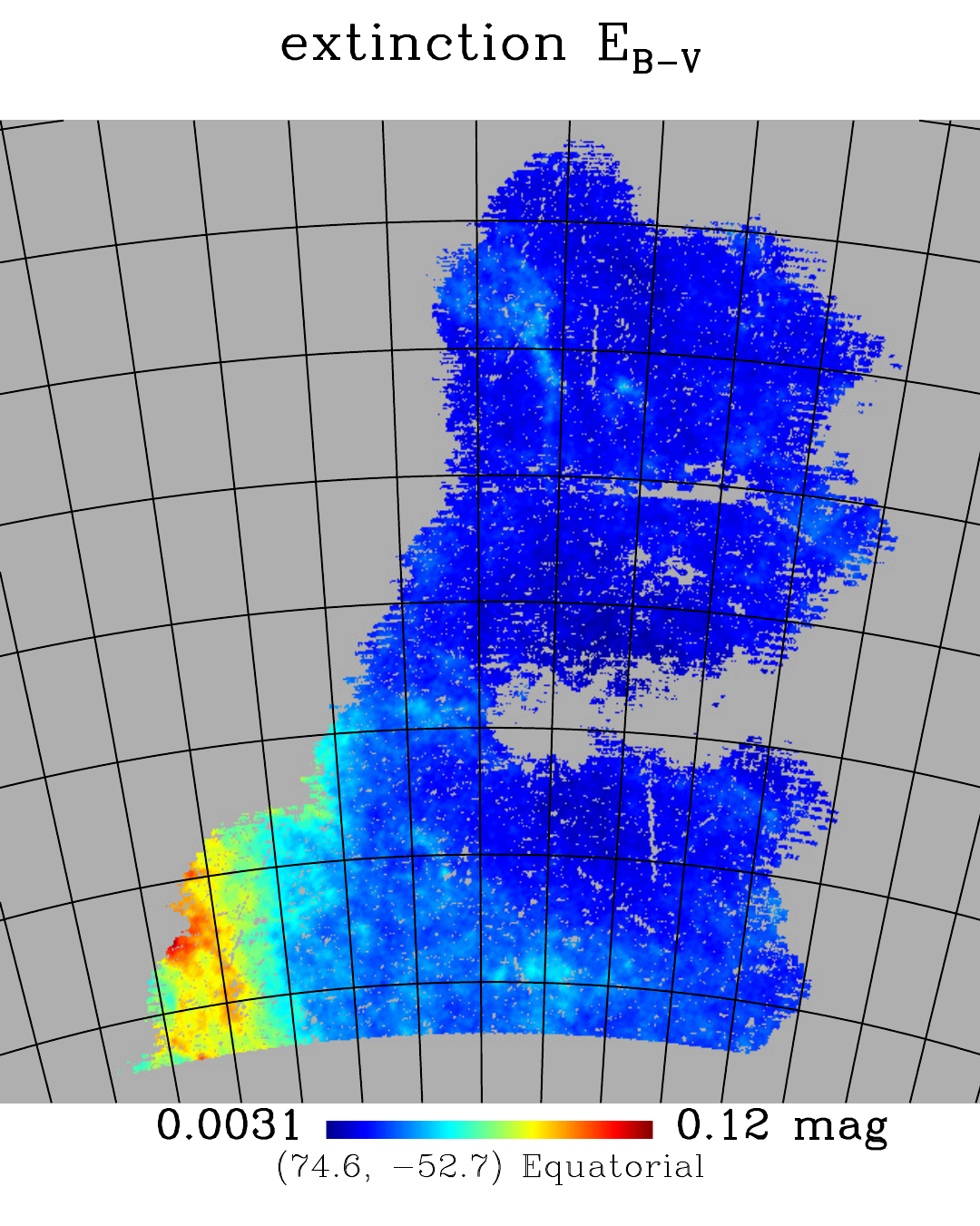}
\includegraphics[width=0.45\linewidth, angle=0]{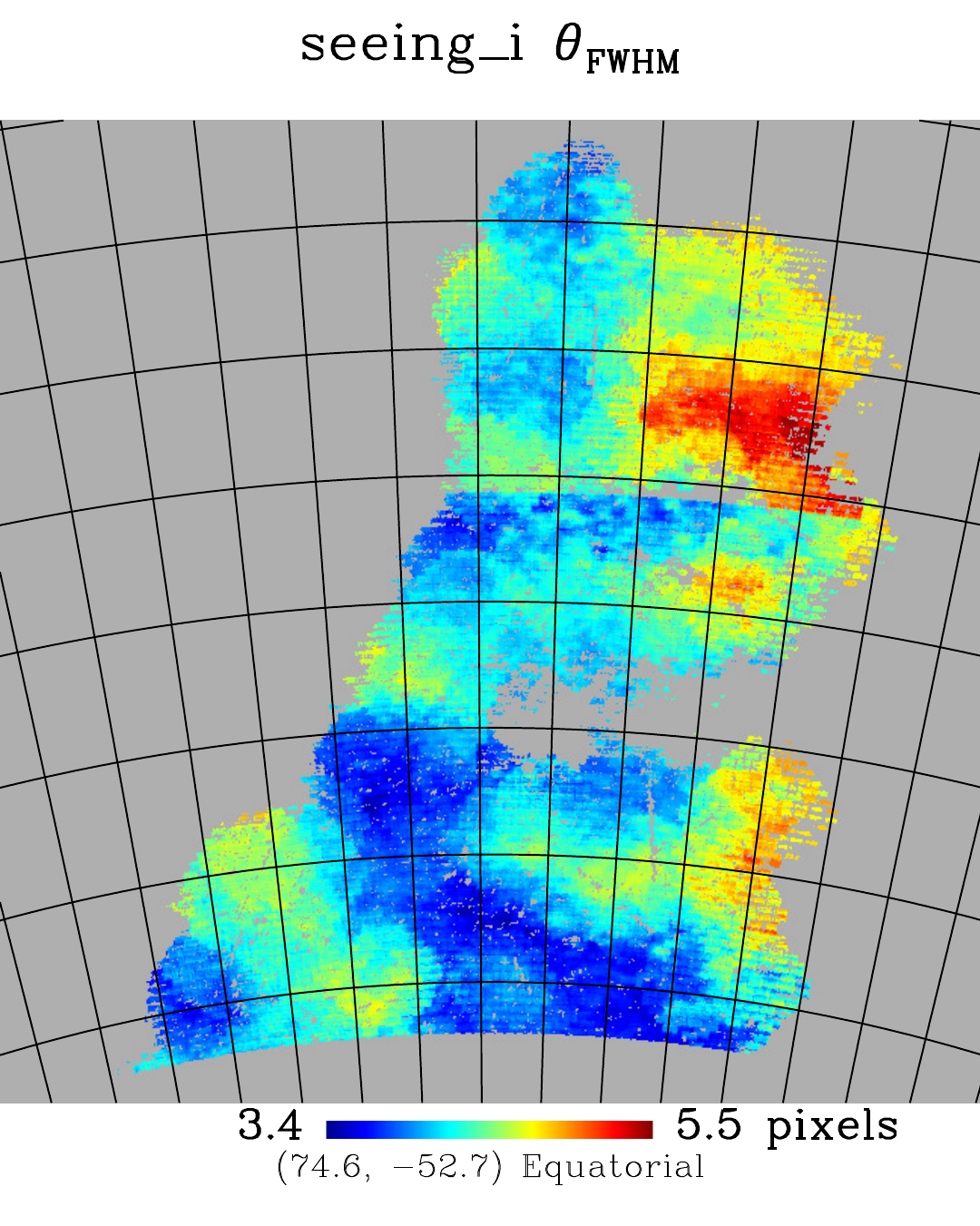}
\includegraphics[width=0.45\linewidth, angle=0]{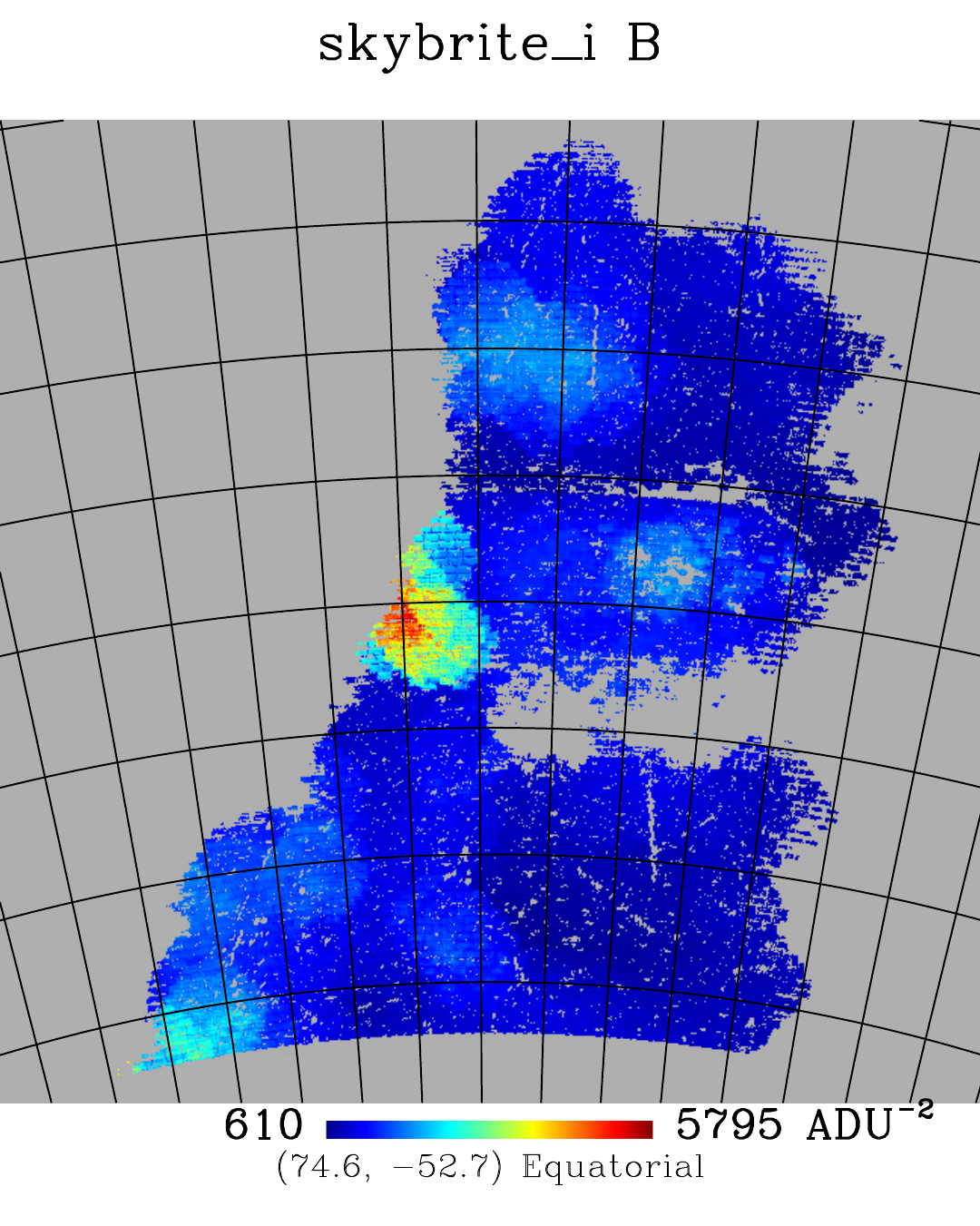}
\includegraphics[width=0.45\linewidth, angle=0]{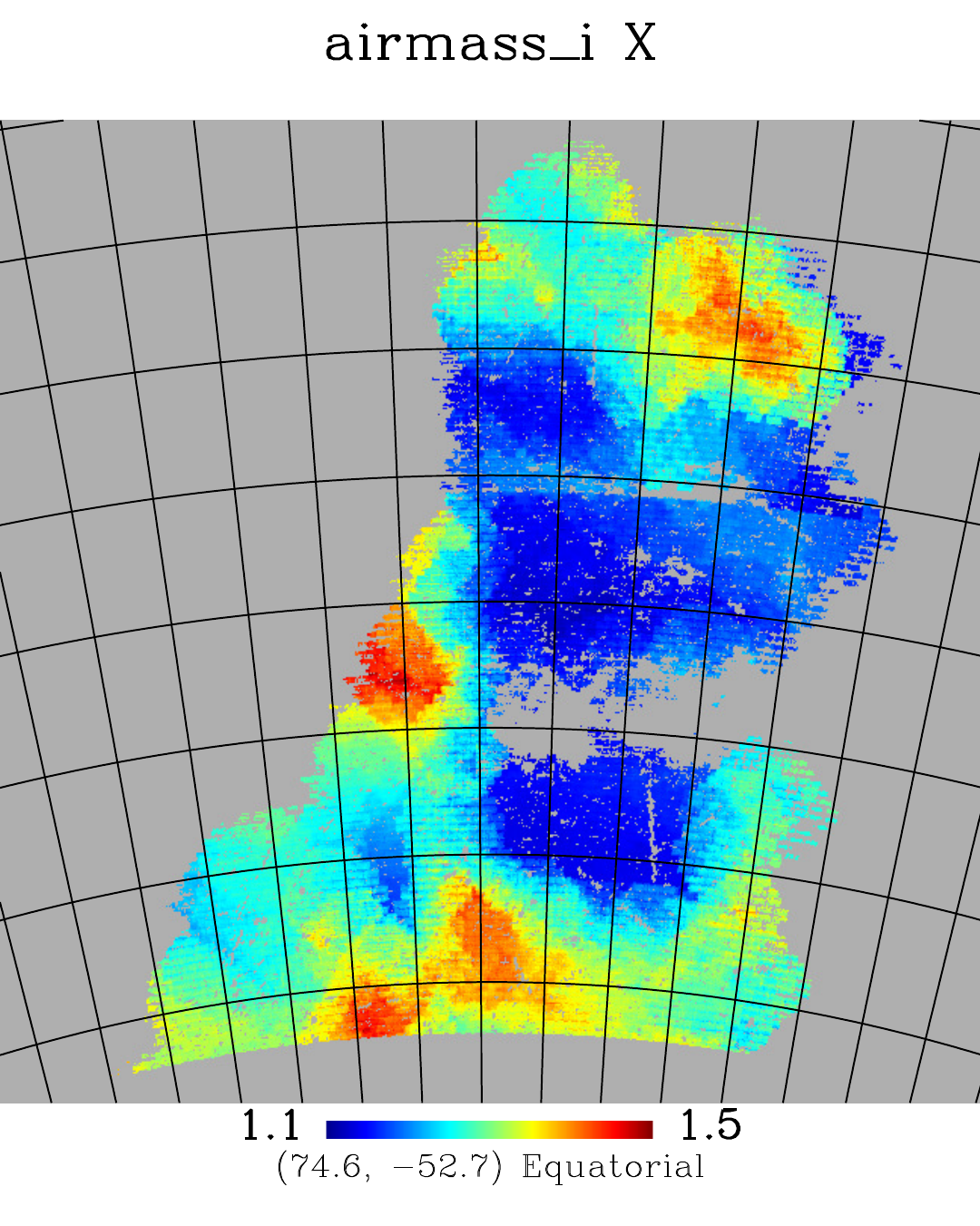}
\caption{Maps of the main potential DES systematics we consider, plotted in the masked region of the SPT-E field we use for our analysis. We show in order extinction as estimated by \Planck \citep{2014A&A...571A..11P}, and seeing, sky brightness and airmass estimated from DES data by \citet{LeistedtMAPS}.}
\label{fig:sys}
\end{center}
\end{figure}
\begin{figure}
\begin{center}
\includegraphics[width=0.49\linewidth, angle=0]{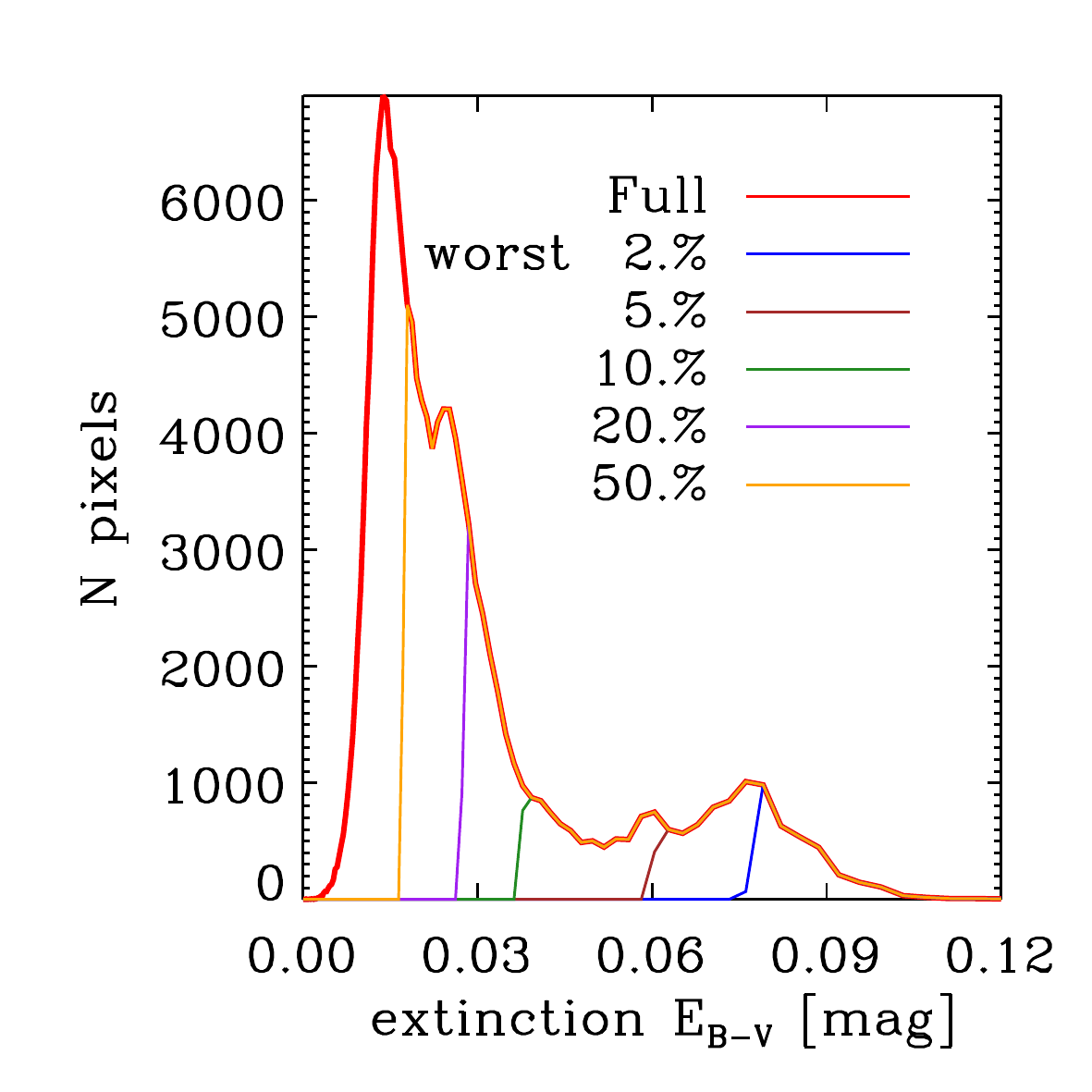}
\includegraphics[width=0.49\linewidth, angle=0]{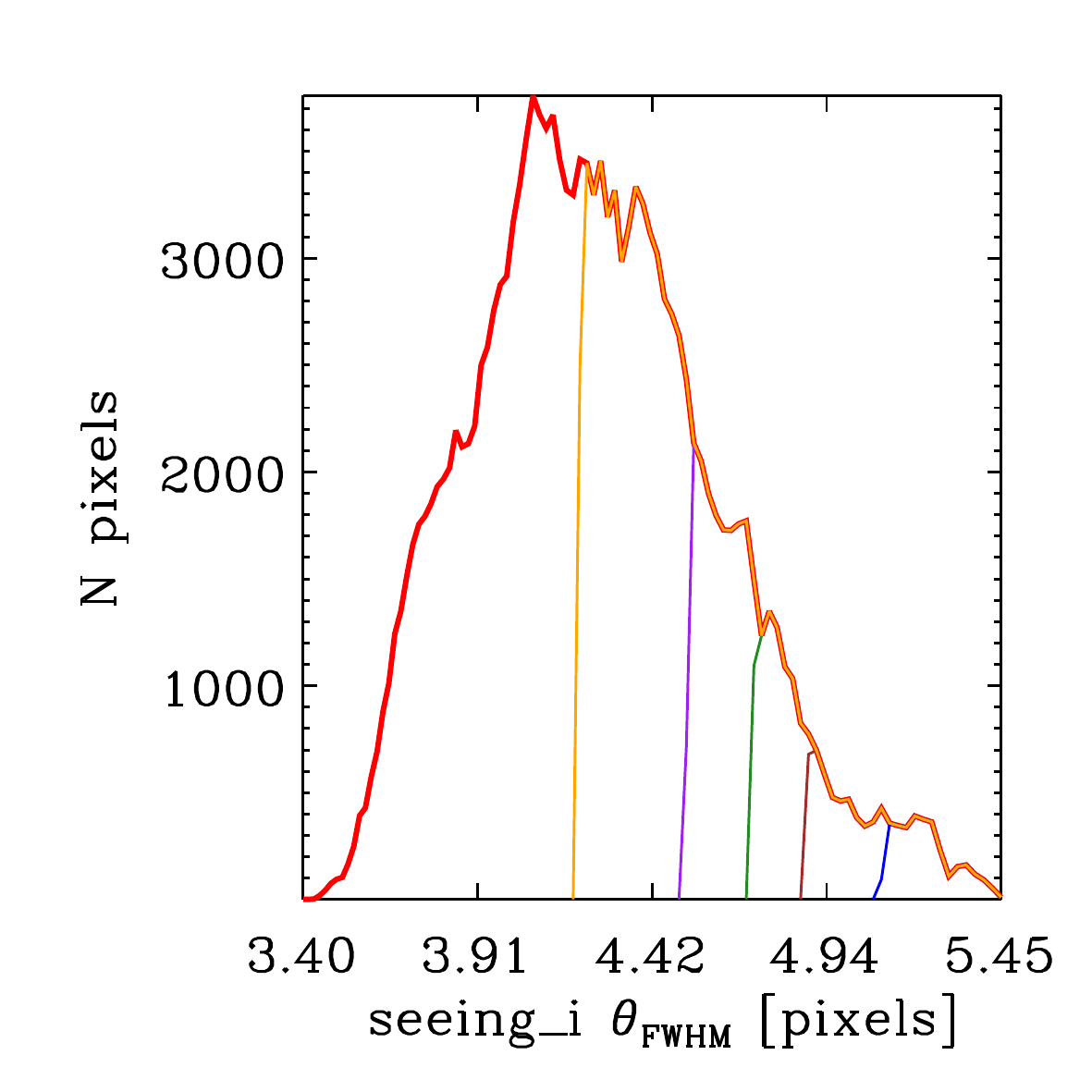}
\includegraphics[width=0.49\linewidth, angle=0]{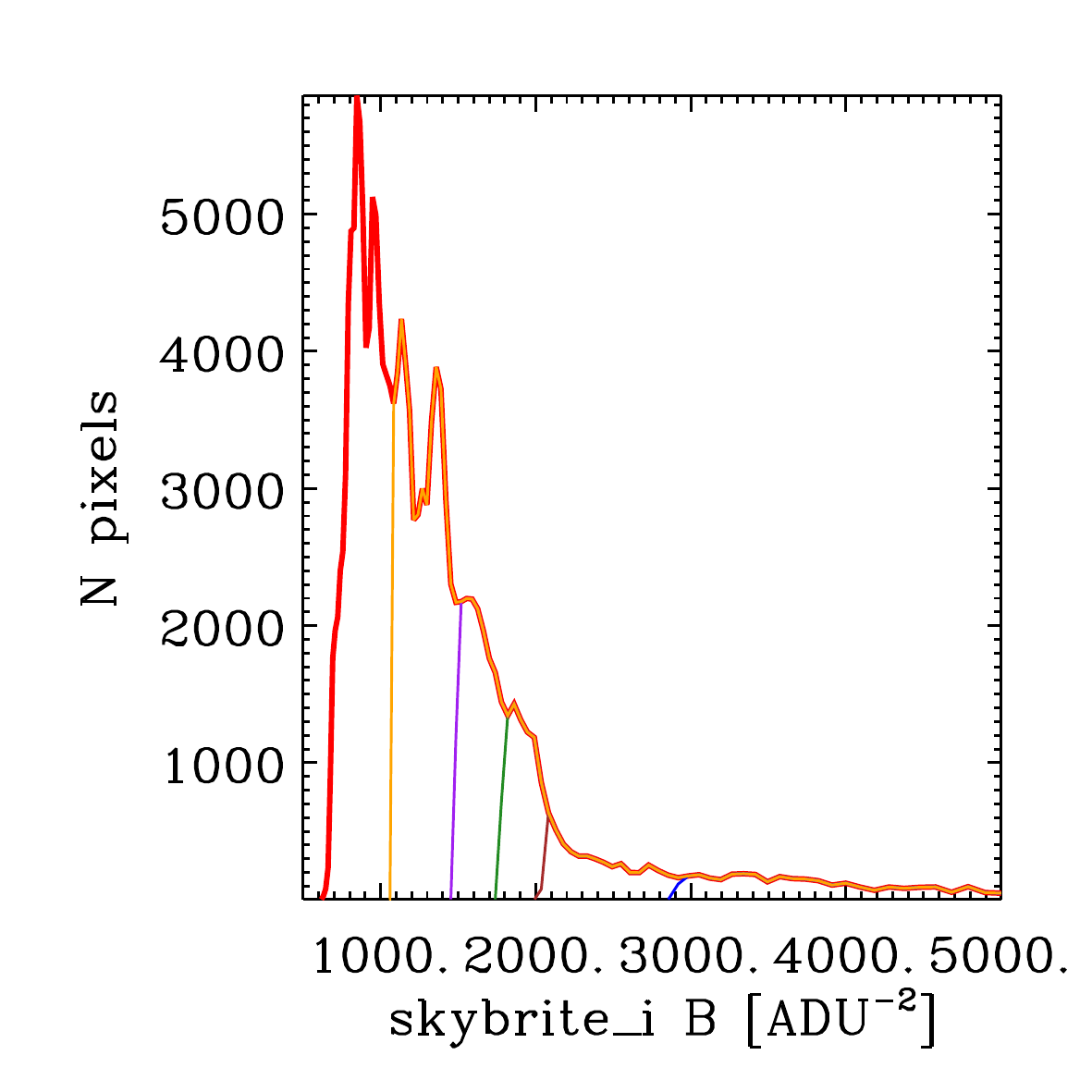}
\includegraphics[width=0.49\linewidth, angle=0]{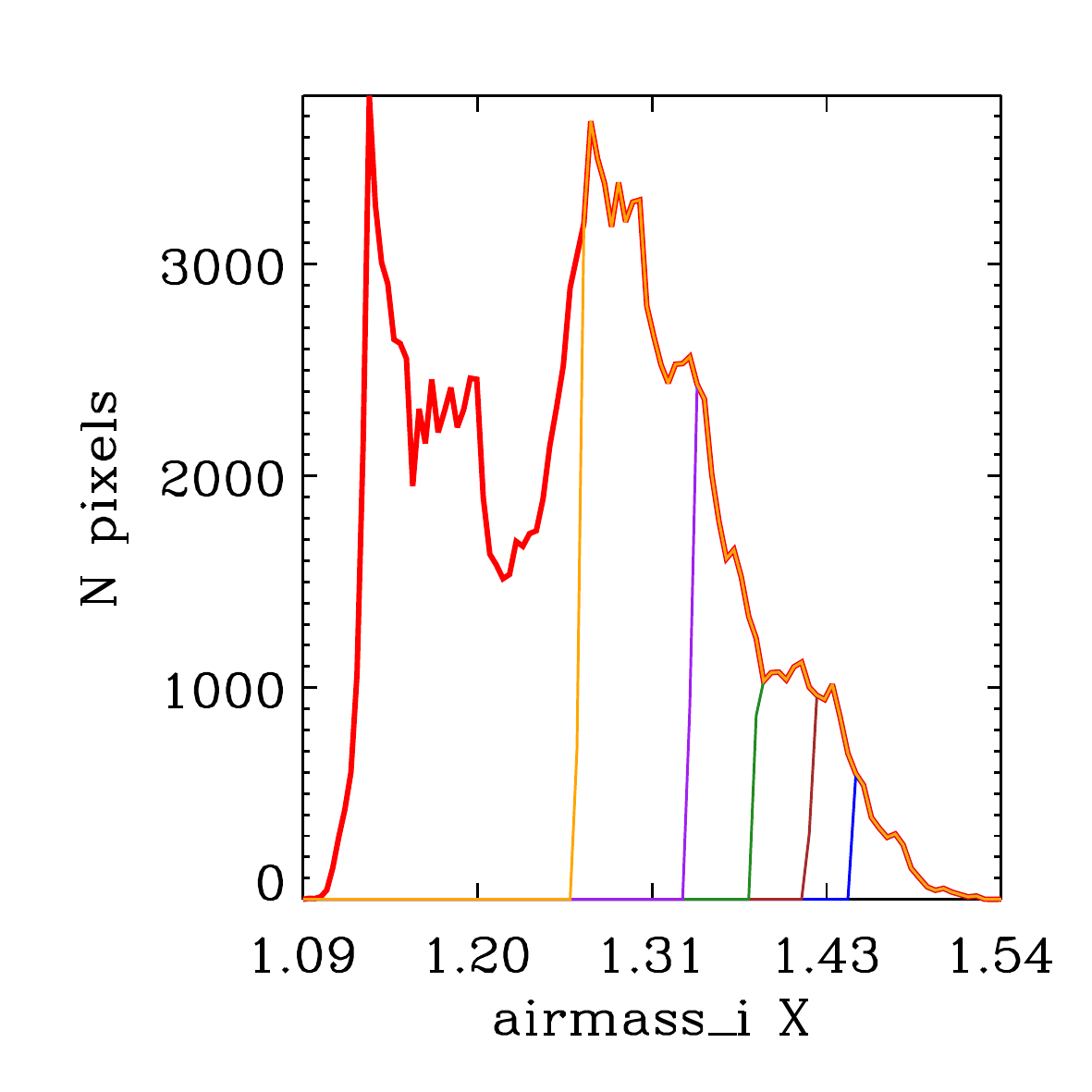}
\caption{Pixel distributions of the potential DES systematics we consider. The histograms show the number of pixels where each systematic assumes the value shown in the abscissa. In addition, five possible cuts of the worst affected areas are shown, ranging from 2 to 50\%.}
\label{fig:syshisto}
\end{center}
\end{figure}

\section {Systematic tests} \label{sec:systematics}

We summarise here a series of tests done to ensure the cross-correlation signal we measure is not significantly affected by systematics.
We first consider the impact of possible DES systematic contaminants at the map and catalogue levels in Section~\ref{sec:syscut}, we then assess the impact of photo-$z$ uncertainties in \ref{sec:photoz}, and we finally test for possible CMB systematics in Section~\ref{sec:sysCMB}.
With the exception of the photo-$z$ case, we perform these tests on the full redshift sample only in the current work with DES-SV data. We will extend the entire systematics analysis to the tomographic bins for future DES data releases.

\subsection{DES systematics} \label{sec:syscut}

We consider a broad range of possible DES systematics, which include potential sources of contamination at both the map and catalogue levels.
The first category includes extinction of distant sources by dust in our galaxy, degradation of image quality due to the observing conditions such as atmospheric seeing,  brightness level of the sky and its fluctuations (sky sigma), and amount of air mass dependent on the distance of the observed field from the zenith, while the second category includes errors on galaxy magnitudes, photo-$z$, amount of nearby bright stars, and goodness of the point spread function (PSF) and magnitude determination. 
The values of all such properties were mapped across the DES-SV area as described in a companion paper \citep{LeistedtMAPS}; considering some of these potential contaminants are mapped in different photometric bands, the total number of maps we can consider amounts to 19.
All were checked and were found to be consistent with the null hypothesis of no significant contamination to the DES-CMB cross-correlations. We describe here for simplicity the null test results of four possible systematics that are likely to bring strong contaminations: extinction from the \Planck colour excess map \citep{2014A&A...571A..11P}, seeing, sky brightness, and airmass in the DES $i$ band. Our aim is to demonstrate that the results are stable with respect to them.

\begin{figure}
\begin{center}
\includegraphics[width=\linewidth, angle=0]{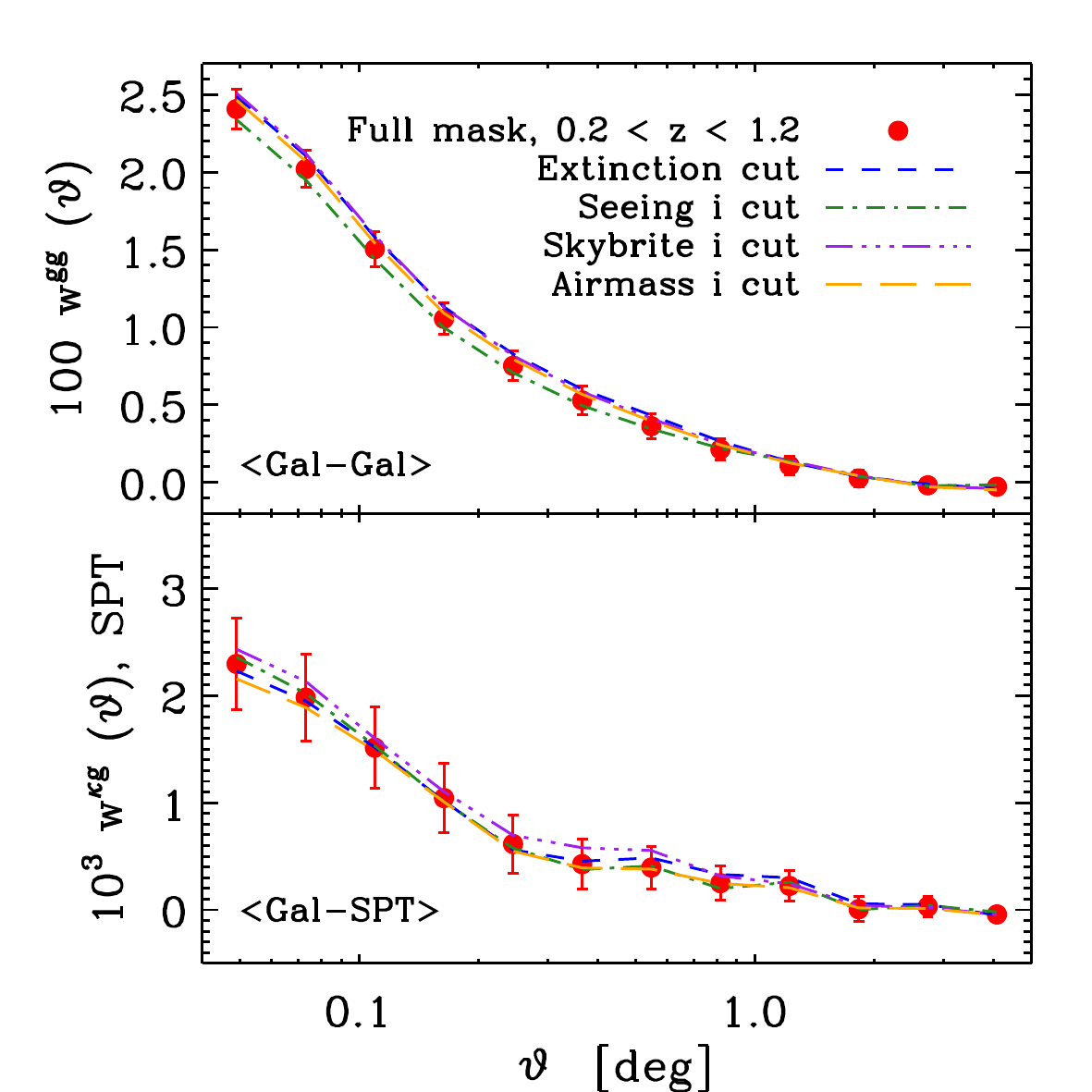}
\caption{Measured DES auto- (\emph{top}) and cross- (\emph{bottom}) correlation functions with SPT lensing obtained from the full mask (\emph{red dots}) and applying cuts in the main contaminants we consider (\emph{coloured lines}). For each potential systematic, we remove 20\% of the area, corresponding to the most affected regions. The results are stable: the correlation functions do not deviate significantly compared with the statistical error bars.}
\label{fig:syscuts}
\end{center}
\end{figure}

We show in Fig.~\ref{fig:sys} the sky maps of these contaminants in the masked region of the SPT-E field that we use for our analysis.
We study the properties of these potential systematics by plotting the histogram of their pixel distributions, which we show in  Fig.~\ref{fig:syshisto}. Here we can see that typical contaminants have a tail in their histogram distribution, corresponding to the most affected areas in the map; for example in the extinction case, the tail at $E(B-V) > 0.05$ mag corresponds to the dusty region in the lower left corner of the map shown in Fig.~\ref{fig:sys}.

A first method for assessing whether any of these potential contaminants has a significant impact on the results of the DES clustering and the DES-CMB lensing correlation is to test whether the results change significantly compared with the statistical uncertainty when the worst-affected areas are masked.
We thus measure the DES auto- and DES-SPT cross-correlation functions applying different cuts in these contaminants, in order to assess the stability of our results. We show in Fig.~\ref{fig:syscuts} the correlation functions we obtain when masking the $20\%$ worst-affected areas for each potential systematic we consider. 
 Here we can see that the results are stable, as the correlation functions of the cut data are consistent with the full data, given the statistical errors. We report in the top section of Table~\ref{tab:bias_syst} how the best-fit bias and cross-correlation amplitude change when the cut maps are used. The same test in harmonic space yields comparable results. These results are indicative of the full set of 19 systematic maps considered. These tests are reassuring and indicate that our claimed detection is not likely to be dominated by this class of systematics.

\begin{figure}
\includegraphics[width=\linewidth, angle=0]{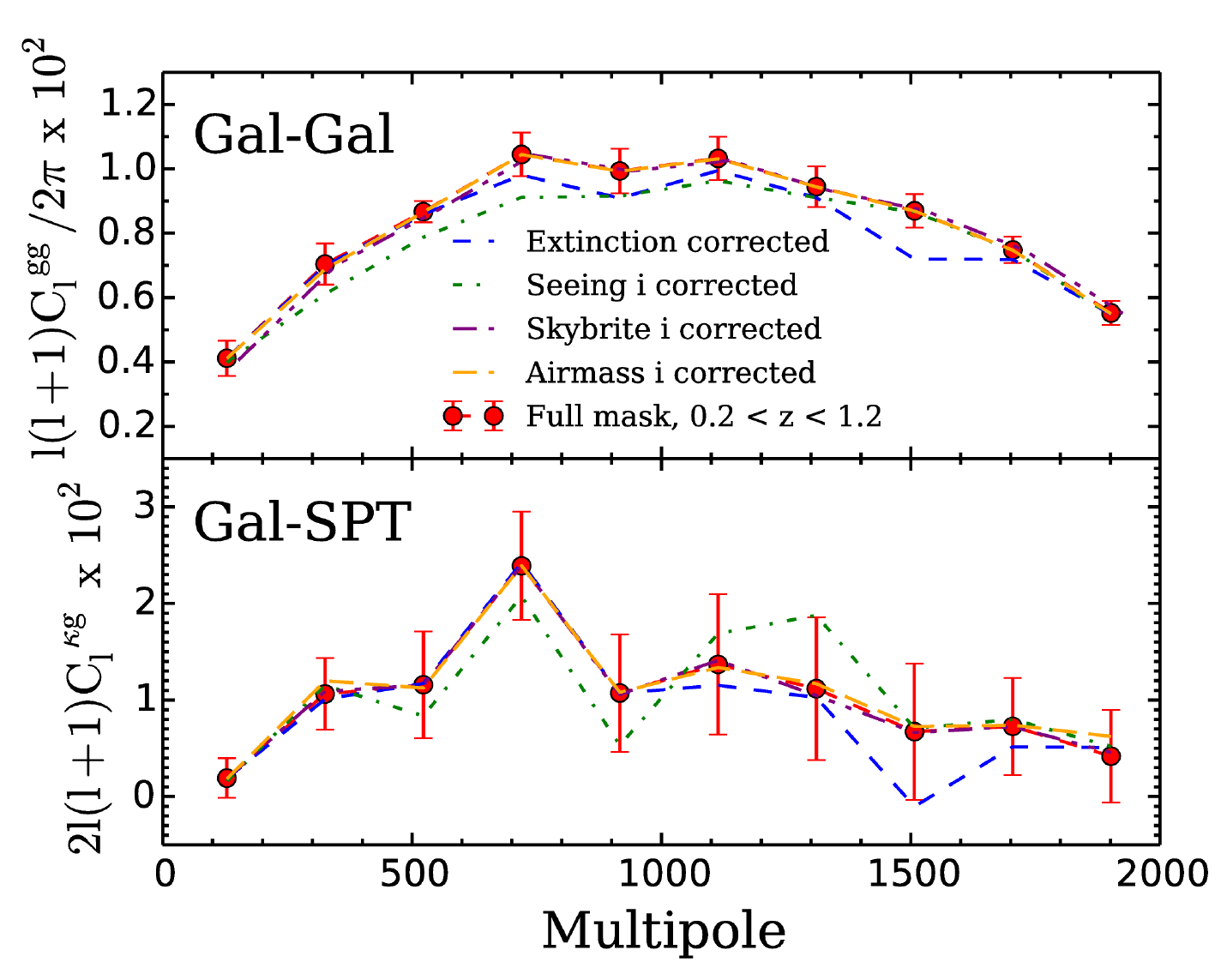}
\caption{The galaxy-galaxy (\emph{top}) and galaxy-SPT lensing (\emph{bottom}) power spectra, including systematic corrections.  Corrections are overall small, and especially so for the cross-spectrum, with all data points within $1 \sigma$ in this case. The best-fit bias and level of detection are negligibly changed when including these corrections.}
\label{fig:syscorr}
\end{figure}

A second method of controlling potential systematics involves measuring cross-correlations with the systematics maps themselves; these cross-correlations can then be used to correct the measurements from contamination. We assume that some systematic source $s$, whose value at a given angular position is given by $\delta_s$, may add a linear contribution to our maps of galaxy overdensity or lensing potential. In the galaxy case this assumption means  \citep{Ross2011, Ho2012, CrocceACF}:
\begin{equation}
\delta_{g,\mathrm{obs}}=\delta_{g,\mathrm{true}}+ \sum\limits_{s} \alpha_s \delta_s  \, ,
\label{syst1}
\end{equation}
if the corrections are small, with an identical treatment for the lensing potential map. If we consider only one possible systematic at a time, the true value of our measurements can be related to the observed correlations between data and systematics. In the cross-correlation case, this is given in harmonic space by
\begin{equation}
C_{\ell,\mathrm{true}}^{g \kappa}= C_{\ell,\mathrm{obs}}^{g \kappa} -  \frac{C_{\ell}^{g s} \; C_{\ell}^{\kappa s}}{C_\ell^{s s}} \, ,
\label{eq:syst2}
\end{equation}
where the last term on the right represents a correction factor to the measurements. We investigated the size of these corrections for all potential systematic maps, consistently finding the corrections to be small compared with the statistical uncertainties on the measurements. We show in Fig.~\ref{fig:syscorr} the corrected power spectra for the four systematic maps of Fig.~\ref{fig:sys}, for both the DES auto- and the DES-SPT lensing cross-power spectra in harmonic space. 
We summarise in the bottom section of Table~\ref{tab:bias_syst} the changes in the best-fit bias and cross-correlation amplitude when we apply the systematic corrections. Also in this case, the results are robust.

\begin{figure}
    \includegraphics[width=\linewidth, angle=0]{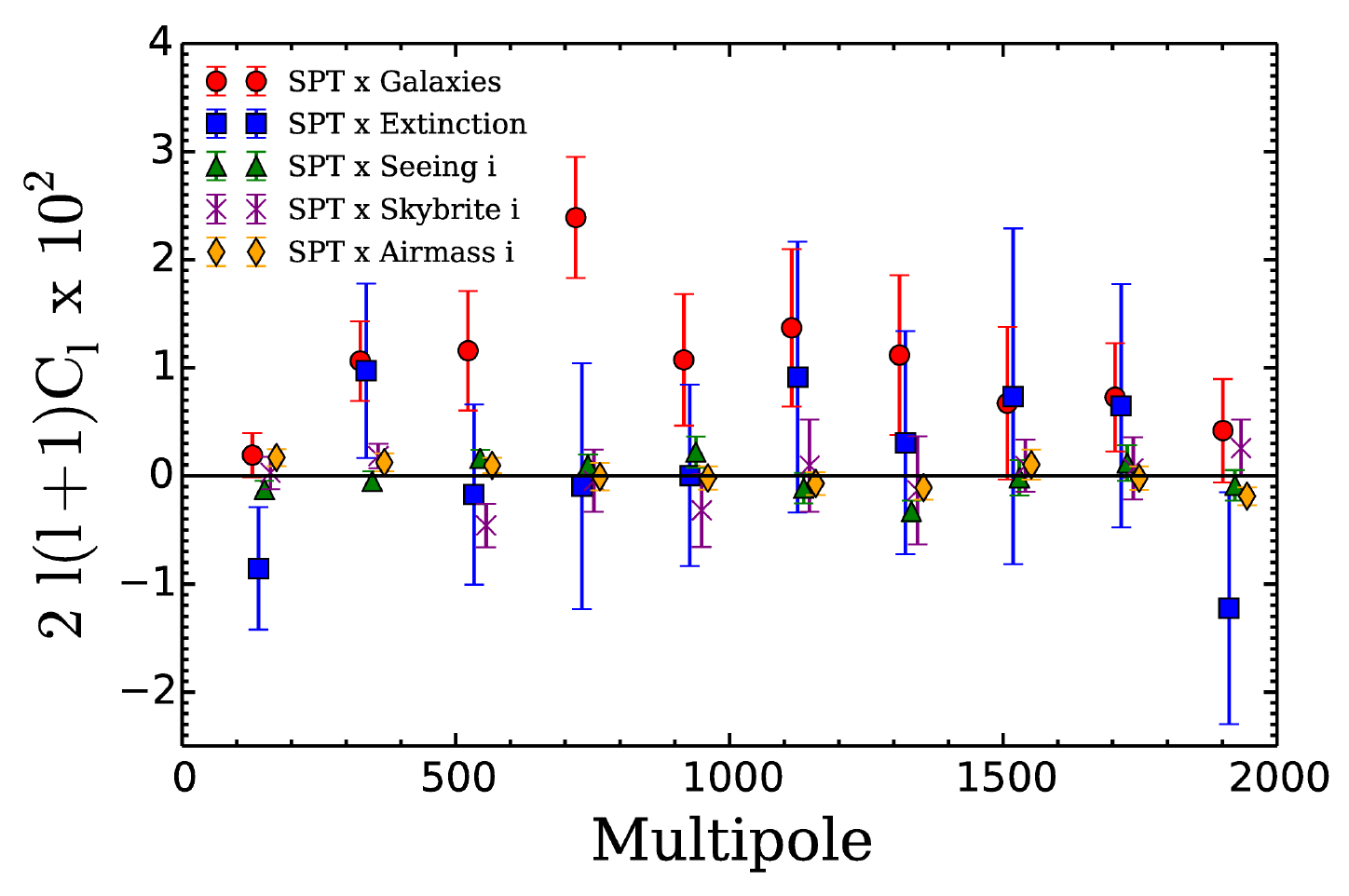}
\caption{Correlations between the SPT convergence map and the potential systematics maps. Data points are offset on the $x$-axis for clarity. The galaxy-lensing correlation (\emph{red circles}) is detected at $ 6 \sigma$, while the majority of the SPT cross systematic maps data points are consistent with zero. Correlating the SPT maps with DES potential systematic maps is expected to produce a null result, which we recover.}
\label{fig:syscorr3}
\end{figure}

\begin{table}
\begin{center}
\begin{tabular}{c c c c c}
\toprule
       Systematic cuts, real space   &  $ b \pm \sigma_b $          & $ A \pm \sigma_A $    \\ 
\midrule
   No cuts        &    $  1.22 \pm 0.03    $   & $  0.84 \pm 0.13   $ \\
\midrule
  Extinction            &    $  1.21 \pm 0.03    $   & $  0.75 \pm 0.13   $ \\
  Seeing                &    $  1.19 \pm 0.03    $   & $  0.85 \pm 0.13    $ \\
  Sky brightness        &    $  1.22 \pm 0.03    $   & $  0.86 \pm 0.13   $ \\
  Airmass               &    $  1.20 \pm 0.03    $   & $  0.74 \pm 0.13    $ \\
\midrule
 Syst. corrections, harmonic space   &   $ b \pm \sigma_b $          & $ A \pm \sigma_A $    \\
 \midrule
 No corrections          &    $  1.22 \pm 0.04    $   & $  0.84 \pm 0.15   $ \\
\midrule
   Extinction          &    $  1.22 \pm 0.04      $   & $  0.79 \pm 0.15    $ \\
   Seeing              &    $  1.18 \pm 0.04      $   & $  0.79 \pm 0.15   $ \\
   Sky brightness      &    $  1.21 \pm 0.04      $   & $  0.84 \pm 0.15    $ \\
   Airmass             &    $  1.22 \pm 0.04      $   & $  0.87 \pm 0.15    $ \\
\bottomrule
\end{tabular}
\caption{Changes in the DES auto- and DES-SPT lensing cross-correlation results when cuts or corrections for potential systematics are applied. The top part of the table shows how the best-fit bias and cross-correlation amplitude from the real-space analysis change when the 20\% worst-affected areas for each potential contaminant are cut. The bottom part of the table shows the corresponding results when the effect of the systematics is corrected, in harmonic space. All results refer to the full galaxy sample at $0.2 < \zphot < 1.2$. The results shown are indicative of the full set of 19 systematic maps we have considered.}
\label{tab:bias_syst}
\end{center}
\end{table}

We finally show in Fig. \ref{fig:syscorr3} the direct cross-correlations of the SPT lensing maps with the contaminant maps, which enter into Equation 
\ref{eq:syst2}. This figure shows that the cross-correlations are consistent with zero, which is a good null test. For further details on the minimisation of potential systematics in the DES galaxy catalogue see \citet{CrocceACF}.

The systematics cuts shift the cross-correlation by less than $1\sigma$ in all cases; thus we do not apply any of these cuts to the main analysis.

A further possible source of systematics is stellar contamination to the galaxy sample, which can potentially alter the measured auto-correlations and dilute the cross-correlations. \citet{CrocceACF} demonstrate by using a spectroscopic subset of DES galaxies derived from the COSMOS survey that the amount of contamination to the `Benchmark' galaxy sample is $< 2\%$ in all redshift bins, so that we can ignore it for the present analysis.

We have also tested the stability of our results with respect to a range of possible choices in the analysis method, finding overall stability in the recovered cross-correlation function. Additional items that we tested include:
 measurement done on Galactic or Equatorial coordinate maps;
using cuts in a different magnitude definition (\texttt{mag\_detmodel\_i} instead of \texttt{slr\_mag\_auto\_i});
 using the intersection of the galaxy and CMB masks versus keeping the two masks distinct;
 reducing the catalogue to a magnitude cut of $18 < i < 22$.

\begin{figure}
\begin{center}
\includegraphics[width=\linewidth, angle=0]{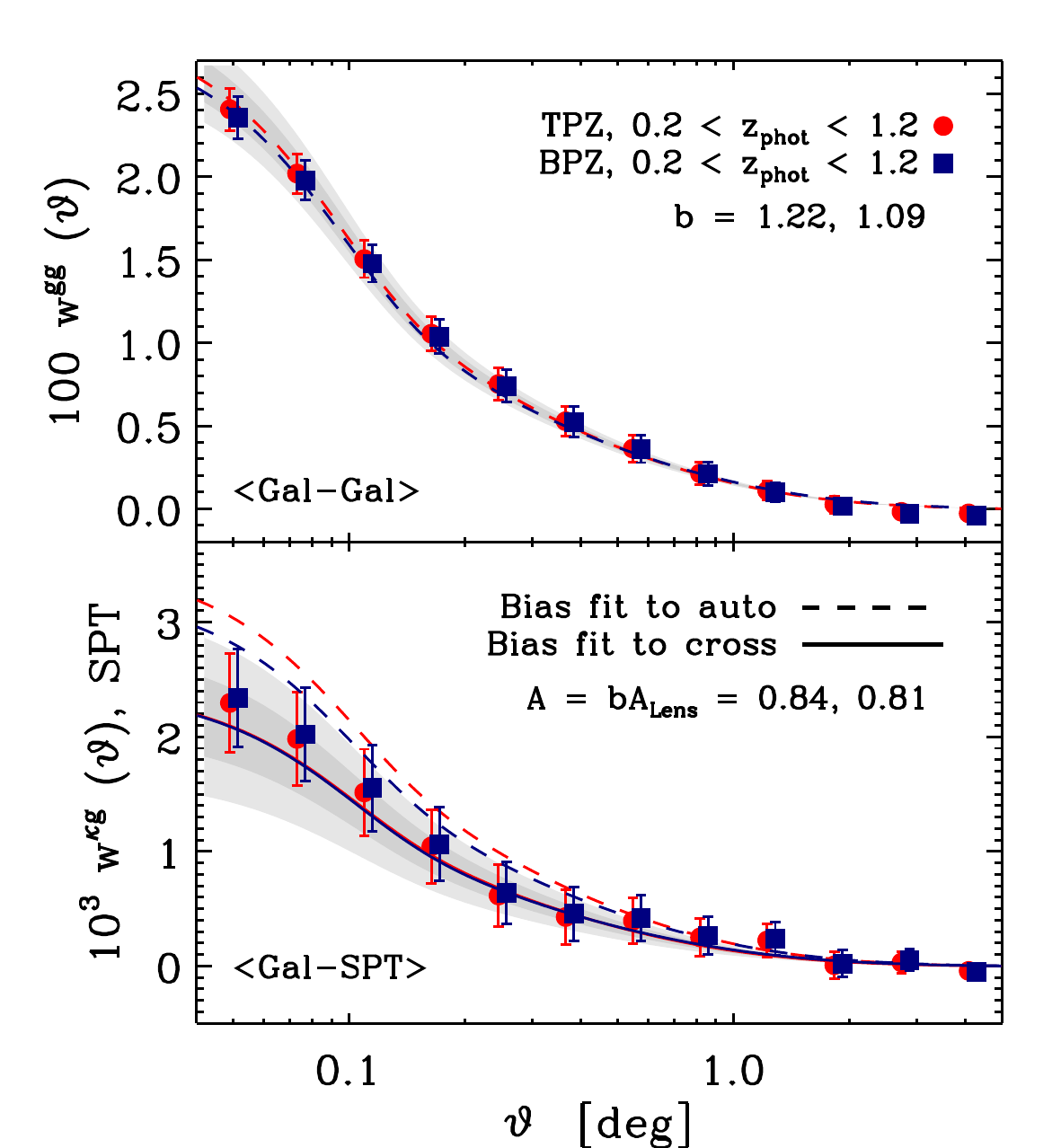}
\caption{Measured DES auto- (\emph{top}) and DES-SPT lensing cross-correlation  (\emph{bottom}) functions for two different choices of photometric redshift estimators: our baseline TPZ choice is shown in red, while the alternative BPZ catalogue is in navy blue. The theory lines are produced accordingly to each catalogue's redshift distribution. The recovered best-fit biases and cross-correlation amplitudes are shown in the caption for both photo-$z$ methods.}
\label{fig:photoz_syst_all}
\end{center}
\end{figure}

\begin{figure}
\begin{center}
\includegraphics[width=\linewidth, angle=0]{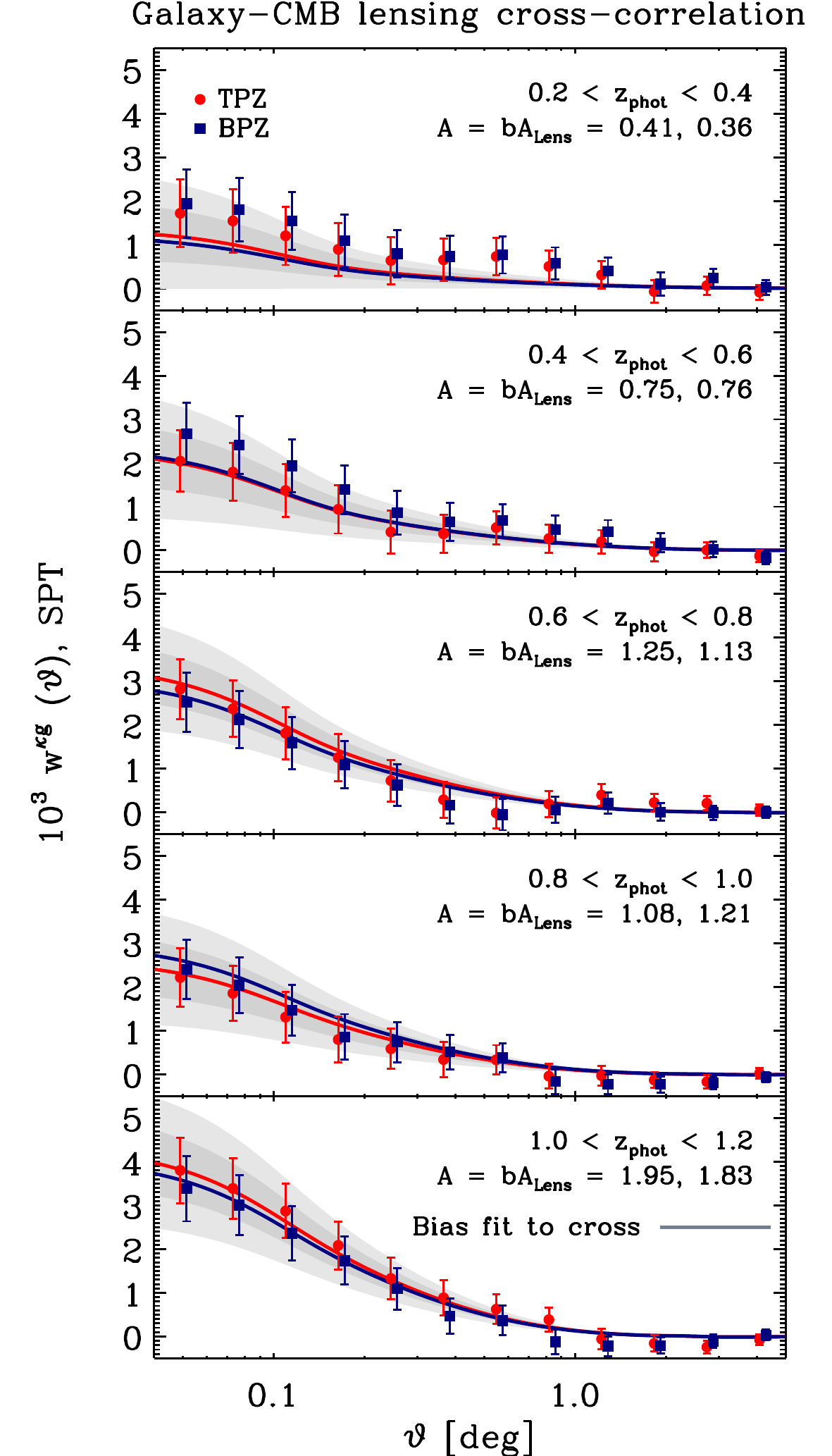}
\caption{Redshift tomography using two different photo-$z$ methods: TPZ (\emph{red}) and BPZ (\emph{navy}). The theoretical curves and best-fit amplitudes for the cross-correlations are also shown for each method. The recovered results agree. See  \citet{CrocceACF} for similar tests on the galaxy auto-correlations. In the first photo-$z$ bin at $0.2 < \zphot < 0.4$ the best fit curves do not trace the data closely, given the mismatch between the measurements and the template shape, and the high covariance between the points.}
\label{fig:wtomo_all}
\end{center}
\end{figure}

\subsection {Photo-$z$ uncertainties}
\label{sec:photoz}

Another source of systematics can be introduced by potential inaccuracies in the photometric redshifts of the galaxy sample.

\subsubsection{Changes in the TPZ photo-$z$ distribution}
\label{sec:warping}  
We first test the effect of smoothing the
photo-$z$ redshift distribution for the full galaxy sample at $0.2 < z_{\text{phot}} < 1.2$.
Smoothing this distribution with a Gaussian kernel
broad enough to remove its oscillations
does not affect the predicted cross-correlations, while it affects the auto-correlation only marginally. This results in an identical value of $A$ and a value of $b$ that is only $\sim 2\%$ higher than our main result, so that our results are not significantly affected.

We further explore how wrong the photo-$z$s would need to be in order to significantly change our results.
We test this by warping the fiducial redshift distribution, which we implement by first fitting the actual TPZ distribution with a Gaussian, and then changing the width of this Gaussian. We consider as two extreme cases a top hat within $0.2 < z < 1.2$, and a narrow distribution centred around the median redshift $z = 0.6$, with $\sigma = 0.1$.

We find that the galaxy-CMB lensing cross-correlations are extremely robust with respect to such warping, due to the broadness of the CMB lensing kernel. For all cases we tested, including the top hat and the narrow Gaussian, the best-fit amplitude $A$ we recover from the cross-correlation is within $5\%$ of the result obtained assuming the TPZ distribution. This highlights that the significance of our detection is robust with respect to changes in the photo-$z$s.

In the case of the auto-correlations we find that whenever the redshift distribution becomes smoother and broader, the expected auto-correlation becomes lower as the galaxies are in average further apart in physical distance; conversely, the auto-correlation increases for a more peaked redshift distribution.
When assuming the top-hat distribution, the recovered bias $b$ increases by $15\%$ compared with the TPZ distribution, while $b$ decreases by $30\%$ if we assume the narrow Gaussian.
Therefore, it is in principle possible to alleviate the observed tension between auto- and cross-correlations by assuming that the true redshift distribution of the DES galaxies is significantly narrower than what is determined with the TPZ method.
However, we find that in order to bring auto- and cross-correlations in full agreement, we need to assume a warping of $\sim 50\%$, i.e. we need to use a Gaussian distribution of width $\sigma = 0.15$, which is twice as narrow as the the TPZ distribution, of $\sigma \simeq 0.3$.
In other words, the stacked probability distribution produced by the TPZ estimator for the full galaxy sample at $0.2 < z_{\text{phot}} < 1.2$ would need to be twice as broad as the true redshift distribution of the DES galaxies. But such a dramatic error is unlikely, as the mean r.m.s. error on the TPZ photo-$z$s was found to by $\bar \sigma_z = 0.078$ by \citet{PhotozDES_2014}; furthermore, the fraction of $3 \sigma$ outliers was found to be $\simeq 2 \%$ only, thus reducing the potential impact of catastrophic redshift errors.
Therefore, we consider our main results to be robust, and we discard the photo-$z$ errors as the main reason of the discrepancy we observe.

\subsubsection{Comparison of two photo-$z$ estimators}

We then demonstrate the robustness of the results with respect to a different choice of photo-$z$ estimator:
besides our baseline choice of TPZ, we also consider here a galaxy catalogue selected on photo-$z$s obtained with the BPZ method.
Given the radical differences between the two methods (TPZ is a machine learning algorithm while BPZ is template-based), it is important to test the robustness of our results with respect to this change.

We therefore change the selection of the galaxy sample according to the alternative BPZ estimator, and we derive modified theoretical predictions with the corresponding BPZ redshift distribution.
We compare in Fig.~\ref{fig:photoz_syst_all} the measured DES auto- and DES-SPT lensing cross-correlation functions with the different photo-$z$ methods for the full sample $0.2 < \zphot < 1.2$.
Here we can see that the results with TPZ (red points and curves) and BPZ (navy) photo-$z$s are generally consistent. The change of the cross-correlation amplitude is consistent with the statistical errors from $A = 0.84$ to 0.81, so that the significance of our measurement remains unaffected. 
On the other hand, the bias from the auto-correlation shifts more significantly from $b = 1.22$ to 1.09. This happens because the BPZ redshift distribution is narrower than TPZ, as BPZ assigns fewer objects to high redshift. As discussed above in Section~\ref{sec:warping}, this causes the predicted auto-correlation to be higher, thus requiring a lower bias to match the nearly identical data.

Using BPZ therefore removes approximately half of the observed tension between auto- and cross-correlations for the full redshift sample. 
However, we think it is unlikely that changes in the photo-$z$ alone can fully remove the tension:
as shown below for the cross-correlation and by \citet{CrocceACF} for the auto-correlation, the recovered bias and amplitude values in the tomographic analysis are significantly more robust. For the tomography, assuming TPZ or BPZ yields consistent results for both auto- and cross-correlation amplitudes. 
As the tension we find between auto- and cross-spectra in the tomography is consistent with the full sample, our main results do not appear to be dominated by the photo-$z$ uncertainty.
Furthermore, the BPZ redshift distribution appears to be a poorer description of the DES-SV galaxies than the TPZ one, given the galaxy-galaxy  cross-correlations observed by \citet{CrocceACF} between different redshift bins are less consistent than what is seen for TPZ.

We then test the robustness of the redshift tomography cross-correlations, which we show in Fig.~\ref{fig:wtomo_all}. 
Here we see once again that the change of photo-$z$ selection method does not change significantly the recovered best-fit cross-correlation amplitudes in any redshift bin. 
We show more quantitatively the resulting best-fit amplitudes for the full sample and tomography in Table~\ref{tab:bias_photoz}. Here we can see that the fluctuations of the results are generally small. 
The cross-correlations are stable, as the variations due to the photo-$z$ differences are small compared with the statistical error bars. 
\citet{CrocceACF} show that the results of the auto-correlations tomography are equally robust for changes between the TPZ and BPZ distributions.  This is because the lack of high-redshift objects in BPZ, which makes the full distribution narrower, has no significant effect on the narrow $dn/dz$ of the five redshift bins. 
In fact, the bias of the full redshift sample $b_{\mathrm{full}}$ is expected to be approximated by a weighted average $b_{\mathrm{avg}}$ over the number of pairs in the $N_{\mathrm{bin}}$ redshift bins:
\begin{equation}
b_{\mathrm{full}} \simeq b_{\mathrm{avg}} = \sum_{i=1}^{N_{\mathrm{bin}}} n_i^2 \, b_i \, ,
\end{equation}
where $n_i$ is the number of galaxies in the bin $i$. We test whether this consistency check is satisfied by the TPZ and BPZ estimators, and we confirm it, by looking at the ratios for the full and averaged biases between the two photo-$z$ methods:
\begin{align}
 {b_{\mathrm{full}}^{\mathrm{TPZ}}} / {b_{\mathrm{full}}^{\mathrm{BPZ}}} &= 1.11 \pm 0.04 \nonumber \\
 {b_{\mathrm{avg}}^{\mathrm{TPZ}}} / {b_{\mathrm{avg}}^{\mathrm{BPZ}}} &= 1.08 \pm 0.03 \, .
\end{align}
This results confirms that 
the larger bias change seen between TPZ and BPZ for the full sample auto-correlation is not in disagreement with the smaller bias changes seen in the tomography, which are shown by \citet{CrocceACF}.
In this sense, the stability of the auto-correlation tomography confirms that the tension we observe between galaxy clustering and CMB lensing correlation is unlikely to be fully removed by changes in the photo-$z$ alone, since switching from TPZ to BPZ removes only half the discrepancy (in the full sample) and leaves the discrepancy unchanged (in the tomography).
We refer to  \citet{CrocceACF}  for a more detailed study of the effect of the photometric redshifts to the determination of galaxy bias.

\begin{table}
\begin{center}
\begin{tabular}{c c c c c}
\toprule
       Photo-$z$ bin       &  $ \left(A \pm \sigma_A \right)^{\mathrm{TPZ}}$  & $ \left(A \pm \sigma_A \right)^{\mathrm{BPZ}}$    \\ 
\midrule
 $0.2 < \zphot < 1.2$    &    $  0.84 \pm 0.13    $   & $  0.81 \pm  0.14   $ \\
\midrule
  $0.2 < \zphot < 0.4$    &    $  0.41 \pm 0.21    $   & $  0.36 \pm 0.22    $ \\
          $0.4 < \zphot < 0.6 $   &    $  0.75 \pm 0.25    $   & $  0.76 \pm 0.24   $ \\
          $0.6 < \zphot < 0.8 $   &    $  1.25 \pm 0.25    $   & $  1.13 \pm 0.25    $ \\
          $0.8 < \zphot < 1.0 $   &    $  1.08 \pm 0.29    $   & $  1.21 \pm 0.29    $ \\
          $1.0 < \zphot < 1.2 $   &    $  1.95 \pm 0.37    $   & $  1.83 \pm 0.34    $ \\
\bottomrule
\end{tabular}
\caption{Real-space comparison of the galaxy-CMB lensing cross-correlations for two different photo-$z$ estimators (TPZ vs. BPZ) for the full sample and the redshift tomography, for the case of $N$-body covariance.  The recovered cross-correlation amplitudes are consistent within the statistical errors.  See  \citet{CrocceACF} for the corresponding results from the galaxy auto-correlations.}
\label{tab:bias_photoz}
\end{center}
\end{table}

\subsection{CMB systematics and cuts in multipole range} \label{sec:sysCMB}
We then investigate the sensitivity of the results to the range of multipoles used. 
This is an important consistency check, and it is especially useful to detect possible systematic contaminations in the CMB lensing maps, as these would typically affect distinct scales differently \citep{2014arXiv1412.4760S}.
For example, a type of possible systematics that could affect our results  would be any residual foreground contamination of the CMB lensing map that is correlated with the galaxies, such as e.g. thermal SZ (tSZ). It was shown by \citet{vanEngelen2014} that using modes out to $\ell=4000$ in the original CMB temperature map used for the lensing reconstruction could lead to more than 5\% bias in the total CMB lensing signal; such bias is more pronounced on the largest angular scales.
\citet{Bleem2012} also found a 5\% bias in their cross-correlation sample based on cross-correlating mock galaxy catalogues with simulated CMB lensing using a tSZ prescription. 
In general any remaining un-subtracted foregrounds will bias the cross-correlation low; such bias will be worse for SPT than \Planck because of the smaller scales used for the lensing reconstruction. The bias could be larger or smaller in our case, so that a more detailed quantification of these effects will be necessary for future work along these lines.
While the most instructive test would be to apply cuts in the range of multipoles used to reconstruct CMB lensing from the temperature map, this is beyond the scope of this work, and we instead apply cuts in the CMB lensing maps themselves.

\begin{table}
\begin{center}
\begin{tabular}{c c c c c c}
\toprule
Correlation & $\ell_{\min}$  &  $\ell_{\max}$   &      $  A \pm \sigma_A $    &  S/N  & $\chi^2 / $ d.o.f.    \\ 
\midrule
Gal-SPT      &  $\mathbf{30}  $ & $ \mathbf{2000} $ & $   \mathbf{0.84 \pm   0.15}  $ & $  \mathbf{5.6} $ & $   \mathbf{8.7 / 19}  $ \\
             &  $30  $ & $ 1000 $ & $   0.93  \pm  0.17  $ & $  5.5 $ & $    5.5 / 9  $ \\
             &  $30  $ & $ 520  $ & $   0.72  \pm  0.23  $ & $  3.1 $ & $   0.84 / 4  $ \\
             &  $230 $ & $ 2000 $ & $   0.93  \pm  0.16  $ & $  5.7 $ & $    7.0 / 17 $ \\
             &  $230 $ & $ 1000 $ & $   1.06  \pm  0.19  $ & $  5.6 $ & $   3.2 / 7   $ \\
             &  $230 $ & $ 520  $ & $   0.88  \pm  0.30  $ & $  2.9 $ & $   0.20 / 2  $ \\
\midrule
Gal-\Planck  &  $   30  $ & $   2000  $ & $  1.08   \pm    0.20  $ & $  5.5  $ & $  31 / 19   $ \\
             &  $ \mathbf{ 30}  $ & $ \mathbf{ 1000} $ & $ \mathbf{ 0.85    \pm   0.21 } $ & $ \mathbf{4.1} $ & $ \mathbf{8.8 / 9 }    $ \\
             &  $  30   $ & $  520   $ & $  1.00    \pm   0.23  $ & $  4.3  $ & $ 1.7 / 4     $ \\
             &  $ 30    $ & $  420   $ & $  0.83  \pm   0.24    $ & $ 3.5 $   & $  1.7/3  $ \\
             &  $  230  $ & $  2000  $ & $  1.10    \pm   0.22  $ & $  4.9  $ & $ 32 / 17    $ \\
             &  $  230  $ & $  1000  $ & $  0.89    \pm   0.25  $ & $  3.6  $ & $ 8.7 / 7      $ \\
             &  $  230  $ & $  520   $ & $  1.14    \pm   0.31  $ & $  3.6  $ & $ 1.2 / 2      $ \\
\bottomrule
\end{tabular}
\caption{Stability of the cross-correlation results with respect to cuts in the range of multipoles considered, for the DES-SPT (\emph{top}) and DES-\Planck correlations (\emph{bottom}). In this case, both maps are smoothed at $5.4'$, to permit the use of the entire multipole range. The cross-correlations are significantly detected in all cases, and the amplitude of the cross-correlations $A = b A_{\mathrm{Lens}}$ is always significantly smaller than the best-fit linear bias $b = 1.22 \pm 0.03$. In the DES-SPT case, we find that the most aggressive choice of including all multipoles at $30 < \ell < 2000 $ is robust, while in the DES-\Planck case this choice leads to a high S/N and a poor $\chi^2$, which is due to the outlying points at $\ell > 1000$. For this reason, we adopt the more conservative cut  $30 < \ell < 1000 $ in this case. Bold font indicates the values used in the main analysis.}
\label{tab:ell_cuts}
\end{center}
\end{table}

We perform this test in harmonic space only, as cuts in the multipole range are easier to implement in this case.
We show the results of this test in Table~\ref{tab:ell_cuts} for both the DES-SPT and DES-\Planck cases (for this test, we smooth both maps at the same scale of $5.4'$).
We first see that both cross-correlations are  detected at high significance (S/N $>3$) in all cases.

Our method for selecting the multipole range to be used in the main analysis is to keep a range as wide as possible, unless there is evidence of inconsistencies such as large deviations of the results ($> 1 \sigma$) or significant outliers leading to a poor reduced $\chi^2$.
More accurately, it was shown by \citet{2015arXiv150702704P} that the variance of a parameter between different data cuts should be approximately given by the difference between the variances obtained when using the two data sets. This criterion is satisfied  in all cases.
For the SPT case, we find that the most aggressive choice of including all multipoles at  $30 < \ell < 2000 $ is robust, as the result only fluctuates within the statistical error when more restrictive choices are made. The $\chi^2$ per degree of freedom is also good in all cases. We therefore adopt this choice for our main DES-SPT results.

In the case of the DES-\Planck correlation, we already noticed in Fig.~\ref{fig:clstpz} the presence of significant outliers at $\ell > 1000$. We also know from the \Planck analysis \citep{2015arXiv150201591P} that, due to the lower sensitivity, the \Planck CMB lensing maps are fully noise-dominated at high multipoles, and the `conservative' \Planck analysis of the lensing auto-spectrum was performed at $\ell < 400$ only, recovering most S/N available over the entire multipole range.
While we do expect both noise and systematics to be less critical in a cross-correlation measurement, we need to take a conservative approach on the higher multipole range of these data.
We see in Table~\ref{tab:ell_cuts} that indeed the results including all $30 < \ell < 2000 $ yield the highest best-fit amplitude and S/N, but this is driven by the significant outlier at $\ell \simeq 1500$; the reduced $\chi^2$ is poor ($\chi^2 /$ d.o.f. $ = 31 / 19$, corresponding to a PTE $= 4\%$). The situation improves significantly if a more conservative cut at  $30 < \ell < 1000 $ is applied, which retains a nearly unchanged error bar while yielding a much more reasonable $\chi^2 / \text{d.o.f.} = 8.8 / 9$. Further cuts, down to the most conservative case at $30 < \ell < 420$, give statistically consistent results, and reasonable $\chi^2$ vaules.
 We therefore adopt the   $30 < \ell < 1000 $ multipole range for our main DES-\Planck results.

\section{Cosmological implications}\label{sec:implications}

While a thorough study of the cosmological implications of the DES CMB lensing tomography is deferred to future work with DES year-1 data, we present here a simple proof of concept of the potential applications of lensing tomography measurements.

\subsection{Bias and growth estimators}
From the theoretical form of the CMB lensing spectra presented  in Section~\ref{sec:theory}, it is evident that CMB lensing tomography is a measurement of structure growth across cosmic time, potentially constraining departures from the standard cosmological model at the linear growth level. Indeed, it is clear that the joint measurement of the auto- and cross-correlations $C_\ell^{gg}, C_\ell^{\kappa g}$ allows one to break the degeneracy that exists between bias and structure growth. 

We use here the simplest possible assumptions, and consider linear, local forms of both the galaxy bias and the growth function, given by $b(z), D(z)$, while keeping the cosmology fixed to the \Planck best-fit fiducial model. A potential caveat of this analysis is that, given the results from Section~\ref{sec:photoz}, the statistical errors on the bias evolution obtained from the galaxy auto-correlations can be comparable with systematic errors due to the uncertainties in the photometric redshifts estimations, which are not taken into account in this section. For a more complete analysis of the bias evolution and a more detailed treatment of the systematics, see  \citet{CrocceACF}.

 Our estimator for the bias in the $i$-th redshift bin is simply $\hat b_i = b_i$, i.e. the best-fit value from each auto-correlation, while a basic estimator for the growth function $D_i$ can be derived from the ratio between the observed (obs) cross-spectrum and a normalising fiducial (the) cross-spectrum:
\be \label{eq:Dest1}
\left( \hat D_0 \right)_i \equiv \left\langle \sqrt{  \frac{  \left( C_\ell^{\kappa g} \right)^i_{\mathrm{obs}}  }  {  \left( \slashed C_\ell^{\kappa g} \right)^i_{\mathrm{the} }  } } \right\rangle_\ell \, ,
\ee
where the expression is averaged over all multipoles considered.
Here we have defined with a slash the normalising power spectrum ${\slashed C}_\ell^{\kappa g}$,
which we define as the usual power spectrum of Section~\ref{sec:theory}, where the kernels had the growth function removed:
\begin{align}
\slashed C_\ell^{\kappa g} \, \, =& \, \,\frac{2}{\pi} \int_0^{\infty} dk \, k^2 \, P(k) \, \slashed W_\ell^{\kappa} (k) \,  \slashed W_\ell^g (k) \,  \label{eq:slashKGa} \\
\slashed W_\ell^{g}(k)\, \, =& \, \, \int_0^{\infty} dz \, b(z) \, \frac{dn}{dz}(z) \, j_\ell[k \chi(z)] \\
\slashed W_\ell^{\kappa}(k) \, \, =& \, \,  \frac{3 \Omega_m H_0^2}{2} \int_0^{\infty} dz \, \frac{\chi_* - \chi}{\chi_* \chi}(z) \,  j_\ell[k \chi(z)] \, . \label{eq:slashKG}
\end{align}
Notice that, while the CMB convergence kernel is formally not bound to the narrow redshift range where $ \frac{dn}{dz} (z) \ne 0$, its overall contribution to the cross-spectrum from redshifts outside this range is negligible; this can be seen more clearly by using the Limber approximation. In this case, as shown e.g. by Eq.~(27) in \citet{Giannantonio:2011a}, the angular power spectrum is given by a single integral over redshift, and when one of the two source terms $W_\ell^{\kappa}, W_\ell^{g}$ vanishes, so does the total $C_l$.
Therefore the $\left( \hat D_0 \right)_i $ estimator correctly recovers the linear growth function in the redshift bin $i$.

In order to estimate the theoretical power spectrum at the denominator of Eq.~(\ref{eq:Dest1}), we still need the galaxy bias. We can remove the dependence on bias by introducing the following estimator:
\be \label{eq:DestE1}
 \left( \hat D_G \right)_i \equiv \left\langle   \frac{  \left( C_\ell^{\kappa g} \right)^i_{\mathrm{obs}}  }  {  \left( \slashed C_\ell^{\kappa g} \right)^i_{\mathrm{the} }   }   \sqrt {  \frac{  \left( \slashed C_\ell^{gg} \right)^i_{\mathrm{the}}  }  {  \left(  C_\ell^{gg} \right)^i_{\mathrm{obs} }   } }  \right\rangle_\ell \, .
\ee
We can see that $D_G$ does not directly depend on the galaxy bias, as its observed and theoretical values simplify exactly in the limit of narrow redshift bins, and that it contains no direct dependence on the theoretical growth function either: we therefore propose this estimator as a novel simplified method for extracting cosmic growth information.
The $D_G$ estimator still includes a dependence on the combination of cosmological parameters  $\Omega_m \, H_0^2 \, \sigma_8$ from the CMB lensing kernel of Eqs.~(\ref{eq:slashKGa},~\ref{eq:slashKG}); this dependence is degenerate with the growth function information in any redshift bin, but the degeneracy can be broken by a multi-bin tomography.

We evaluate $D_G$ directly using the harmonic space bandpowers and the real-space correlation functions; we further improve the estimator of Eq.~(\ref{eq:DestE1}) by weighting the averages with the diagonal errors on the power spectra and correlation functions respectively. While the expectation value is $\langle D_G \rangle = D$ on linear scales, we note that the dependence on non-linearities will largely cancel between the theoretical and observed parts of the estimator.
We nonetheless use scales at $\ell < 1000$ only, to reduce potential contamination by non-linear contributions.
 We estimate the errors on $D_G$ and the full covariance matrix between the redshift bins by repeating the $D_G$ calculation for our set of 100 $N$-body realisations of the galaxy density and CMB lensing data.

Our estimator $D_G$ is related to, but different from, the $E_G$ estimator introduced by \citet{Zhang2007}, used to confirm GR with observations by \cite{Reyes2010}, and studied for projections with future surveys by \citet{Pullen2015}. This alternative estimator is defined as
\be
E_G \propto \frac {C_{\ell}^{\kappa g}} {C_{\ell}^{\theta g}} = \frac {C_{\ell}^{\kappa g}} {\beta \, C_{\ell}^{gg}} \, ,
\ee
where $\theta$ indicates the linear velocity perturbations, given by $\theta = f \delta$, where $f = d \ln D / d \ln a$ is the linear growth rate, and $\beta = f / b$ is observable from redshift space distortions (RSD).
Both $E_G$ and $D_G$ have the advantage of being independent from galaxy bias by construction. $E_G$ has the additional bonus of being more easily related to modified gravity theories, as it can be directly connected to departures from the Poisson equation and the anisotropic stress; furthermore it is scale-independent in GR.
On the other hand, $E_G$ can only be accurately measured from a spectroscopic survey.

In the case of photometric data, such as DES, a further possible alternative to $E_G$ would be to simply test the ratio $C_{\ell}^{\kappa g} / C_{\ell}^{gg}$, which would retain many of the desirable features of $E_G$, as this is still scale-independent in GR and easily related to modified gravity theories. However, this simple ratio requires external information on the galaxy bias, which is a serious drawback.
For this reason, we propose to use the $D_G$ estimator as an alternative for photometric surveys.

\subsection{Results and interpretation}

\begin{figure}
\begin{center}
\includegraphics[width=\linewidth, angle=0]{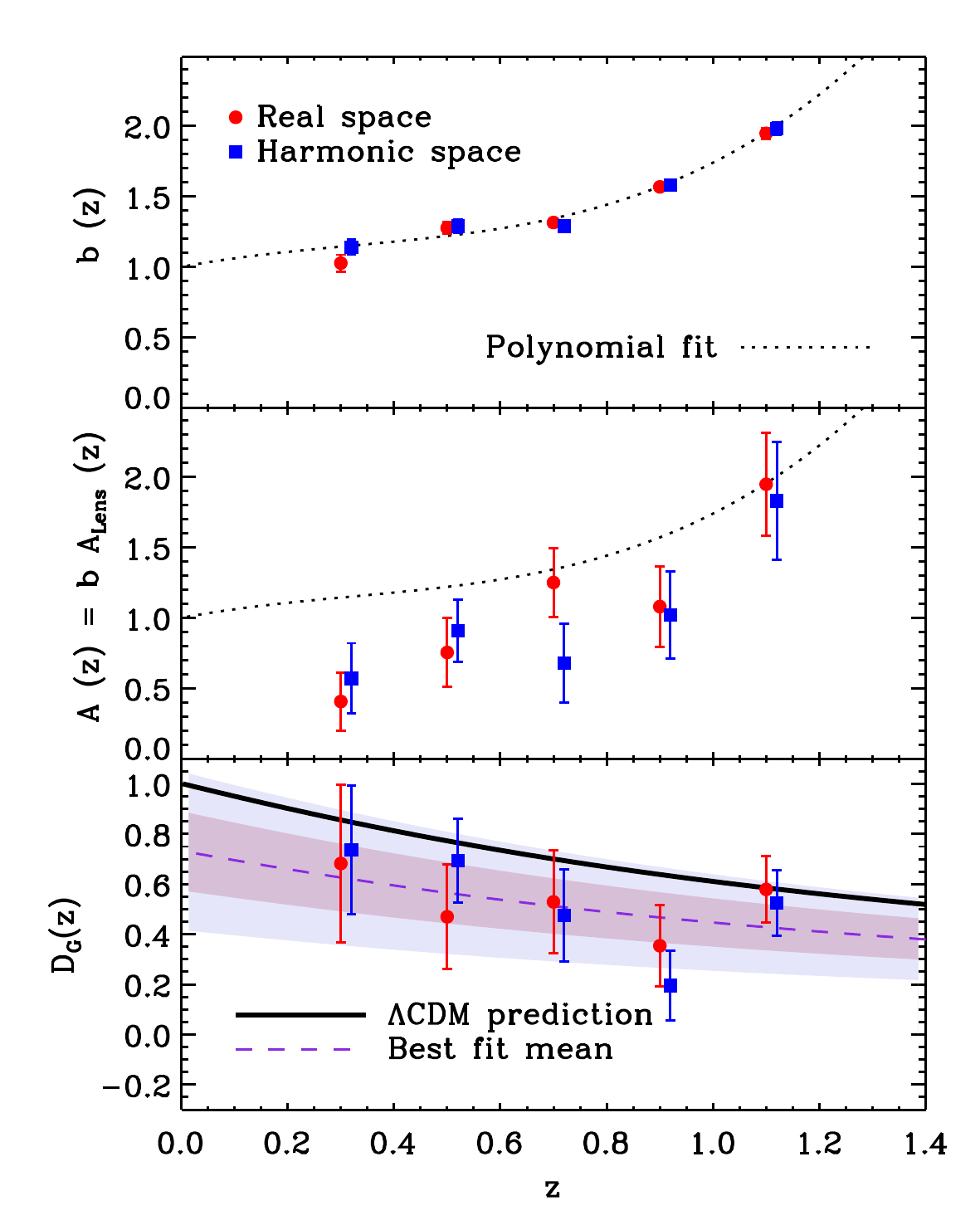}
\caption{Reconstructed measurements of the redshift evolution of linear bias $b(z)$ from galaxy auto-correlations, as also presented by \citet{CrocceACF} (\emph{top panel}), galaxy-CMB lensing cross-correlation amplitudes $A(z)$ from the cross-correlations (\emph{central panel}) and linear growth function from the $D_G(z)$ estimator (\emph{bottom panel}) from the combined tomography of galaxy clustering and galaxy-CMB lensing correlations. The red (round) points are derived from the correlation functions, while the blue (square) points are from the angular power spectra. The purple dashed line shows the mean best fit amplitude to $D_G$ with 1 and $2\sigma$ uncertainty bands.  We also show for comparison the best-fit bias model of Eq.~(\ref{eq:bzmodel}) in the top and central panels (\emph{dotted lines}), and the theoretical growth function for the \Planck fiducial cosmology in the bottom panel (\emph{thick solid line}). The low values of $A$ we observe translate into a preference for a lower $D_G$ in most redshift bins.}
\label{fig:bD}
\end{center}
\end{figure}

By applying the $D_G$ estimator described above to our tomographic data in real and harmonic space, we obtain the results shown in Fig.~\ref{fig:bD}. Here we plot the redshift evolution of linear bias (top panel), galaxy-CMB lensing cross-correlation amplitude (central panel) and the linear growth function derived with the $D_G$ estimator (bottom panel). 

The evolution of galaxy bias is presented and discussed in more detail by \citet{CrocceACF}; we follow this study, and compare the bias with a simple third-order polynomial fit, which was shown  in Appendix A by \citet{CrocceACF} to be in good agreement with results from the MICE $N$-body simulations:
\be \label{eq:bzmodel}
b(z) = 1 + a_1 z + a_2 z^2 + a_3 z^3 \, .
\ee
We show in the top panel of Fig.~\ref{fig:bD} that the best-fit model by \citet{CrocceACF}, of parameters
$a_1 = 0.87$, $a_2 = -1.83$, $a_3 = 1.77$
is also an excellent fit to our measurements in both real and harmonic spaces, further validating both analyses.

We show in the central panel of  Fig.~\ref{fig:bD} the redshift evolution of the galaxy-CMB lensing correlation amplitude $A = b A_{\mathrm{Lens}}$: as shown above in Table~\ref{tab:summary}, $A$ is in most cases lower than the expected value given the auto-correlations. We can see once again that real- and harmonic-space results agree well, with the one exception of the third bin cross-correlation, as discussed above in Section~\ref{sec:tomoCl}.

We then focus on the linear growth function: we show in the bottom panel of Fig.~\ref{fig:bD} the results from the $D_G$ estimator of Eq.~(\ref{eq:DestE1}) for real and harmonic spaces, where we use scales at $\ell < 1000$ only. We see that the data prefer a smaller growth of structure than what is expected in the fiducial \Planck \LCDM model: this result is driven by the lower than expected values of the observed galaxy-CMB lensing correlations.  The estimators in real and harmonic space agree well in most bins.

If we assume the template shape of $D_G(z)$ to be fixed by the fiducial \Planck cosmology and we fit its amplitude $ A_D $, so that  
\be
D_G(z) =  A_D \left[D_G(z) \right]_{\mathrm{fid}} \, ,
\ee
we find $A_D  = 0.76 \pm 0.17 $ from the real-space analysis and $ A_D  = 0.70 \pm 0.15 $ in harmonic space. As the two results are consistent and there is no reason to prefer one over the other, we take their mean as our main result:
\begin{equation}
A_D  = 0.73 \pm 0.16 \, ,
\end{equation}
where the error is also the mean of the errors, as the two methods are based on the same data.
This result includes the full covariance between the photo-$z$ bins, which is typically $30 \%$ between neighbours.
We note that, as discussed above in Section~\ref{sec:photoz}, if the real redshift distribution of the galaxies in all bins is narrower than our assumption, the tension could be alleviated, but the photo-$z$ alone are unlikely to be responsible for this discrepancy in full. In particular we have tested that, if we use the alternative BPZ photo-$z$s, we obtain $A_D = 0.70 \pm 0.16$, in agreement with the TPZ results.

We can then assess the significance of the discrepancy with respect to the fiducial \Planck cosmology. From the point of view of template fitting, the mean best-fit value is $1.7\sigma$ away from the fiducial value $A_D = 1$.
 Alternatively, we perform a null hypothesis test and find that the $\chi^2$ difference between the best fit and the fiducial model is $\Delta \chi^2 = 7.2 $ in real space (10.5 in harmonic space) for 4 degrees of freedom, corresponding to a PTE = 13\% in real space (3.3\% in harmonic space). 
We therefore conclude that the observed tension is only weakly significant.
We discuss however in the following what the implications could be, if the lower $A_D$ persists with more accurate measurements.

The $D_G$ estimator retains a dependence on the ratio between the real and the fiducial values of the background parameters $\Omega_m h^2 \sigma_8 \equiv \omega_m \sigma_8$; it is thus  in principle possible to attribute the observed mismatch to a preference for different parameter values.
 The parameter shift required is large compared with the current CMB constraints from \Planck \citep{2015arXiv150201589P}: 
in order to shift the amplitude $A_D$ from its best-fit value $0.73 \pm 0.16$ to 1, would require a fractional decrease in $\omega_m \sigma_8$ of 27\%.

It is worth mentioning that in the last few years several independent measurements of LSS probes have hinted at low significance towards low growth in recent times, including measurements of $\sigma_8$ from galaxy clusters \citep{Bocquet2015}, weak lensing \citep{MacCrann2014}, redshift-space distortions \citep{Beutler2014}, and a combination of probes \citep{Ruiz2015}. It is important to stress that, in most cases, alternative analyses showing weaker or no tension do exist, e.g. by \citet{Samushia2014} for RSD, and by \citet{Mantz2015} for galaxy clusters.
Only better data in the near future will clarify whether statistical flukes, systematic effects or new physics are behind these observations; we prefer for the moment to avoid over-interpreting the results, and we defer to the upcoming DES year-1 data a more detailed study that will include a more rigorous quantification of the photo-$z$ and SZ systematic uncertainties, varying cosmological parameters and the full covariance between all data. 

\subsection{Relaxing cosmology}
\begin{figure}
\begin{center}
  \includegraphics[width=\linewidth, angle=0]{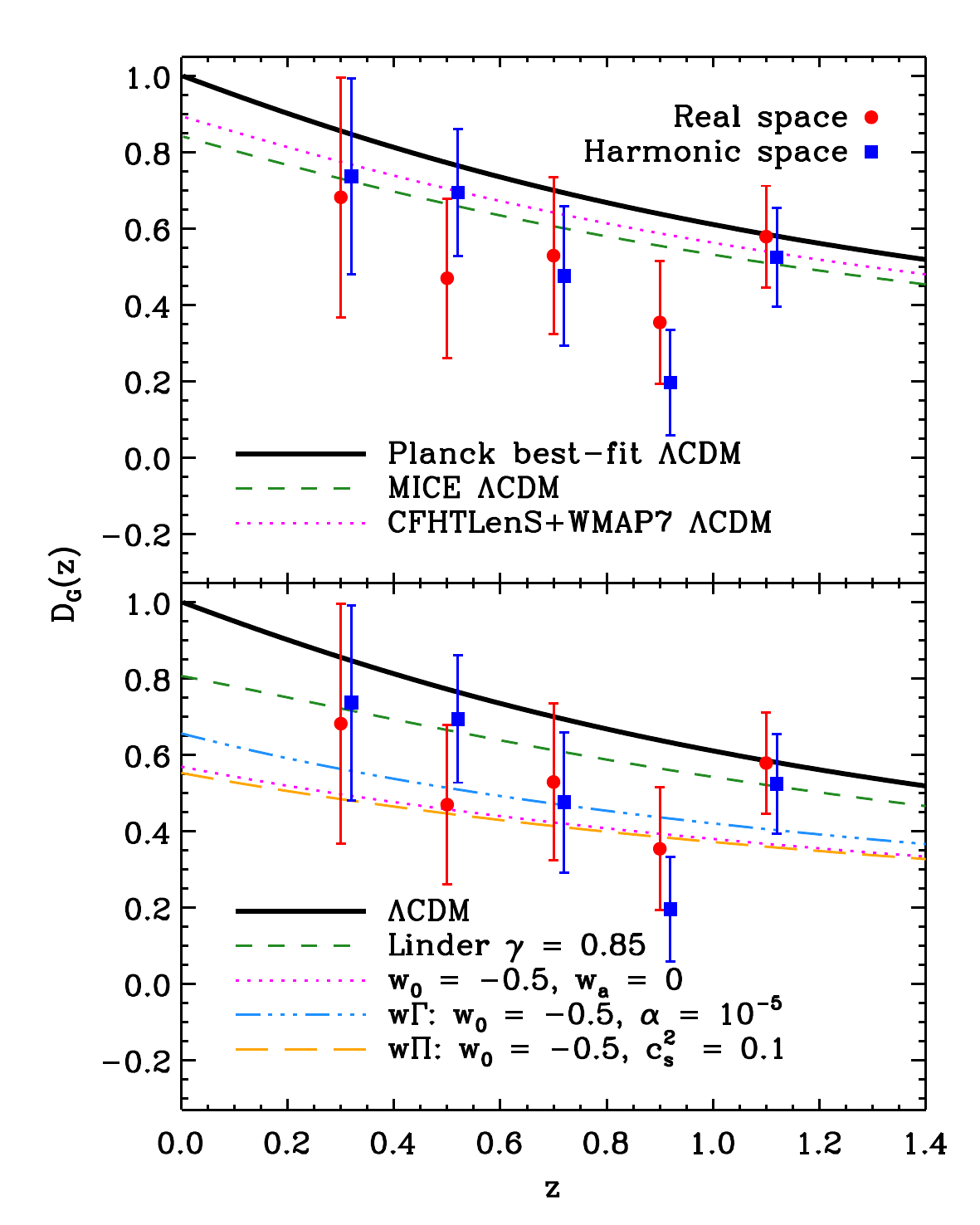}
  \caption{
    The red circles and blue squares show our growth function measurements with the $D_G$ estimator, compared with the fiducial \Planck best fit \LCDM prediction (thick black line),    
    different choices of the \LCDM parameters (\emph{top panel}), and a selection of dark energy and modified gravity models (\emph{bottom panel}).
\emph{Top panel}: The green dashed line shows the prediction for the MICE cosmology, while the orange dot-dashed line refers to the best-fit \LCDM model to the CFHTLenS + \WMAP7 data by \citet{Heymans2013}.
\emph{Bottom panel}:  The coloured lines display in the order: a Linder $\gamma$ model  \citep{Linder2007}, a dark energy model parameterised by $w_0, w_a$ \citep{Chevallier2001}, and two models of modified gravity at the perturbative level: entropy perturbation ($w\Gamma$) and anisotropic stress models ($w\Pi$), as described by \citet{Battye2013}.}
\label{fig:D_MG}
\end{center}
\end{figure}

Motivated by the results of the previous section, we test how the interpretation of our results changes when we assume a different fiducial cosmology. We first adopt the baseline MICE cosmology defined above in Section~\ref{sec:mocks}; notably, in this case $\Omega_m = 0.25$, so that a significant reduction of the tension between auto- and cross-correlations is expected.
We repeat the amplitude fitting of Section~\ref{sec:results} to the measured auto- and cross-correlations in real and harmonic spaces, and we find for the full redshift sample the best-fit values of Table~\ref{tab:results_MICE}. Here we can see that indeed the change in the fiducial cosmology relieves most of this tension: the remaining differences are at the $1 \sigma$ level only.
We further proceed to a revised interpretation of the growth function estimator $D_G$, based on the MICE cosmology.
We find that as expected the tension is significantly alleviated: we obtain $A_D = 0.86 \pm 0.19$, which is consistent within $1\sigma$ with the MICE cosmology expectations. In order to shift the best-fit value to $A_D = 1$ would require in this case  a fractional decrease in $\omega_m \sigma_8$ by 14\%.
In the upper panel of Fig.~\ref{fig:D_MG} we illustrate how shifting from the \Planck best fit to other \LCDM cosmologies could bring the theoretical model closer to the observations. We consider here the MICE cosmology used in our $N$-body simulations ($\Omega_m = 0.25$, $h = 0.70$, $\sigma_8 = 0.80$) and the  best-fit \LCDM model to the CFHTLenS + \WMAP7 data by \citet{Heymans2013} ($\Omega_m = 0.255$, $h = 0.717$, $\sigma_8 = 0.794$). 
Note that for the \Planck cosmology we normalise $D_G = 1$ today, while for any other model $i$, $D_G$ is rescaled by the factor $\left( \omega_m \sigma_8 \right)_i / \left( \omega_m \sigma_8 \right)_{\mathit{Planck}} $, as the fiducial \Planck value for $\omega_m \sigma_8$ was assumed in the measured $D_G$.

\begin{table}
\begin{center}
\begin{tabular}{c c c c c}
\toprule
   \multicolumn{5}{c}{MICE cosmology, full sample, $0.2 < \zphot < 1.2$}  \\
\midrule  
 Correlation & Space      & $b \pm \sigma_b$ &  S/N  & $\chi^2 / $ d.o.f. \\ 
\midrule
Gal-Gal      & harmonic   & $ 1.27 \pm 0.04  $  &  34 & 1.6 / 3\\
             & real       & $ 1.27 \pm 0.03  $  & 41  & 4.2 / 8  \\
\midrule  
 Correlation & Space      & $A \pm \sigma_A$ &  S/N  & $\chi^2 / $ d.o.f. \\ 
\midrule    
Gal-SPT      &  harmonic  &  $ 1.06 \pm 0.19  $   & 5.5 & 9.3 / 19 \\
             &     real   & $ 1.06 \pm 0.17  $    & 6.3 & 8.2 / 11 \\
\midrule    
Gal-\Planck  & harmonic   &   $ 0.98 \pm 0.25  $   & 4.0 &  7.6 / 9 \\
             &    real    &  $ 1.03 \pm 0.30  $    & 3.4 & 7.1 / 10 \\
 \bottomrule
\end{tabular}
\caption{Summary of the results obtained when assuming the low-matter density MICE cosmology in real and harmonic spaces. We use the $N$-body covariance matrix in all cases. Assuming this fiducial model relieves most of the tension: the disagreement between auto- and cross-correlation best-fit amplitudes is in this case at the $\sim 1\sigma$ level only.}
\label{tab:results_MICE}
\end{center}
\end{table}

A further interesting possibility is to use the growth function measurement to constrain modified gravity theories. We compare in the lower panel of Fig.~\ref{fig:D_MG} our data with a selection of parameterised departures from the \LCDM model. In order to avoid the ambiguities related to scale-dependent growth for simplicity, we only consider models where the growth function remains approximately scale-independent. These include Linder's $\gamma$ parameterisation \citep{Linder2007}, in which the growth of structure evolves as $f(z) \propto \Omega_m^{\gamma}$, where $\gamma \simeq 0.55$ in \LCDMc; a dark energy model with equation of state $w(z) = w_0 + w_a z / (1 + z)$ \citep{Chevallier2001}; and two modifications of gravity at the perturbative level as described by \citet{Battye2013} and recently constrained by \citet{Soergel2015}, in which the dark fluid is described by an entropy  perturbation ($w\Gamma$ model) or anisotropic stress ($w\Pi$ model).
 For all models $i$, their growth function $D$ is normalised to recover the \LCDM behaviour at early times; in addition, $D_G$ is rescaled by the factor $\left( \omega_m \sigma_8 \right)_i / \left( \omega_m \sigma_8 \right)_{\mathit{Planck}} $. We can see that some of these models succeed in explaining the low-growth behaviour at low redshifts, although clearly the current data are not accurate enough for a solid model selection, which we defer to future DES data releases.

\subsection{Stochasticity} \label{sec:stochasticity}
\citet{CrocceACF} demonstrate that bias non-linearities can be excluded for the DES-SV `Benchmark' galaxy sample on the scales we consider.
In this case, as discussed in Section~\ref{sec:theory} above, it is possible to interpret our results by assuming that any tension between auto- and cross-correlations is due to stochasticity. 
If we do so and assume cosmology is fixed to our fiducial model, we can directly interpret our constraint on $D_G$ as a constraint on $r$, as this quantity defined in Eq.~(\ref{eq:stoc}) can be simply estimated as $ r = b_{\text{cross}} / b_{\text{auto}} $. Thus, under the assumption of the \Planck fiducial cosmology, our measurement at face value translates to $r = 0.73 \pm 0.16$.
Such result would indicate a $1.7\sigma$ preference for non-negligible stochasticity in our sample; this appears to be close to the early results by \citet{Hoekstra2002}, but in disagreement with the more recent work by \citet{Jullo2012}.

Nonetheless, an analysis of stochasticity from the galaxy-matter correlation function of the MICE-GC simulations, which were shown to reproduce most aspects of the DES-SV data correctly, find $r = 1$ to  1\% precision on all scales of interest \citep{CrocceACF}, which strongly suggests that the mismatch between auto- and
cross-correlation amplitudes can not be entirely due to stochasticity.

\section {Conclusions} \label{sec:conclusion}
We have detected the cross-correlation between the matter overdensities in the Universe as traced by the DES-SV galaxies and the CMB lensing maps reconstructed by the SPT and \Planck collaborations. 
The total significance of the detections is $6 \sigma$ for the SPT case and $4 \sigma$ for \Planck when using the DES main galaxies in the SPT-E field over 130 square degrees.  

Given the sufficient signal to noise available, and the well-tested photometric redshifts for our galaxy sample, we have studied the redshift evolution of the cross-correlation signal. Ours is the first study to examine this evolution from a single survey. 
 We divided the DES main galaxies into five photometric redshift bins of width $\Delta z = 0.2$. We found that the auto- and cross-correlations evolve in redshift as expected, recovering a significant detection at $> 2 \sigma$ in all bins and $> 3 \sigma$ in all but the lowest redshift bin.
We have finally applied these tomographic measurements of auto- and cross-correlations to reconstruct the evolution of galaxy bias and the linear growth of structure in our redshift range.

While the results are overall consistent with the \LCDM expectations, we do find a $\sim2 \sigma$  tension (including statistical errors only) between the observed amplitudes of the auto- and cross-correlations when using the full galaxy sample at $0.2 < \zphot < 1.2$, which we confirm with two fully independent analyses in real and harmonic space. This tension is observed when using either the DES-SPT or DES-\Planck cross-correlations.  
When dividing the galaxy sample into five redshift bins, we also found the amplitude of the DES-SPT cross-correlations is consistently lower than expected from the DES auto-correlations.

We then introduced a new linear growth estimator, $D_G(z)$, which 
combines auto- and cross-correlations, so that it is independent of galaxy bias on linear scales. Using this new estimator, we measured the evolution of the linear growth function in five redshift bins. 
We then compared the $D_G(z)$ measurements with a template, based on the fiducial \LCDM cosmology with a free constant amplitude $A_D$, obtaining
 $A_D = 0.73 \pm 0.16$, which is the final result of this work. 
This result shows a weak ($1.7\sigma$) tension with the fiducial \LCDM cosmology based on \Planckc.

We have quantified the impact of photo-$z$s on our results by repeating the analysis with two photo-$z$ estimators: TPZ and BPZ. We have found that using either method leaves the significance of the cross-correlation detections unaffected. If assuming BPZ, the inferred tension between auto- and cross-correlations of the full galaxy sample is reduced by $\sim 50 \%$, but the results are nearly unchanged in the tomography. In particular, our final result on the growth function estimator $D_G$ is unaffected by the choice of BPZ, as in this case we find  $A_D = 0.70 \pm 0.16$.
Further work with the upcoming DES and SPT data of extended coverage and sensitivity will be accompanied by more thorough tests of the possible systematics, including a quantitative estimation of the systematic errors from photometric redshifts and from foreground contamination by the SZ effect.

If taken at face value, the mild tension we observe can be interpreted as the data favouring a lower growth of structure  in the late universe than expected from the fiducial model, or equivalently a lower  value of $\omega_m \sigma_8$ with respect to what is fixed by the CMB at recombination.
An alternative possibility that would eliminate the tension we observe is a significant stochastic component in the galaxy density of the DES-SV sample; this interpretation leads at face value to a correlation coefficient $r = 0.73 \pm 0.16$, assuming non-linear bias can be safely ignored on the scales of interest see \citep[see][for a companion analysis supporting this assumption]{CrocceACF}. However, this is at variance with the most recent results on the subject from observations \citep{Jullo2012} and $N$-body simulations \citep{CrocceACF}.

The inferred low amplitude of the cross-correlation signal
 can be compared with the literature that reports a wide range of $A$ values. Some authors \citep[e.g.][]{vanEngelen2014} found $A$ to be consistent with the expectations, while others have found values of the CMB lensing amplitude that are $< 1$ with modest statistical significance, such as \citet{2015arXiv150405598L}, who cross-correlated the CMB lensing map from \Planck and the cosmic shear map from CFHTLenS, and \citet{Omori2015}, who correlated \Planck lensing and CFHTLens galaxy density data.

We have tested that our significance levels are reliable by running two independent analysis pipelines in real and harmonic spaces, and by
estimating the covariances with four different methods. We have checked that the results are robust by estimating the impact of nineteen possible DES systematics, by exploring the stability of the signal with a broad range of cuts in the scales considered, and with different estimators of the photometric redshifts, showing that their impact on our measurements is not statistically significant.

The CMB lensing tomography with DES will improve dramatically in the upcoming years. As shown in Fig.~\ref{fig:SN}, the area increase alone from SV to the full survey (5000 deg$^2$) is expected to boost the signal-to-noise to $\sim 30\sigma$ with either \Planck or SPT data, as the lower level of noise in SPT is compensated by the larger overlap between DES and \Planckc. Notably, this projection does not account for improvements in the CMB lensing data. If we include the expected advances from the upcoming SPT-3G survey, we obtain a signal-to-noise of $\sim 90\sigma$ with the final DES data.
 The ACT survey and its successor are also of interest. There is modest overlap with the DES footprint already, and with the Advanced ACT survey we expect to have close to complete overlap, allowing for promising cross-correlation studies similar to the SPT-3G survey. Similar sensitivity will also be achievable with the Simons Array \citep{2014SPIE.9153E..1FA}.
Looking forward to the future, it is most likely that an optimal reconstruction of the CMB lensing-matter correlation will be accomplished by a multi-probe, multi-wavelength approach: optical galaxy surveys (such as DES now, the Dark Energy Spectroscopic Instrument, the \textit{Euclid} satellite and the Large Synoptic Survey Telescope in the future) will probe the full LSS up to a redshift $\sim 2$, while the higher-redshift matter distribution will be reconstructed with other techniques, such as CIB, and later 21cm radiation intensity mapping. This multi-probe approach will eventually allow a full reconstruction of the process of structure formation across cosmic time, and determine the nature of dark energy and gravity on cosmological scales.

\section*{Acknowledgements}
TG thanks Anthony Challinor and George Efstathiou for comments on a draft version of this paper, and James Fergusson, Martin Kilbinger and Ariel S\'{a}nchez for useful discussions.
TG acknowledges support from the Kavli Foundation, STFC grant ST/L000636/1, and from the
Excellence Cluster `Universe' of Garching, Germany, as well as the
Institut de Ci\`encies de l'Espai, IEEC-CSIC, Universitat Aut\`{o}noma de
Barcelona, for hospitality. 
PF acknowledges support from the MareNostrum supercomputer (BSC-CNS,
\href{http://www.bsc.es}{www.bsc.es}), grants AECT-2008-1-0009 to 
  2010-1-0007, Port d'Informaci\'o Cient\'ifica 
 (\href{http://www.pic.es>}{www.pic.es}), and the CosmoHUB portal 
 (\href{http://cosmohub.pic.es}{cosmohub.pic.es}), where the MICE simulations 
 were run, stored, and distributed, respectively.
 PF is funded by MINECO, project ESP2013-48274-C3-1-P.
FE, BL and HVP were partially supported by the European Research Council under the European Union's Seventh Framework Programme (PP7/2007-2013) / ERC grant agreement no 306478-CosmicDawn.
CR acknowledges support from the University of Melbourne and from the Australian Research Council's Discovery Projects scheme (DP150103208).

This paper has gone through internal review by the DES collaboration.

We are grateful for the extraordinary contributions of our CTIO colleagues and the DECam Construction, Commissioning and Science Verification
teams in achieving the excellent instrument and telescope conditions that have made this work possible.  The success of this project also 
relies critically on the expertise and dedication of the DES Data Management group.

Funding for the DES Projects has been provided by the U.S. Department of Energy, the U.S. National Science Foundation, the Ministry of Science and Education of Spain, 
the Science and Technology Facilities Council of the United Kingdom, the Higher Education Funding Council for England, the National Center for Supercomputing 
Applications at the University of Illinois at Urbana-Champaign, the Kavli Institute of Cosmological Physics at the University of Chicago, 
the Center for Cosmology and Astro-Particle Physics at the Ohio State University,
the Mitchell Institute for Fundamental Physics and Astronomy at Texas A\&M University, Financiadora de Estudos e Projetos, 
Funda{\c c}{\~a}o Carlos Chagas Filho de Amparo {\`a} Pesquisa do Estado do Rio de Janeiro, Conselho Nacional de Desenvolvimento Cient{\'i}fico e Tecnol{\'o}gico and 
the Minist{\'e}rio da Ci{\^e}ncia, Tecnologia e Inova{\c c}{\~a}o, the Deutsche Forschungsgemeinschaft and the Collaborating Institutions in the Dark Energy Survey. 
The DES data management system is supported by the National Science Foundation under Grant Number AST-1138766.

The Collaborating Institutions are Argonne National Laboratory, the University of California at Santa Cruz, the University of Cambridge, Centro de Investigaciones En{\'e}rgeticas, 
Medioambientales y Tecnol{\'o}gicas-Madrid, the University of Chicago, University College London, the DES-Brazil Consortium, the University of Edinburgh, 
the Eidgen{\"o}ssische Technische Hochschule (ETH) Z{\"u}rich, 
Fermi National Accelerator Laboratory, the University of Illinois at Urbana-Champaign, the Institut de Ci{\`e}ncies de l'Espai (IEEC/CSIC), 
the Institut de F{\'i}sica d'Altes Energies, Lawrence Berkeley National Laboratory, the Ludwig-Maximilians Universit{\"a}t M{\"u}nchen and the associated Excellence Cluster Universe, 
the University of Michigan, the National Optical Astronomy Observatory, the University of Nottingham, The Ohio State University, the University of Pennsylvania, the University of Portsmouth, 
SLAC National Accelerator Laboratory, Stanford University, the University of Sussex, and Texas A\&M University.

The DES participants from Spanish institutions are partially supported by MINECO under grants AYA2012-39559, ESP2013-48274, FPA2013-47986, and Centro de Excelencia Severo Ochoa SEV-2012-0234.
Research leading to these results has received funding from the European Research Council under the European Union's Seventh Framework Programme (FP7/2007-2013) including ERC grant agreements 
240672, 291329, and 306478.

The South Pole Telescope program is supported by the National Science Foundation through grant PLR-1248097. Partial support is also provided by the NSF Physics Frontier Center grant PHY-0114422 to the Kavli Institute of Cosmological Physics at the University of Chicago, the Kavli Foundation, and the Gordon and Betty Moore Foundation through Grant GBMF\#947 to the University of Chicago.

\appendix

\section{Mocks generation and validation} \label{sec:appendix_rotations}

From the full sky projection of the MICE-GC $N$-body simulations described in Section~\ref{sec:nbody}, we produce 100 non-overlapping rotations of the SPT-E mask.
We then rotate instead the simulated overdensity map 100 times into the real DES SV mask, thus generating 100 independent realisations of the data. We do the same for the MICE CMB lensing map, which also covers the full sky and includes the lensing effect of all sources at $z < 100$. Onto each CMB lensing mock we then add one mock CMB lensing noise realisation, as provided by the SPT or \Planck collaboration. In the \Planck case, what is provided are actually 100 realisations of the full observable signal, which include both cosmological signal and noise, and the corresponding 100 realisations of the cosmological signal only, so that we reconstruct 100 noise-only maps by taking the difference between the two.

\begin{figure}
 \begin{center}
 \includegraphics[width=\linewidth, angle=0]{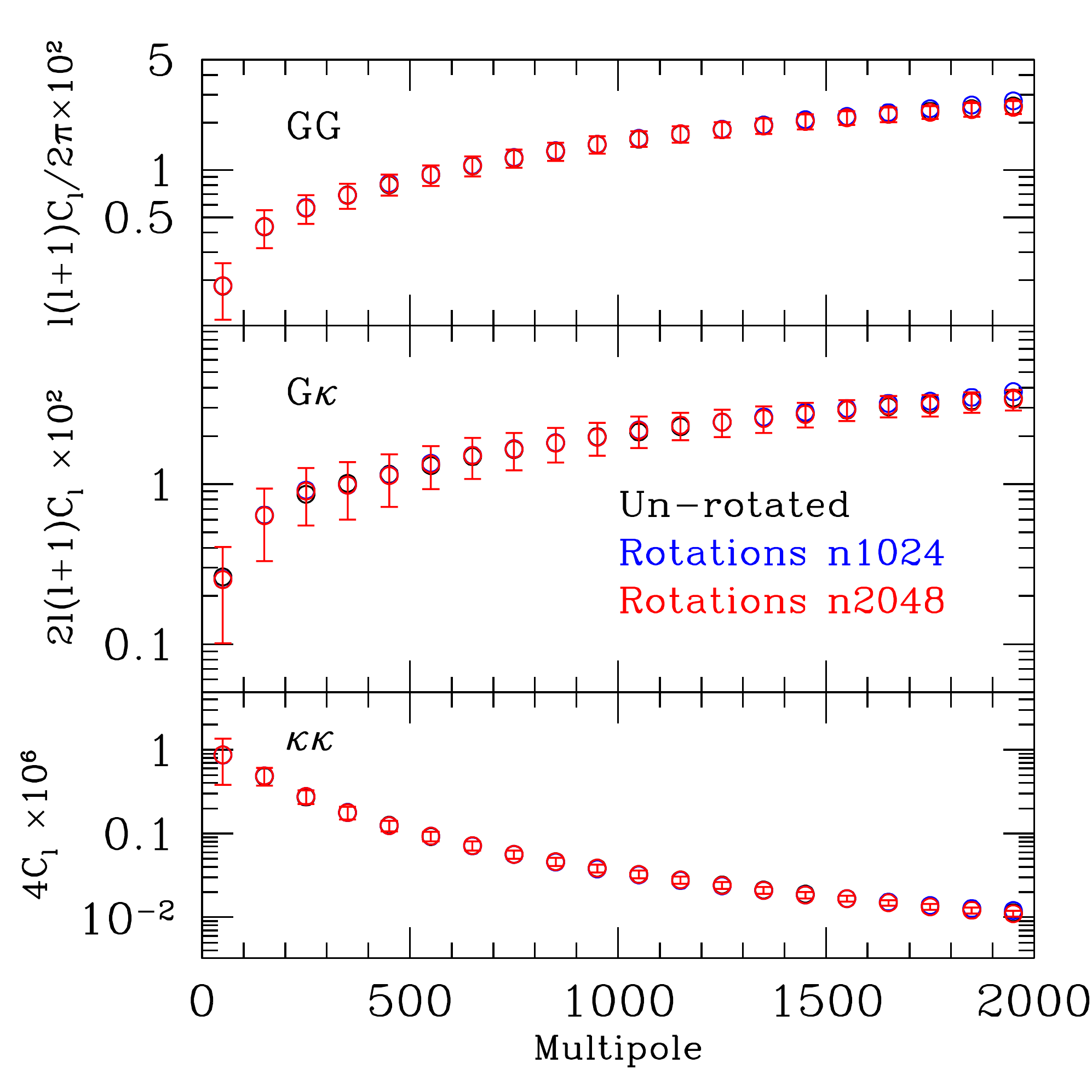}
 \caption{Comparison between the rotated and un-rotated (all-sky) power spectra (\emph{from top to bottom:} galaxy-galaxy, galaxy-CMB lensing, and lensing-lensing). Rotated mocks (\emph{in blue and red for $\nside = 1024, 2048$ respectively}) yield unbiased results with respect to the un-rotated maps (\emph{black}). The Gaussian smoothing was not applied for this particular test.}
 \label{fig:clsrotmock}
 \end{center}
 \end{figure}

We have checked using the Monte Carlo realisations that this method of all-sky map rotations yields the same covariance matrix as a statistically independent set of realisations. Furthermore the method yields an unbiased estimate of the auto- and cross-correlations. By this we mean that the average correlations (or power spectra) of the suite of rotated mocks are equal to those of the un-rotated all-sky map, as shown in Fig. \ref{fig:clsrotmock}. 
We have performed a similar validation procedure in real space, also obtaining consistent results.

\begin{figure}
 \begin{center}
 \includegraphics[width=\linewidth, angle=0]{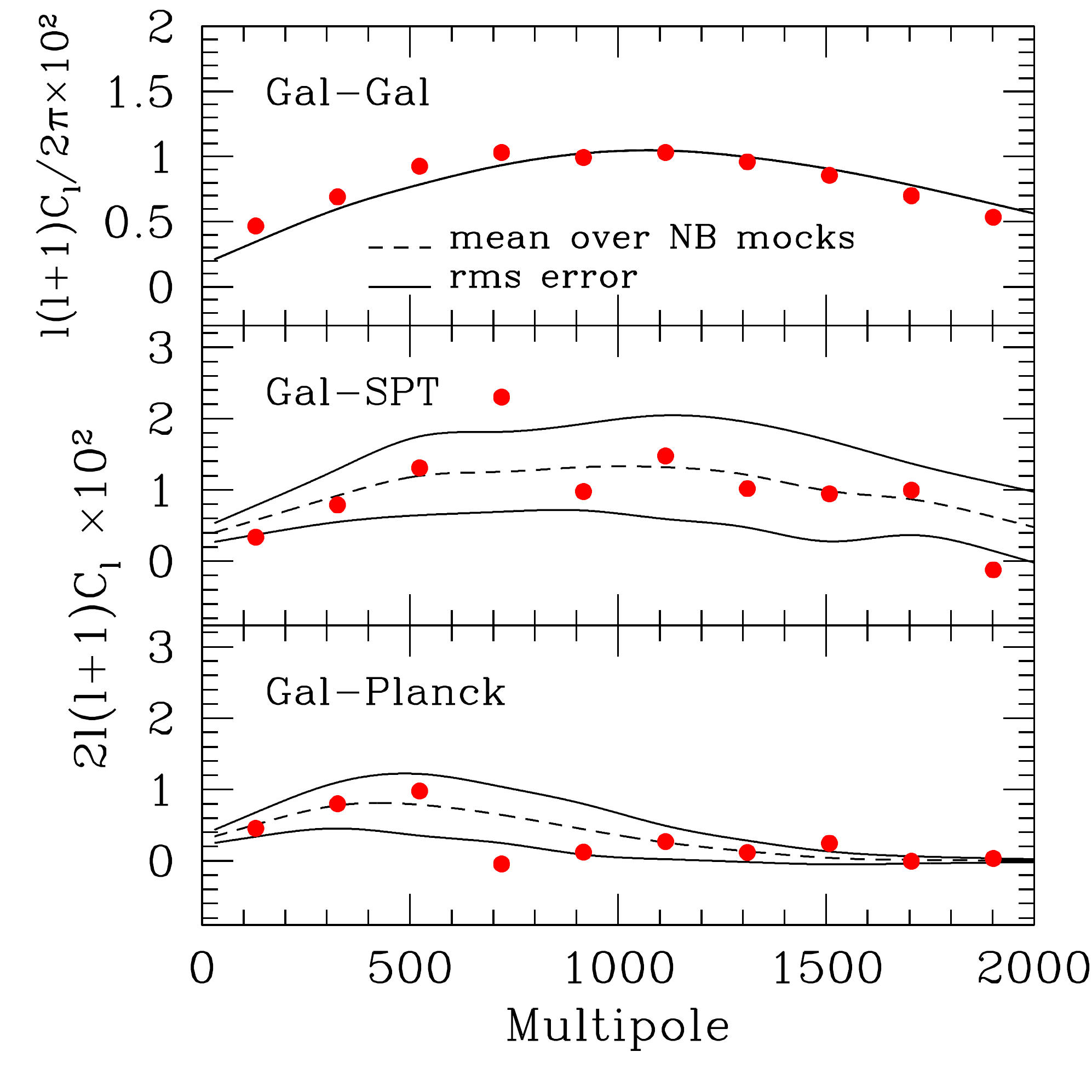}
 \caption{Comparison between the data and the mocks. The red points show the measured auto- and cross-spectra of the real DES, SPT and \Planck data, while the black dashed and solid lines describe the mean value and the $1\sigma$ scatter of the same spectra measured on our $N$-body mocks. The different shapes of the SPT and \Planck cross-spectra are due to the different smoothing we apply.}
 \label{fig:clsdatamocks}
 \end{center}
 \end{figure}

After generating the mock galaxy maps, we add Poisson noise on each pixel by randomly resampling each pixel number density from a Poisson distribution. Finally, we smooth the mock maps with the same Gaussian beam we apply to the real data.
We demonstrate in Fig.~\ref{fig:clsdatamocks} that the mean auto- and cross-power spectra of the mocks and their scatter agree well with the properties of the real data.

\section{Shot noise}
\label{sec:shotnoise}

\begin{figure}
\begin{center}
\includegraphics[width=\linewidth, angle=0]{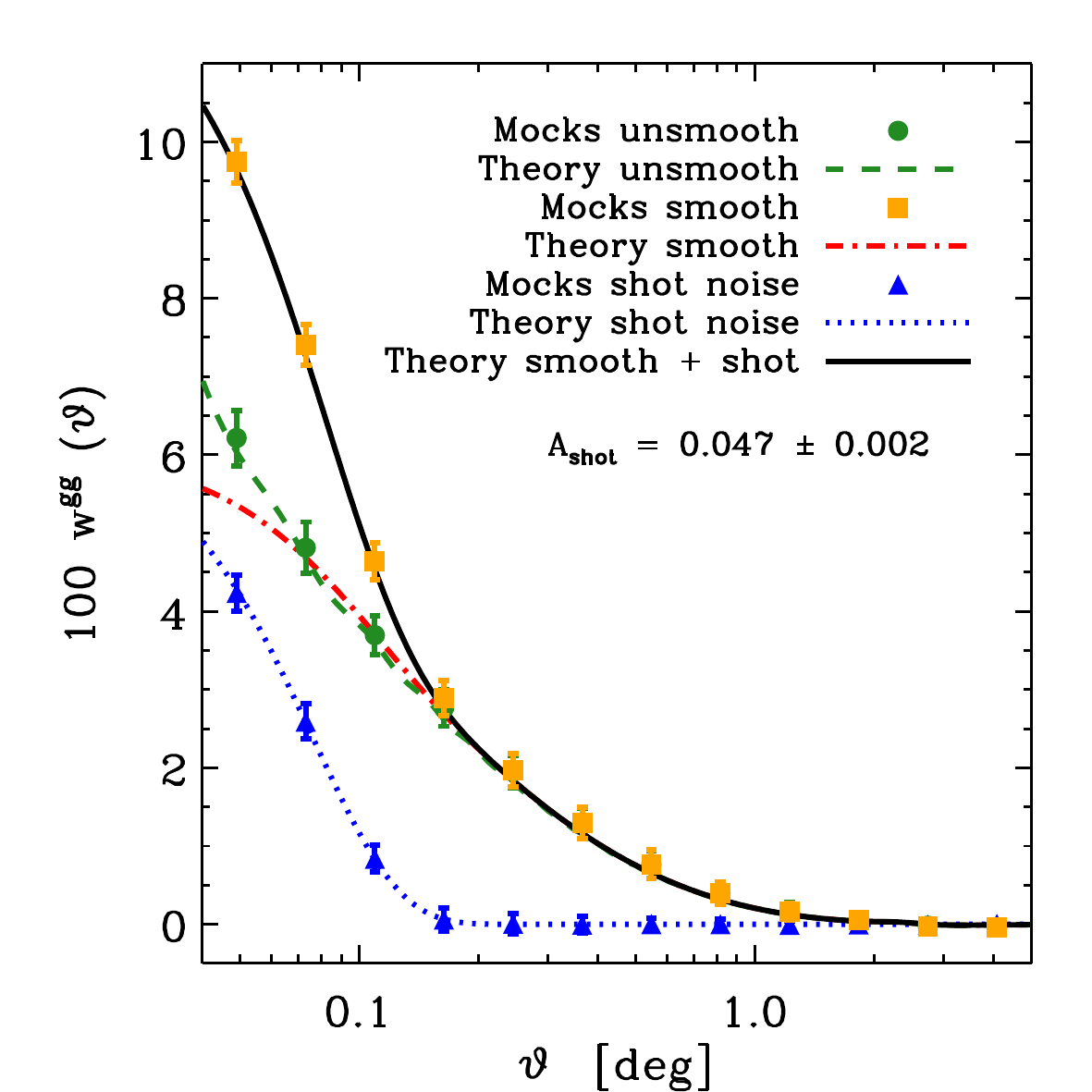}
\caption{Modelling the shot noise contribution to a smoothed galaxy map. The mean auto-correlation function of 100 unsmoothed galaxy mocks  (\emph{green circles}) in the redshift bin $1.0 < \zphot < 1.2 $ is in good agreement with the fiducial cosmological model $w^{gg} (\vartheta)$ (\emph{green dashed line}); the effect of shot noise is limited to an additional contribution at zero separation, not shown in the plot. When the mocks are smoothed, the shot noise component spreads to non-zero angles: the observed mean auto-correlation (\emph{orange squares}) does not match the smoothed cosmological model $w^{gg}_{\mathrm{smooth}} (\vartheta)$ (\emph{dot-dashed red curve}), as the shot noise contribution is missing from the model. We plot with blue triangles the shot noise component measured from the mocks with the estimator of Eq.~(\ref{eq:shotnoise}), which is well fit by the model $ w^{\mathrm{shot}}_{\mathrm{smooth}} (\vartheta$) with an amplitude $ A_{\mathrm{shot}} = 0.047$  (\emph{dotted blue line}). By adding this to the smoothed cosmological theory, we obtain the full model of Eq.~(\ref{eq:fullshotmodel}) (\emph{black solid line}), which is a good match to the smoothed mocks.}
\label{fig:shotnoise}
\end{center}
\end{figure}

As we estimate the matter overdensity via the number of observed galaxies per unit area, shot noise is introduced in the analysis. In harmonic space, this is described in the ideal case by the contribution $N_\ell^{gg} = 1/ n$ mentioned in Section~\ref{sec:theory}; this is constant on the full sky, but it is affected in the same way as the cosmological signal $C_\ell^{gg}$ by the effects of survey mask, pixellation, and any additional smoothing applied to the map.

In a real-space analysis of a pixellated map, shot noise only affects the auto-correlation function at zero lag, by adding to the cosmological signal $w^{gg} (\vartheta)$ a contribution 
\be
w^{\mathrm{shot}} (0 \deg) = 1/\bar{n}_{\mathrm{pix}} \, ,
\ee
where  $\bar{n}_{\mathrm{pix}}$ is the number density of galaxies per pixel.
However, as in our analysis we apply an additional Gaussian smoothing, the effect of shot noise is diluted onto angular separations $\vartheta > 0$ deg. The effective auto-correlation can be written as
\be \label{eq:fullshotmodel}
w^{gg + \mathrm{shot}}_{\mathrm{smooth}} (\vartheta) = w^{gg}_{\mathrm{smooth}} (\vartheta) + w^{\mathrm{shot}}_{\mathrm{smooth}} (\vartheta) \, ,
\ee
where $w^{gg}_{\mathrm{smooth}} (\vartheta)$ is the smoothed galaxy auto-correlation of cosmological origin, while the shot noise contribution is \citep{2002ApJ...580..672B}
\be \label{eq:shotmodel}
 w^{\mathrm{shot}}_{\mathrm{smooth}} (\vartheta) = A_{\mathrm{shot}} \,  \frac{1}{\bar{n}_{\mathrm{pix}}} \, e^{-\frac{\vartheta^2}{4 \sigma^2}} \, .
\ee
The constant $ A_{\mathrm{shot}} $ depends on the relative size of the pixel $d_{\mathrm{pix}}$ and the smoothing beam: for $\sigma \ll d_{\mathrm{pix}} $, $ A_{\mathrm{shot}} \to 1$ and the shot noise is returned to the zero-lag limit, while the dilution will affect larger scales, and $ A_{\mathrm{shot}} \to 0 $, if $\sigma \gg d_{\mathrm{pix}} $.

We determine $ A_{\mathrm{shot}} $ from our set of $N$-body simulations as follows. 
We first estimate the shot noise contribution in each angular bin by averaging over the 100 mock maps as 
\be \label{eq:shotnoise}
\hat w^{\mathrm{shot}}_{\mathrm{smooth}} (\vartheta) = \left \langle \hat w^i_{\mathrm{smooth}} (\vartheta) - \hat w^{i} (\vartheta) \frac { w^{gg}_{\mathrm{smooth}} (\vartheta)} { w^{gg} (\vartheta)} \right \rangle_i \,  ,
\ee
where $ \hat w^i_{\mathrm{smooth}} (\vartheta)$ is the measured auto-correlation from the smoothed mock $i$,  $ \hat w^i (\vartheta) $ is the measured auto-correlation from the unsmoothed mock $i$, and  $ w^{gg}_{\mathrm{smooth}} (\vartheta) /  w^{gg} (\vartheta) $ is the ratio between the smoothed and unsmoothed theoretical predictions for the cosmological signal.
We focus on the highest redshift bin $1.0 < \zphot < 1.2 $, which has the lowest number density, and thus the highest shot noise.
We then derive $ A_{\mathrm{shot}} $ with a one-parameter likelihood fit, minimising the $\chi^2$ between the mock data of Eq.~(\ref{eq:shotnoise}) and the model of Eq.~(\ref{eq:shotmodel}), using the full covariance matrix from the same mocks.
We thus obtain $ A_{\mathrm{shot}} = 0.047 \pm 0.002 $.

As we can see in Fig.~\ref{fig:shotnoise}, this model is in good agreement with  the measured auto-correlations of the smoothed mocks. We have confirmed that the same value of $ A_{\mathrm{shot}} $ is accurate for all redshift bins as expected. We use this model to subtract the shot-noise contribution from all measured real-space auto-correlations. The cross-correlations are naturally unaffected.

\section{Robustness of the covariance matrix estimation}  \label{sec:appendcovariance}

\begin{table*}
\begin{center}
\begin{tabular}{c c c c c c c c}
\toprule
               \multicolumn{2}{c} {Full sample,   $0.2 < \zphot < 1.2$}  &     \multicolumn{3}{c} {Real space }              &   \multicolumn{3}{c} {Harmonic space}    \\
\midrule 
 Correlation & Covariance  &  $b \pm \sigma_b$ &  S/N  & $\chi^2 / $ d.o.f.  & $b \pm \sigma_b$ &  S/N  & $\chi^2 / $ d.o.f.  \\ 
 \midrule
 Gal-Gal         &$N$-body& $1.22 \pm 0.03 $  & 41  &  3.8 / 8  &  $ 1.22 \pm 0.04  $  & 34 & 2.7 / 3 \\
                 & Theory  & $1.23 \pm 0.02 $  & 51  &  5.8 / 8  &  $ 1.26 \pm 0.03  $  & 51 & 1.3 / 3 \\
                 & MC      & $1.23 \pm 0.03 $  & 47  &  5.8 / 8  &  $ 1.26 \pm 0.04  $  & 33 & 0.54 / 3 \\
                 &  JK      & $1.22 \pm 0.03 $  & 48  &  5.4 / 8  & ---  & --- & --- \\
\midrule 
 Correlation & Covariance  &  $A \pm \sigma_A$ &  S/N  & $\chi^2 / $ d.o.f.   &  $A \pm \sigma_A$ &  S/N  & $\chi^2 / $ d.o.f.   \\ 
\midrule
Gal-SPT         & $N$-body& $0.84 \pm 0.13  $  & 6.3  &  8.4 / 11  &  $ 0.84 \pm 0.15  $   & 5.6 & 8.7 / 19  \\
                 & Theory  & $0.86 \pm 0.13 $  & 6.6 &  13 / 11 &  $ 0.85 \pm 0.13  $  & 6.6 & 11  / 19 \\
                 & MC      & $0.91 \pm 0.13 $  & 6.9 &  9.2 / 11 &  $ 0.81 \pm 0.15  $  & 5.4 & 15 / 19 \\
                 &  JK      & $0.91 \pm 0.14 $  & 6.5 &  5.3 / 11 &  ---  & --- & --- \\
\midrule
Gal-\Planck     &  $N$-body & $0.78 \pm 0.21  $  & 3.7  &  11 / 10 &  $0.81 \pm 0.20  $   & 3.8 & 7.7 / 9  \\
                 & Theory  & $0.86 \pm 0.24 $  & 3.6 &  25 / 10 &  $ 0.82 \pm 0.21  $  & 3.8 & 8.3 / 9 \\
                 & MC      & $0.77 \pm 0.20 $  & 3.8 &  10 / 10 &  $ 0.82 \pm 0.25  $  & 3.3 & 5.3 / 3 \\
                 &  JK      & $0.77 \pm 0.18 $  & 4.4 &  7.8 / 10 &  ---  & --- & --- \\
\bottomrule
\end{tabular}
\caption{Summary of the results for the main galaxy sample for real (\emph{left}) and harmonic (\emph{right}) spaces:  best-fit linear bias $b$ and correlation amplitudes $A = b A_{\mathrm{Lens}}$ for the three correlation functions and four covariance estimators. The results are consistent between each other and with respect to the theoretical expectations for our fiducial model, but the cross-correlation amplitude is lower than the auto-correlation by $2-3 \sigma$. The recovered $\chi^2$ per degree of freedom indicates the models and covariance estimators are in all cases appropriate for the data, with the only exception of the DES-\Planck theoretical covariance in real space.}
\label{tab:Aall}
\end{center}
\end{table*}

\begin{figure}
\begin{center}
\includegraphics[width=\linewidth, angle=0]{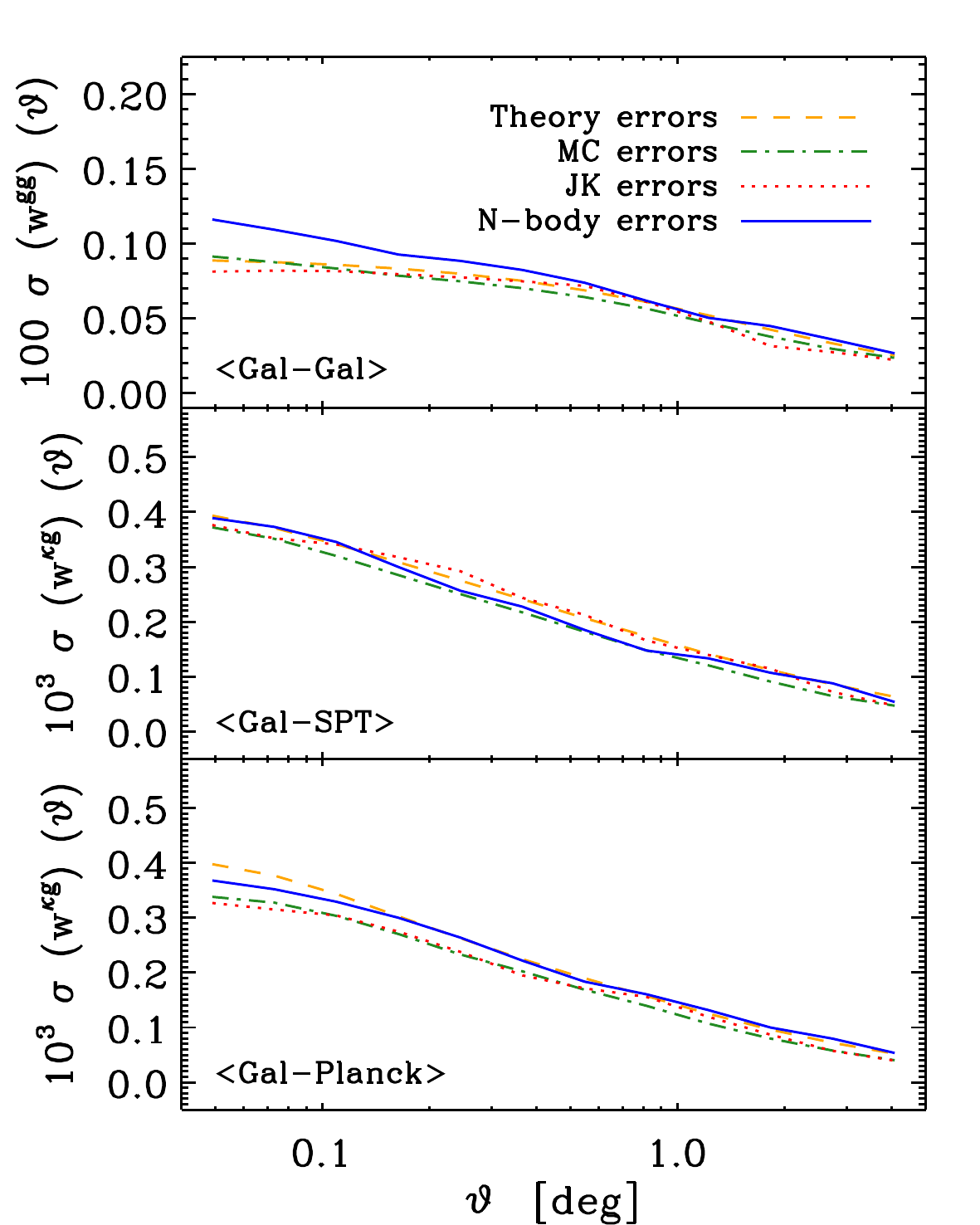}
\caption{Comparison of the real-space diagonal error bars for our four estimators of the covariance matrix: theoretical (\emph{orange dashed}), MC (\emph{green dot-dashed}), JK (\emph{red dotted}) and from $N$-body simulations (\emph{blue solid}). The three panels refer from top to bottom to galaxy auto-correlation, and the cross-correlations with SPT and \Planckc. The different methods agree for the cross-correlations, while the $N$-body covariance yields marginally larger error bars for the auto-correlation as expected due to the effect of non-Gaussianities. We use the $N$-body errors for our main results.}
\label{fig:diag_errs}
\end{center}
\end{figure}

\begin{figure}
\begin{flushleft}
\includegraphics[width=0.32\linewidth, angle=0]{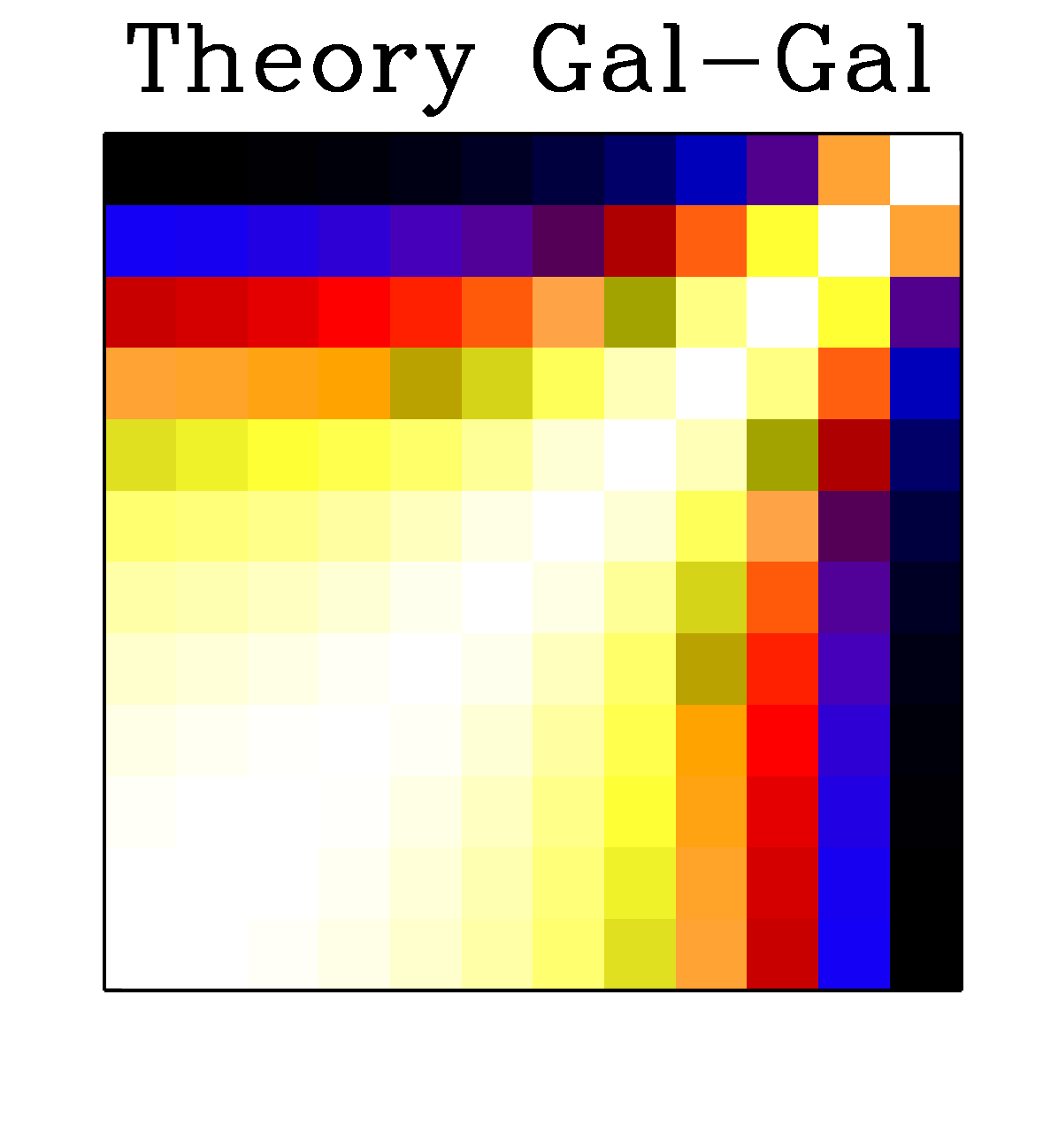}
\includegraphics[width=0.32\linewidth, angle=0]{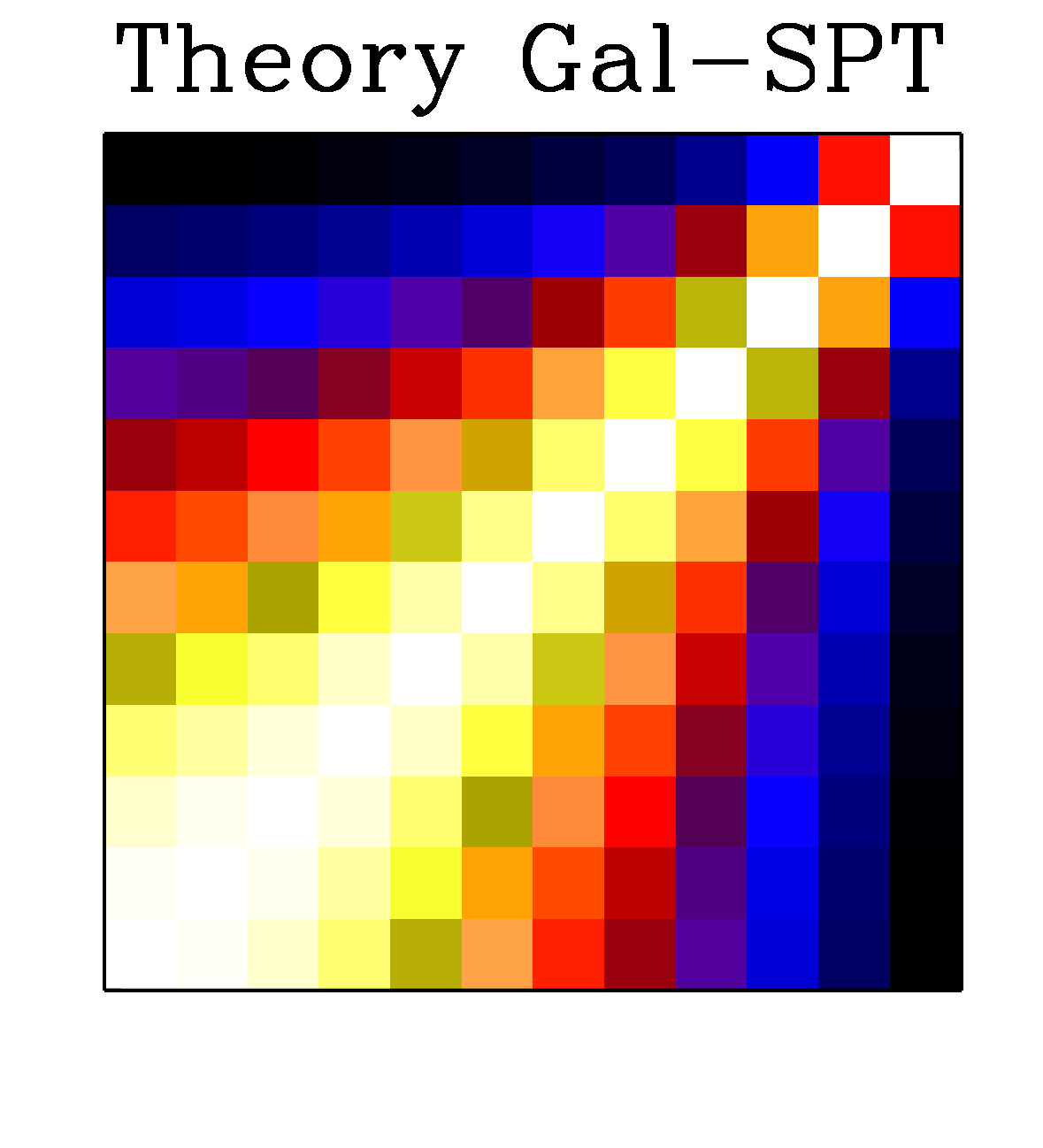}
\includegraphics[width=0.32\linewidth, angle=0]{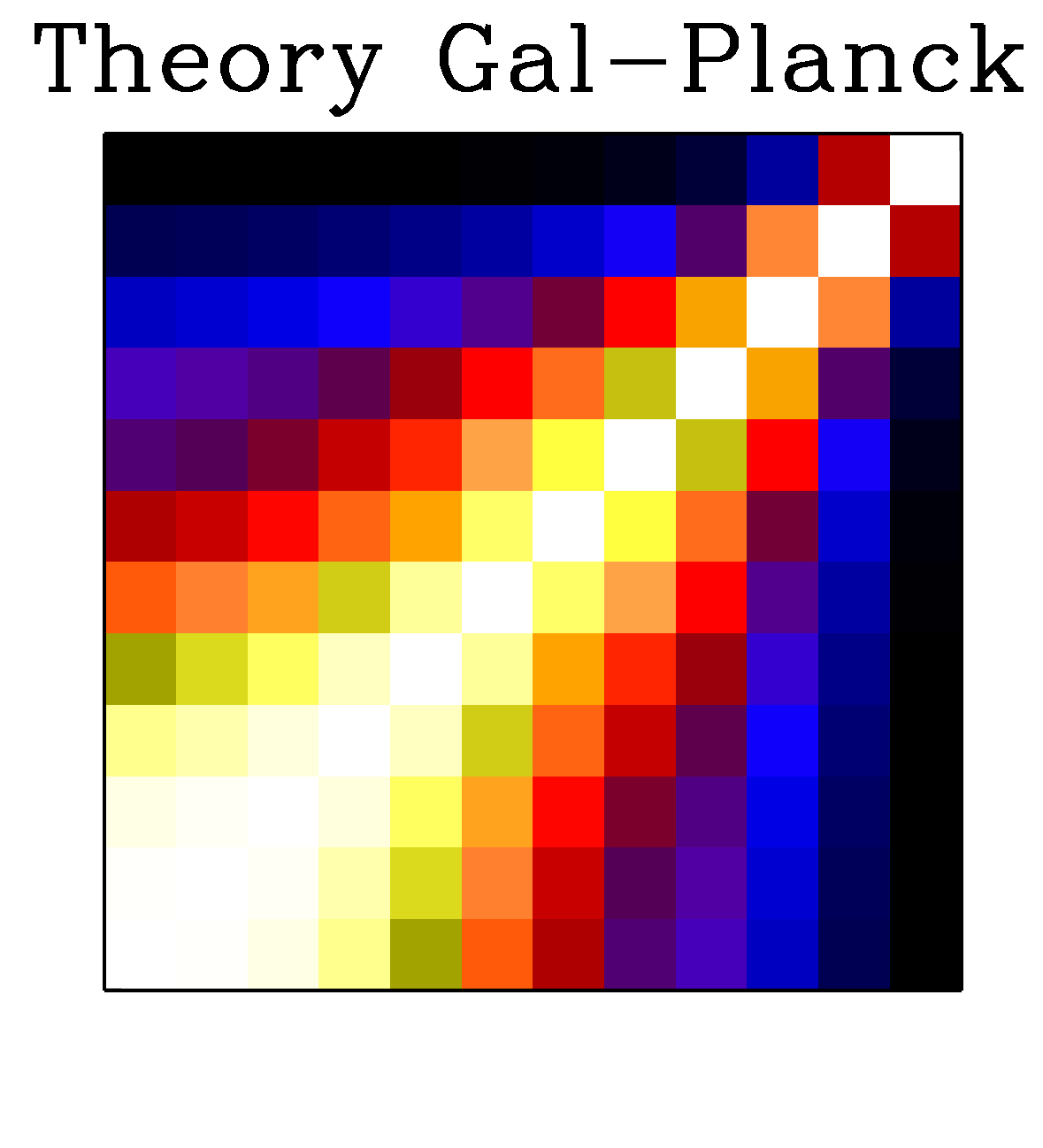}
\includegraphics[width=0.32\linewidth, angle=0]{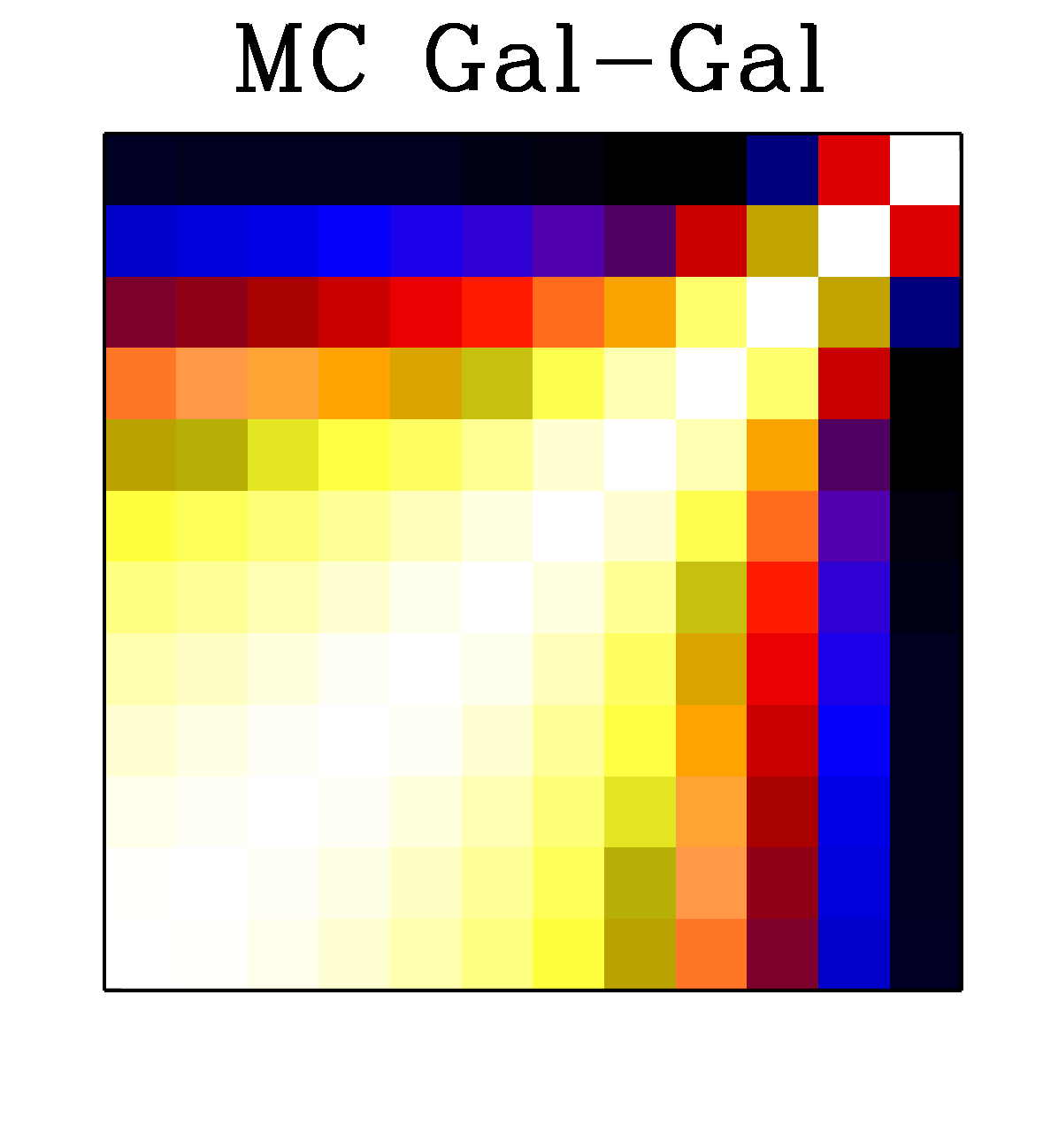}
\includegraphics[width=0.32\linewidth, angle=0]{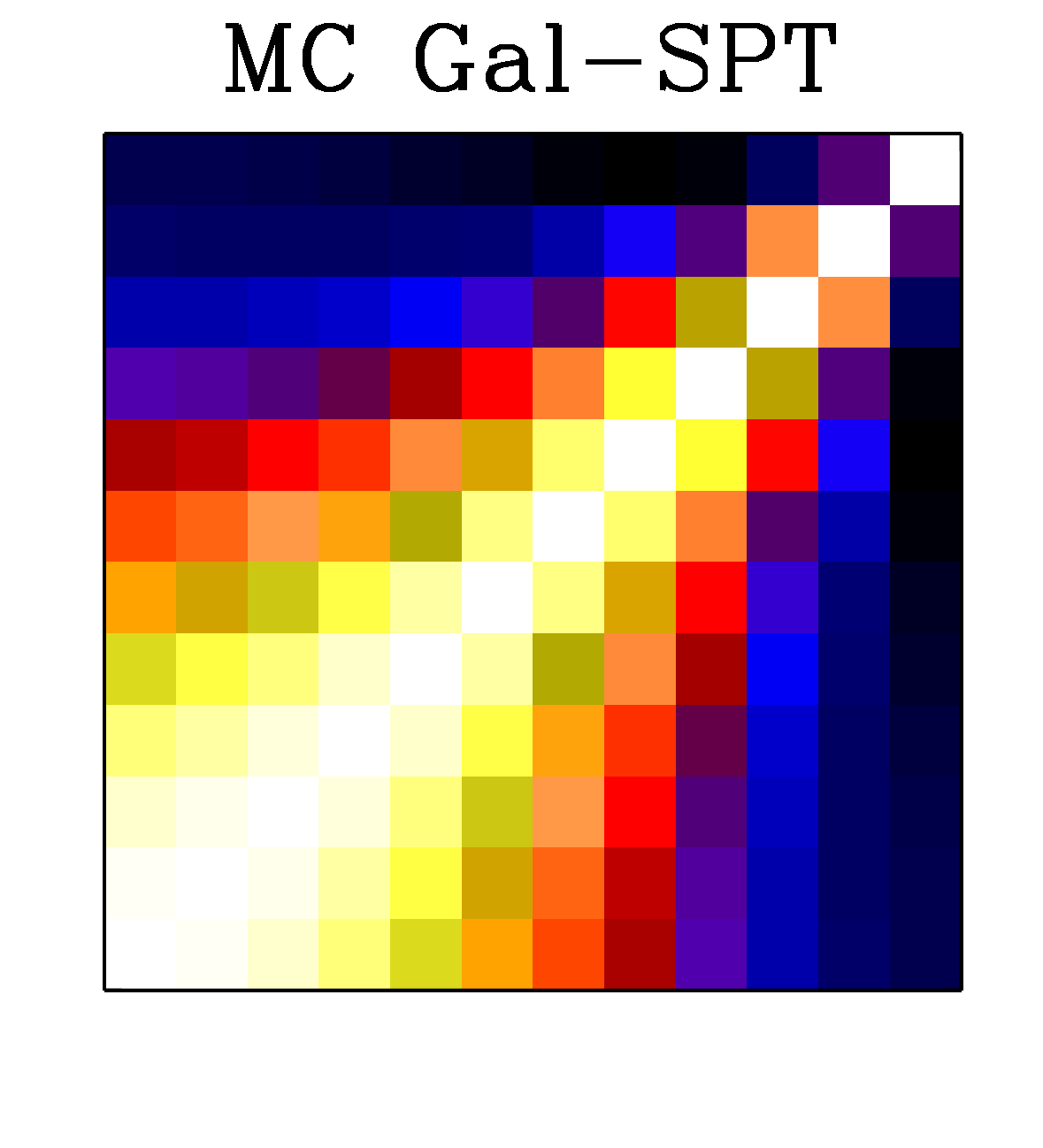}
\includegraphics[width=0.32\linewidth, angle=0]{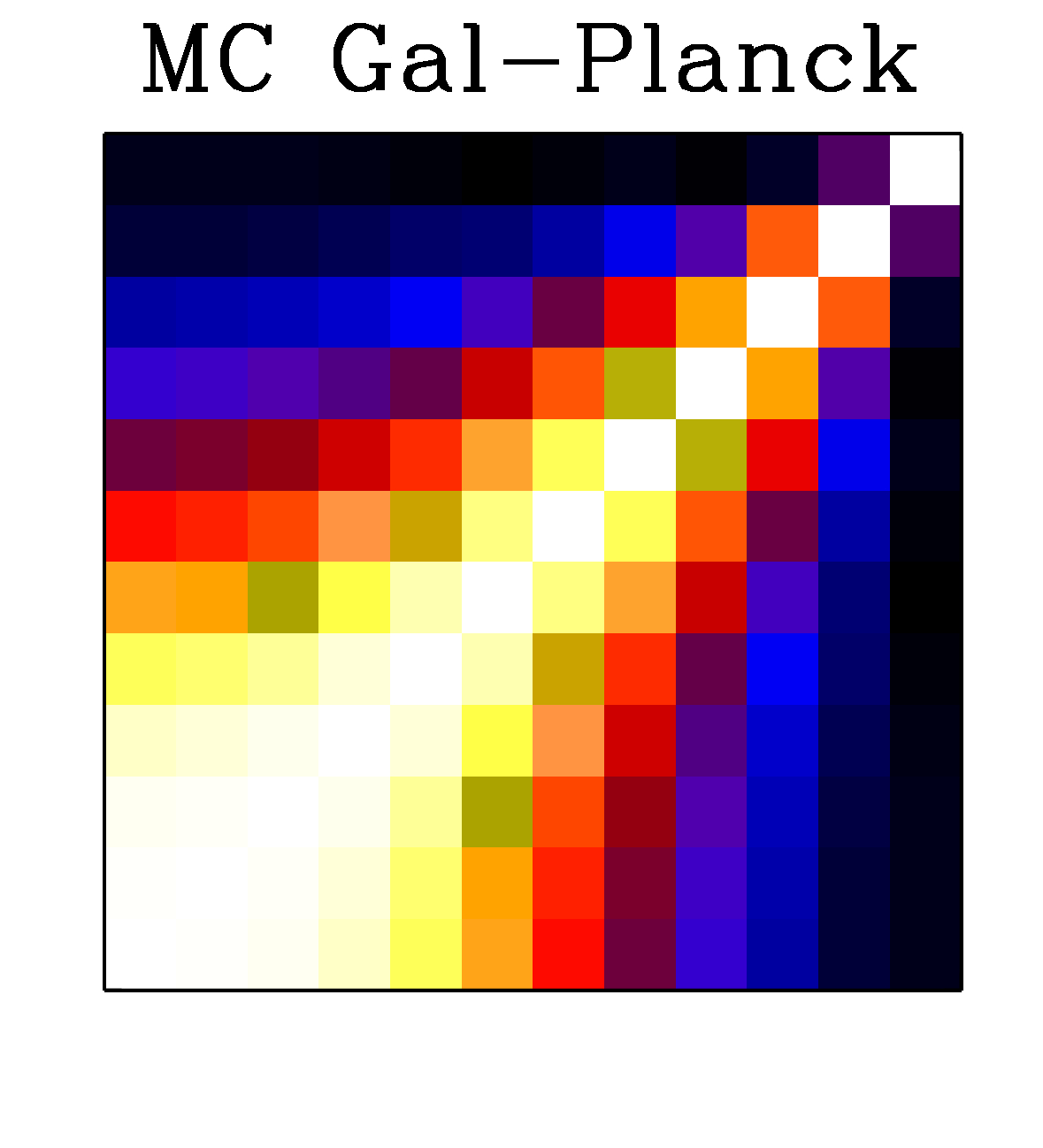}
\includegraphics[width=0.32\linewidth, angle=0]{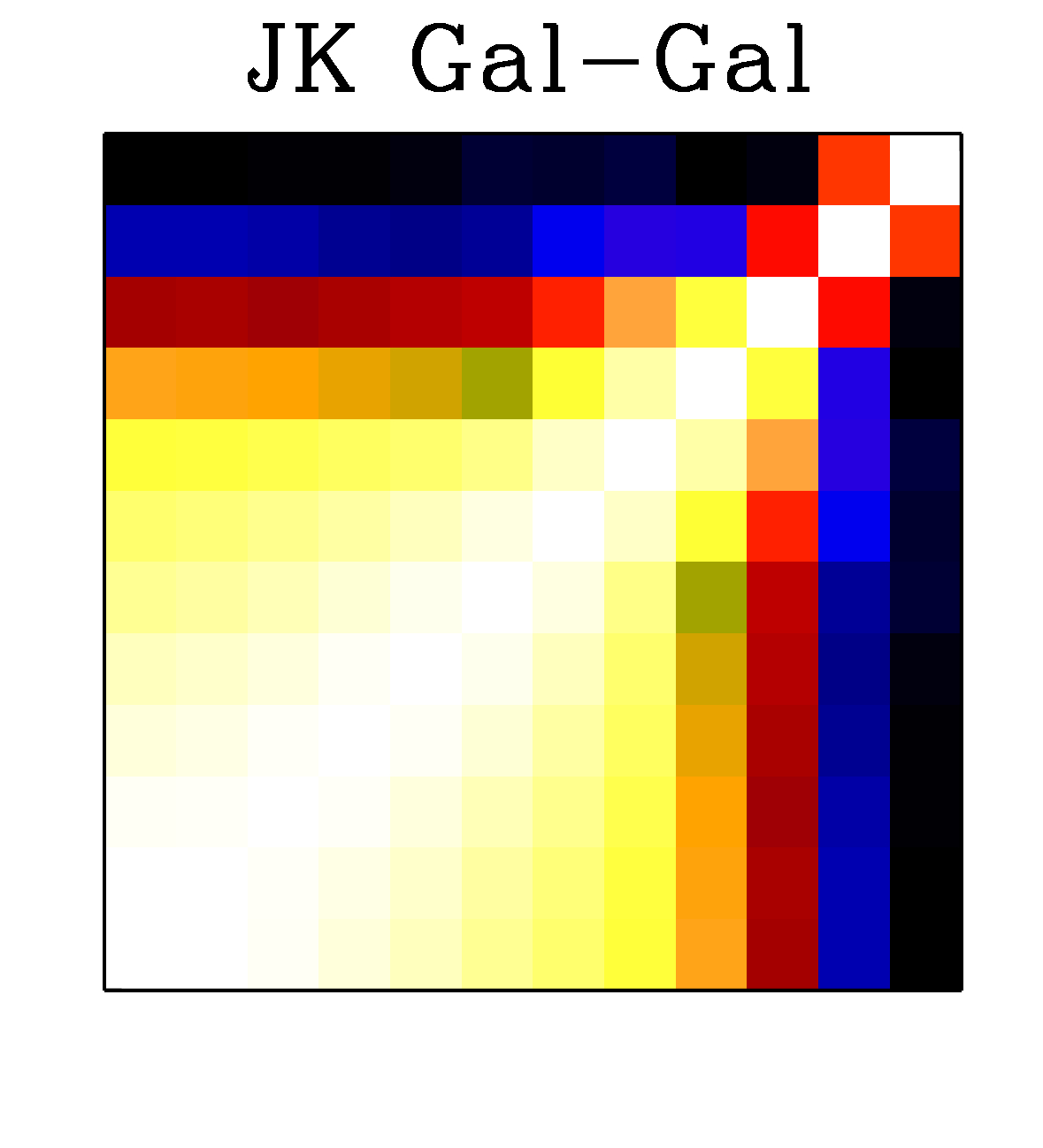}
\includegraphics[width=0.32\linewidth, angle=0]{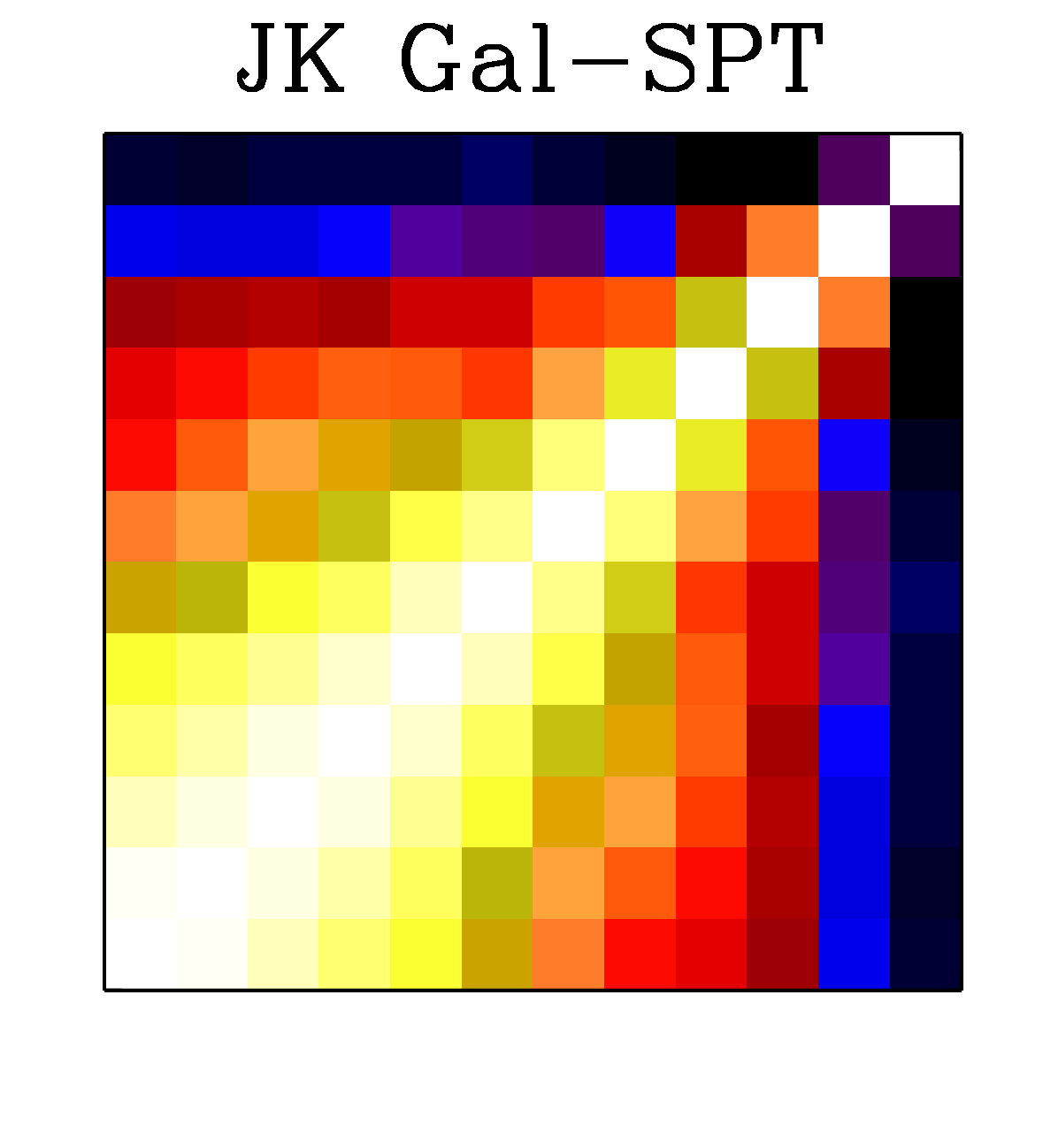}
\includegraphics[width=0.32\linewidth, angle=0]{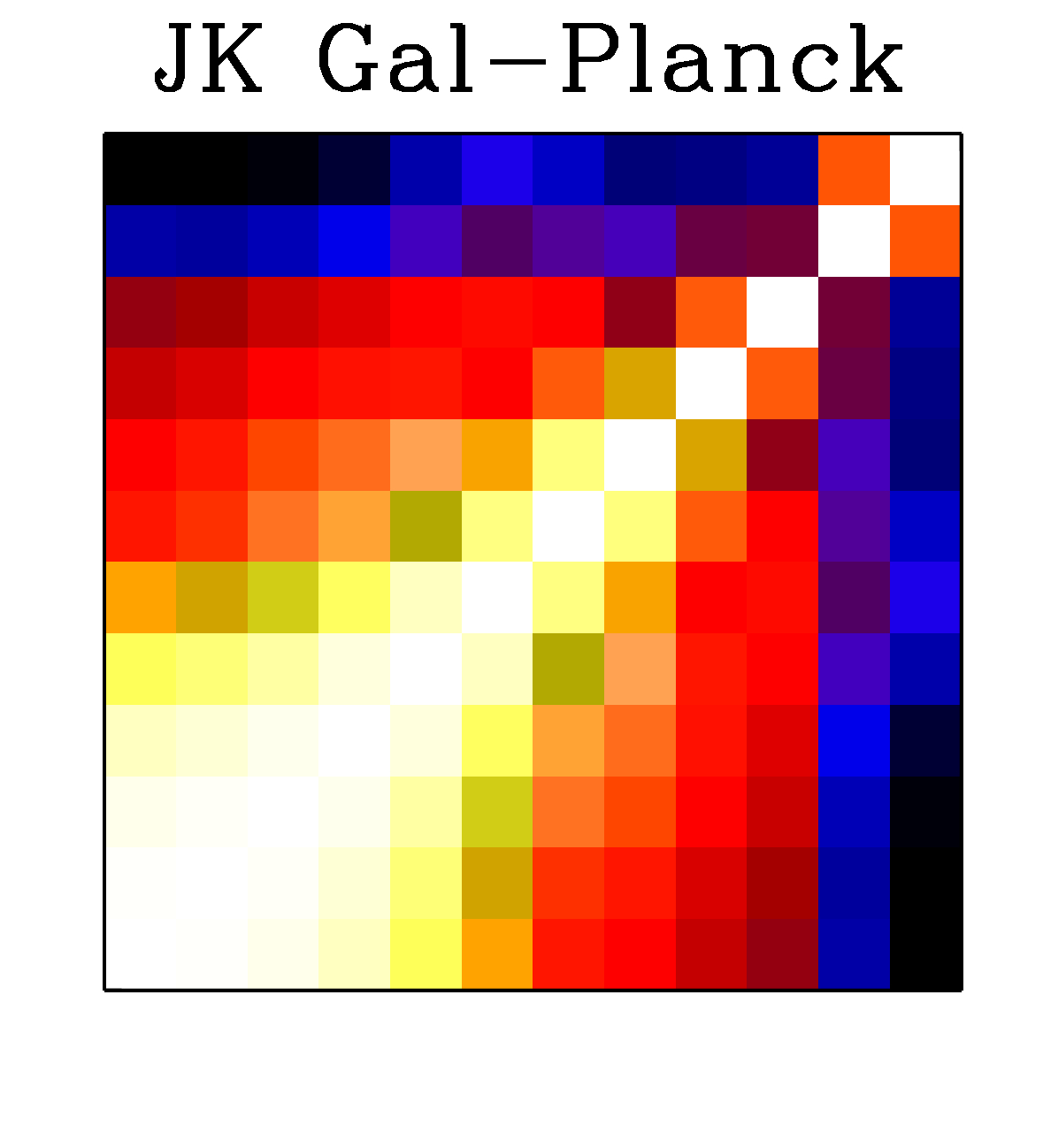}
\includegraphics[width=0.32\linewidth, angle=0]{figures/SIM100_corrmat_gg_N2048_twosmooth.jpg}
\includegraphics[width=0.32\linewidth, angle=0]{figures/SIM100_corrmat_kg_SPT_N2048_twosmooth.jpg}
\includegraphics[width=0.32\linewidth, angle=0]{figures/SIM100_corrmat_kg_Planck_N2048_twosmooth.jpg}
\caption{Comparison of four correlation matrix estimators in real space, for the three correlations we consider. The first row refers to the theoretical covariance, the second row is obtained from 1000 MC realisations, the third row is the JK method (100 regions), and the final row shows the 100 $N$-body realisations. The three columns refer to galaxy-galaxy, galaxy--SPT and galaxy--\Planck lensing convergence correlations respectively. The angular range is from 2.4 arcmin to 5 deg as in Fig.~\ref{fig:results_w}. The different methods produce consistent covariances; by comparing the different correlation matrices, we see that the galaxy-CMB lensing correlation matrices are more diagonal than the galaxy-galaxy case, which is related to the auto-correlation theory being more non-linear, and thus more non-Gaussian and covariant, at these scales. Furthermore, all matrices become more covariant in the first few angular bins due to the introduction of the Gaussian smoothing to the maps, which effectively blurs information on scales  $\vartheta < \vartheta_{\mathrm{FWHM}} = 5.4'$ (DES-SPT) and  $\vartheta < \vartheta_{\mathrm{FWHM}} = 10.8'$ (DES-\Planckc).}
\label{fig:corrmatFull}
\end{flushleft}
\end{figure}

\begin{figure}
 \begin{center}
\includegraphics[width=\linewidth, angle=0]{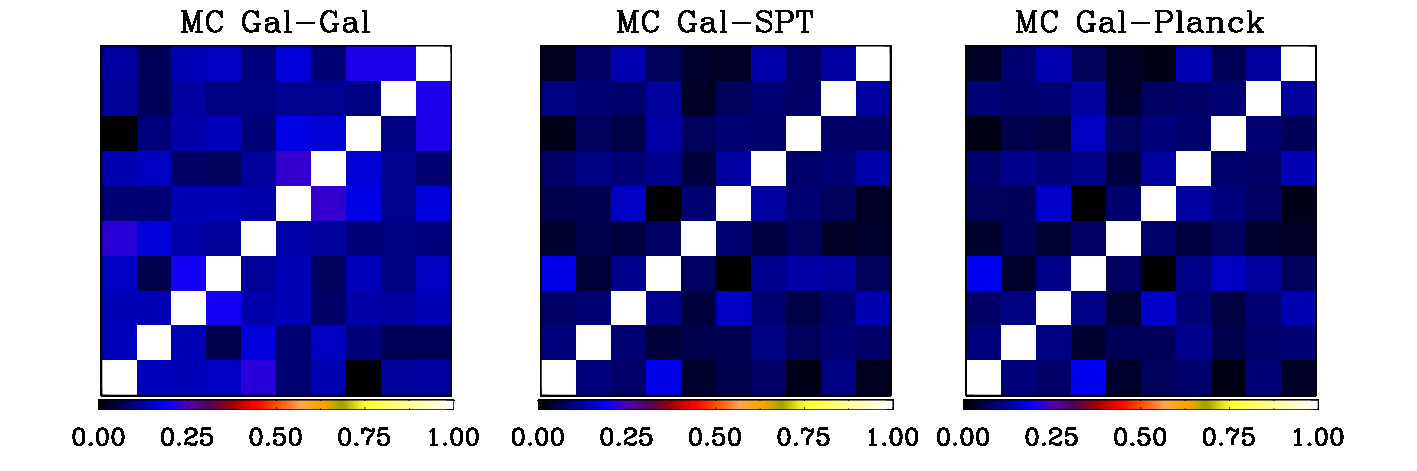}
%bincovtl_1.eps}
   \includegraphics[width=\linewidth, angle=0]{figures/bincovtl_combined_10nbins_dl197_spte_micetpz1newpsm_bin1_SPT+PLANCK_100sims_n2048_lmin30_lmax2000_red.jpg}
%{figures/bincovtl_2.eps}
\caption{Comparison of two different correlation matrix estimators in harmonic space: the first row shows the results from MC realisations, while the second row refers to 100 $N$-body realisations. We show correlations among $C_\ell$ band-powers for the galaxy auto-, galaxy-SPT lensing and galaxy-\Planck cross-correlations, from left to right, respectively. 
We use ten linear multipole band-powers from $\ell_{min}=30$ to $\ell_{max}=2000$, with $\Delta_{\ell} = 197$, matching the bins of Fig.~\ref{fig:clstpz}.} 
 \label{fig:ClcovFull}
 \end{center}
 \end{figure}

We demonstrate in this section the robustness of our covariance matrix estimation in real and harmonic spaces.

We estimate the covariance matrix of the results in four different ways.
In addition to the MC and $N$-body methods described above in Section~\ref{sec:covar}, we first use an analytic approach: we (optimistically) assume a diagonal covariance in harmonic space, including the uncertainties from cosmic variance, shot noise in the galaxy counts, and CMB lensing noise, as described in Eq.~(\ref{eq:errcl}). The galaxy shot noise is determined by the observed galaxy number density in our sample, and we use for the CMB lensing noise its level as determined from the CMB lensing maps auto-spectrum, as discussed below in Section~\ref{sec:pseudo_cl}.  For any pair of maps $a,b$, the harmonic space covariance $\sigma^2 \left( C_\ell^{ab} \right) $ defined above in Section~\ref{sec:theory} is readily transformed to real space \citep{CrocceCabreGaztanaga2011, 2011MNRAS.415.2193R}:
\begin{multline} \label{eq:covwtheta}
\left( \mathcal{\hat C}_{ij}^{ab} \right)_{\mathrm{TH}} \equiv \mathrm{Cov} [w^{ab}] (\vartheta_i,\vartheta_j) = \\
= \sum_{l=0}^{\infty} \frac{(2 \ell + 1)^2}{(4 \pi)^2} P_\ell (\cos \vartheta_i) P_\ell (\cos \vartheta_j) \, \sigma^2 \left( C_\ell^{ab} \right) \, .
\end{multline}
Notice the sum goes in principle up to infinity, but it is in practice possible to truncate it to a finite value given the Gaussian smoothing we apply.

We also estimate the covariance matrices with a jack-knife (JK) technique. This consists of removing  in turn $N_{\mathrm{JK}}$ subsets of the data, to obtain $N_{\mathrm{JK}}$ pseudo-random realisations of the correlations, whose scatter can be used to estimate the covariance as in Eq.~(\ref{eq:covmat}), but with an additional factor:
\be \label{eq:covmatJK}
\left( \mathcal{\hat C}_{ij}^{ab}\right)_{\mathrm{JK}} =\frac{N_{\mathrm{JK}} - 1}{N_{\mathrm{JK}}} \sum_{\alpha=1}^{N_{\mathrm{JK}}} \left( \hat w_{\alpha, i}^{ab} - \bar w_{i}^{ab} \right) \left( \hat w_{\alpha, j}^{ab} - \bar w_{j}^{ab} \right) \, .
\ee
The advantage in this case is that the method is completely model-independent; it is nevertheless not uniquely defined, as the number of patches that it is possible to remove is limited, and it typically yields different results depending on the particular procedure chosen. Also in this case we use the $\beta$ correction of the inverse covariance (Eq. \ref{eq:hartlap}); while not mathematically exact in the case of non-independent realisations, it was shown by \citet{2007A&A...464..399H} to yield accurate results also in this case.
We tested several JK methods; we show below the results for
a scheme where we have divided the galaxy mask into $N_{\mathrm{JK}} = 100$ mostly contiguous patches with the same number of pixels. We have achieved this by selecting 100 sets of pixels whose pixel ID is contiguous in the \textsc{Healpix} \emph{nested} scheme; this ensures the patches are mostly contiguous.

We summarise in Table~\ref{tab:Aall} the best-fit results obtained using the four covariance estimators in real and harmonic spaces, where we can see that all methods agree, and they all yield realistic reduced $\chi^2$ values. The only exception is the DES-\Planck case with the theoretical covariance estimator (PTE: <1\%); the $N$-body covariance yields anyway a typical $\chi^2$ (PTE: 35\%).

We then show in Fig.~\ref{fig:diag_errs} the real-space diagonal error bars obtained with the four estimators we consider: theoretical prediction, jack-knife, Monte Carlo, and full $N$-body. The error bars obtained with all methods are in excellent agreement: we can see that all methods fully agree for the cross-correlations, while in the auto-correlation the $N$-body errors are larger than the others; this is reasonable as this is the only method, besides the less stable JK, that incorporates the non-Gaussian variance produced on small scales by non-linear structure formation.
In order to also compare the off-diagonal part of the covariance matrices, we show in Fig.~\ref{fig:corrmatFull} the real-space correlation matrices of the three correlations we study, obtained with all four methods we consider. We can see that the agreement between the methods is excellent; the JK results are marginally noisier, but the general behaviour consistently shows a high off-diagonal covariance for the galaxy auto-correlation on small scales, and lower covariance for the cross-correlations.

Finally, we show in Fig.~\ref{fig:ClcovFull} a comparison of the harmonic-space correlation matrices. In this case, the theoretical correlation matrix is fully diagonal, as the effect of the survey mask is not included, so we do not show it, but we limit the comparison to the MC and $N$-body estimators. We find that the MC covariances are still mostly diagonal, while more significant off-diagonal contributions emerge in the case of the full $N$-body estimator, especially for the auto-spectrum case. This is expected, and it is due to the significant non-Gaussianities produced by non-linearities, which are non-negligible on these scales.

\section{The optimal quadratic $C_\ell$ estimator}
\label{sec:oqe_appendix}

\subsection{Implementation}

In the following, we briefly reiterate the basic equations of the
optimal quadratic estimator proposed by \citet{1997PhRvD..55.5895T}.
We then summarise the necessary extensions for power spectrum
estimation in wide multipole bins, and discuss a simple regularisation
approach for bandpass filtered data.

We start by defining the $\npix \times \npix$ data covariance matrix in
pixel space as a sum of contributions from signal and noise,
$\mathbf{C} = \mathbf{S} + \mathbf{N}$. For an isotropic signal with
power spectrum $C_\ell$,
\be
\label{eq:oqe_signal_cov}
\mathbf{S}_{n n^\prime} = \sum_{\ell = \lmin}^{\lmax} \frac{2\ell + 1}{4 \pi}
C_\ell P_\ell(\hat{n} \cdot \hat{n}^\prime) \, ,
\ee
where we have introduced the Legendre polynomials $P_\ell$ with the argument given by the
dot product between the normal vectors of pixels $n$ and $n^\prime$.
Then, the optimal power spectrum estimate $\hat{C}_\ell$ is given by a
quadratic combination of the data vector $\mathbf{d}$,
\be
\hat{C}_\ell = \frac{1}{2} \sum_{\ell^\prime} \left( F^{-1}
\right)_{\ell \ell^\prime} \mathbf{d}^\dagger \mathbf{E}_{\ell^\prime} \mathbf{d} \, ,
\ee
where
\be
\label{eq:oqe_estimation_matrix}
\mathbf{E}_\ell = \mathbf{C}^{-1} \frac{\partial \mathbf{C}}{\partial
  C_\ell} \mathbf{C}^{-1} \, ,
\ee
and the Fisher matrix is
\be
F_{\ell \ell^\prime} = \frac{1}{2} \mathrm{tr} \left( \mathbf{C}^{-1}
\frac{\partial \mathbf{C}}{\partial C_\ell} \mathbf{C}^{-1}
\frac{\partial \mathbf{C}}{\partial C_{\ell^\prime}} \right) \, .
\ee

For unbiased power spectrum estimation in multipole bins $b$, we now
extend Eq.~(\ref{eq:oqe_signal_cov}) by introducing weight functions
$w_\ell$,
\be
\mathbf{S}_{n n^\prime} = \sum_{b} C_b \sum_{\ell \in b} \frac{2\ell +
  1}{4 \pi} \frac{1}{w_\ell} P_\ell(\hat{n} \cdot \hat{n}^\prime) \, ,
\ee
where the equality holds if $C_b = w_\ell C_\ell = \mathrm{constant}$ within each
bin. We therefore choose the weights $w_\ell \propto 1 / C_\ell$,
normalised to unity. The derivative with respect to individual
power spectrum elements in Eq.~(\ref{eq:oqe_estimation_matrix}) is
then simply replaced by $\partial / \partial C_b$. We note that to
compare power spectrum estimates to a theoretical model, the model has
to be binned with the same weight function $w_\ell$.

For a signal covariance matrix with a smaller number of Fourier modes
than pixels, Eq.~(\ref{eq:oqe_signal_cov}) will return a rank
deficient matrix. In case the same holds true for the noise covariance
matrix (e.g., for bandpass filtered noise), the inverse of the
covariance matrix does not exist. One way to solve this
problem is to restrict all computations to the non-singular subspace
of $\mathbf{C}$. Here, we adopt the simpler approach of
regularisation. We multiply the diagonal elements of the covariance
matrix, $\tilde{C}_{n n} = (1 + \epsilon) \cdot C_{n n}$, where
typically $\epsilon \approx 10^{-7}$. In the next section, we
demonstrate that our pipeline as described here produces reliable
results.

\subsection{Verification}

We generated 1000 realisations of a Gaussian random field at $\nside =
2048$ and $\lmax = 2500$ from the theoretical model of the SPT-E
galaxy $\times$ galaxy power spectrum. We added an isotropic
contribution of white noise at a level of $N_\ell = 2.1 \cdot 10^{-8}$
to the maps, roughly consistent with the observed level of shot noise
in this field.

For each simulated map, we computed auto power spectra at a downgraded
resolution of $\nside = 512$ using the SPT-E mask in 10 uniform
multipole bins up to the applied bandpass limit of $\lmax = 1500$. We
further calculated the $C_\ell$ Fisher matrix using its mathematically
exact analytical formula.

\begin{figure}
  \centerline{\resizebox{\hsize}{!}{\includegraphics*{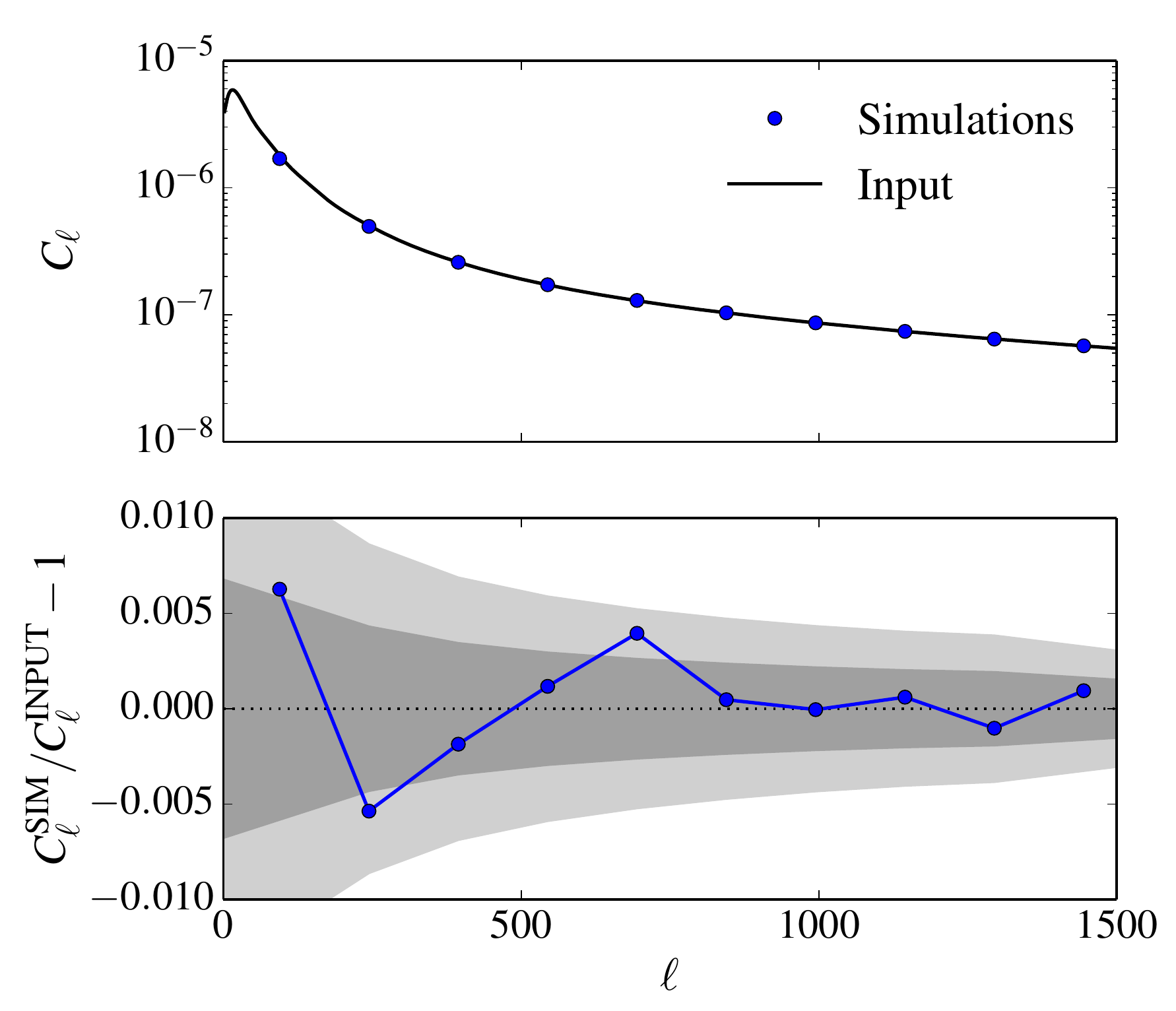}}}
  \caption{The optimal quadratic estimator is unbiased. We compare the
    power spectra estimates, averaged over all 1000 simulations, to
    the input power spectrum (\emph{top panel}). We also plot the
    residuals after dividing by the fiducial model (\emph{bottom
      panel}) together with the 1 and 2-$\sigma$ error on the mean
    (\emph{grey regions}), i.e., the error bars of an individual
    simulation divided by $\sqrt{1000}$. We find no evidence for a
    detectable bias, even in the highest multipole bins close to the
    resolution limit of a \Healpix~map at $\nside = 512$.}
  \label{fig:oqe_sim_test_spt_e}
\end{figure}

In Fig.~\ref{fig:oqe_sim_test_spt_e}, we show the results of our
comparison. We plot the averaged power spectrum estimates and compare
them to the inputs and find no evidence for a significant bias.

\bibliography{ms}
\section*{Affiliations}
\textit{$^1$Kavli Institute for Cosmology Cambridge, Institute of Astronomy, University of Cambridge, Madingley Road, Cambridge CB3 0HA, United Kingdom\\
$^2$Centre for Theoretical Cosmology, DAMTP, University of Cambridge, Wilberforce Road, Cambridge CB3 0WA, United Kingdom\\
$^3$Universit\"ats-Sternwarte, Fakult\"at f\"ur Physik, Ludwig-Maximilians Universit\"at M\"unchen, Scheinerstr. 1, D-81679 M\"unchen, Germany\\
$^4$Institut de Ci\`encies de l'Espai, IEEC-CSIC, Campus UAB, Facultat de
Ci\`encies, Torre C5 par-2, Barcelona E-08193, Spain\\
$^5$Astrophysics Group, Department of Physics and Astronomy, University College London, 132 Hampstead Road, London, NW1 2PS, United Kingdom\\
$^6$Department of Astronomy and Astrophysics, University of Chicago, Chicago, IL 60637, USA\\
$^7$Institute of Cosmology and Gravitation, University of Portsmouth, Dennis Sciama Building, Burnaby Road, Portsmouth, PO1 3FX, United Kingdom\\
$^8$Center for Cosmology and AstroParticle Physics, Ohio State University, 191 West Woodruff Avenue, Columbus, OH 43210, USA\\
$^9$Department of Physics, McGill University, 3600 rue University, Montreal, QC, H3A 2T8, Canada\\
$^{10}$Department of Physics \& Astronomy, University of Pennsylvania, 209 South 33rd Street, Philadelphia, PA 19104-6396, USA\\
$^{11}$Kavli Institute for Particle Astrophysics and Cosmology, Stanford University, Physics Astrophysics Building, 452 Lomita Mall, Stanford, CA 94305, USA\\
$^{12}$Kavli Institute for Cosmological Physics, 933 East 56th Street, Chicago, IL 60637, USA\\
$^{13}$School of Physics, University of Melbourne, Parkville, VIC 3010, Australia\\
$^{14}$Department of Astronomy, University of Illinois at Urbana-Champaign, MC-221, 1002 W. Green Street, Urbana, IL 61801, USA\\
$^{15}$Harvard Smithsonian Center for Astrophysics, 60 Garden St, MS 12, Cambridge, MA, 02138, USA\\
$^{16}$Fermi National Accelerator Laboratory, Batavia, IL 60510-0500, USA\\
$^{17}$Excellence Cluster Universe, Boltzmannstr. 2, D-85748 Garching bei M\"unchen, Germany\\
$^{18}$Max Planck Institute for Extraterrestrial Physics, Giessenbachstr., D-85748 Garching, Germany\\
$^{19}$Department of Physics, University of Michigan, Ann Arbor, MI 48109-1040, USA\\
$^{20}$Department of Physics, University of Chicago, 5640 South Ellis Avenue, Chicago, IL 60637, USA\\
$^{21}$Physics Department, Center for Education and Research in Cosmology and Astrophysics, Case Western Reserve University, Cleveland, OH 44106, USA
\\
$^{22}$Argonne National Laboratory, Argonne, IL 60439, USA\\
$^{23}$Cerro Tololo Inter-American Observatory, National Optical Astronomy Observatory, Casilla 603, La Serena, Chile\\
$^{24}$Department of Astrophysical Sciences, Princeton University, Peyton Hall, Princeton, NJ 08544, USA\\
$^{25}$Carnegie Observatories, 813 Santa Barbara St., Pasadena, CA 91101, USA\\
$^{26}$SLAC National Accelerator Laboratory, Menlo Park, CA 94025, USA\\
$^{27}$Laborat\'orio Interinstitucional de e-Astronomia - LIneA, Rua Gal. Jos\'e Cristino 77, Rio de Janeiro, RJ - 20921-400, Brazil\\
$^{28}$Observat\'orio Nacional, Rua Gal. Jos\'e Cristino 77, Rio de Janeiro, RJ - 20921-400, Brazil\\
$^{29}$National Center for Supercomputing Applications, 1205 West Clark St., Urbana, IL 61801, USA\\
$^{30}$George P. and Cynthia Woods Mitchell Institute for Fundamental Physics and Astronomy, and Department of Physics and Astronomy, Texas A\&M University, College Station, TX 77843,  USA\\
$^{31}$Jet Propulsion Laboratory, California Institute of Technology, 4800 Oak Grove Dr., Pasadena, CA 91109, USA\\
$^{32}$Institut de F\'{\i}sica d'Altes Energies, Universitat Aut\`onoma de Barcelona, E-08193 Bellaterra, Barcelona, Spain\\
$^{33}$Australian Astronomical Observatory, North Ryde, NSW 2113, Australia\\
$^{34}$Department of Astronomy, The Ohio State University, Columbus, OH 43210, USA\\
$^{35}$Department of Physics and Astronomy, Pevensey Building, University of Sussex, Brighton, BN1 9QH, UK\\
$^{36}$Centro de Investigaciones Energ\'eticas, Medioambientales y Tecnol\'ogicas (CIEMAT), Madrid, Spain\\
$^{37}$Department of Physics, University of Illinois, 1110 W. Green St., Urbana, IL 61801, USA\\
$^{38}$Jodrell Bank Center for Astrophysics, School of Physics and Astronomy, University of Manchester, Oxford Road, Manchester, M13 9PL, UK\\
$^{39}$CNRS, UMR 7095, Institut d'Astrophysique de Paris, F-75014, Paris, France\\
$^{40}$Sorbonne Universit\'es, UPMC Univ Paris 06, UMR 7095, Institut d'Astrophysique de Paris, F-75014, Paris, France\\
$^{41}$Department of Physics, The Ohio State University, Columbus, OH 43210, USA\\
$^{42}$Departamento de F\'{\i}sica Matem\'atica, Instituto de F\'{\i}sica, Universidade de S\~ao Paulo, CP 66318, CEP 05314-970 S\~ao Paulo, SP Brazil\\
$^{43}$Department of Physics, University of California, Berkeley, CA 94720, USA\\
$^{44}$Department of Physics and Electronics, Rhodes University, PO Box 94, Grahamstown, 6140, South Africa\\
}

\label{lastpage}

\end{document}